\documentclass[12pt]{iopart}
\usepackage[utf8]{inputenc}
\usepackage{graphicx}
\usepackage{xcolor}
\usepackage{soul}

\begin{document}

\title[]{Nonlinear MHD simulations of external kinks in quasi-axisymmetric stellarators using an axisymmetric external rotational transform approximation}

\author{R Ramasamy$^{1, 2}$, M Hoelzl$^1$,
E Strumberger$^1$ and K Lackner$^1$ and S G\"unter$^1$}
\address{$^1$ Max-Planck Institut f\"ur Plasmaphysik, Boltzmannstraße 2, 85748 Garching bei München}
\address{$^2$ Max-Planck Princeton Center for Plasma Physics, Princeton, New Jersey 08544 USA} 
\ead{rohan.ramasamy@ipp.mpg.de}

\begin{abstract}
    Reduced magnetohydrodynamic (MHD) equations are used to study the nonlinear dynamics of external kinks in a quasi-axisymmetric (QA) stellarator with varying fractions of external rotational transform. The large bootstrap currents associated with high beta plasmas \textcolor{black}{may make} QA configurations susceptible to low n external modes, limiting their operational space. The violence of the nonlinear dynamics, and, in particular, when these modes lead to a disruption, is not yet understood. In this paper, the nonlinear phase of external kinks in an unstable QA configuration with an edge safety factor below two is simulated. An axisymmetric approximation of this stellarator is constructed in the nonlinear MHD code, JOREK, capturing the influence of the external rotational transform. \textcolor{black}{The use of this approximation for the considered stellarator is validated by comparing the linear dynamics against the linear viscoresistive MHD code, CASTOR3D. The nonlinear dynamics of this stellarator approximation are compared with an equivalent tokamak to understand the influence of a relatively small external rotational transform. While the external rotational transform does have a stabilising influence on the MHD activity, it remains violent. To explore the first order influence of a larger external rotational transform, this equilibrium parameter is artificially increased for the considered stellarator, reducing the effective plasma current.} The violence of the kink instability is quantified, and shown to reduce with the increasing external rotational transform. At the same time, the external kink triggers internal modes that exacerbate the loss in confinement during the nonlinear phase, such that it remains large over much of the parameter space. It is only with a significant fraction of external rotational transform that these subsequent modes are stabilised.  
\end{abstract}


\submitto{\NF}
\maketitle


\section{Introduction}
In recent years, there have been ongoing efforts to produce a compact, quasi-axisymmetric (QA) stellarator that can be scaled up to a fusion reactor \cite{henneberg2019}. An important question that remains unanswered in this work is the susceptibility of such devices to disruptions. 

Previous studies of 3D equilibria with Ohmic current profiles have shown that stellarators can be considered vertically stable, and free from tearing mode driven disruptions when a sufficiently large external rotational transform\textcolor{black}{, $\iota_{ext}$,} is applied \cite{team1980, weller2001, archmiller2014}. However the potential instabilities for QA configurations are different, and at the time of writing, no experimental QA devices have been constructed to provide insight into their nonlinear stability properties. For this reason, the current understanding of these configurations relies on existing theory and numerical codes. 

Linear ideal magnetohydrodynamic (MHD) codes have been applied in the optimisation of QA stellarators \cite{fu2007ideal, fu2000magnetohydrodynamics}. When these devices make use of a significant bootstrap current to maintain the equilibrium conditions, they are typically susceptible to similar MHD instabilities as ITER advanced operational scenarios \cite{hender2007}. Such candidate instabilities include infernal modes, double tearing modes, and external kinks. These instabilities will be milder than those in tokamaks, which do not have the stabilising influence of an external rotational transform, magnetic well, and 3D shaping, but the violence of the nonlinear dynamics that they can drive is still unclear.

The linear, viscoresistive MHD code, CASTOR3D \cite{Strumberger2016, Strumberger2019}, was recently applied to a QA configuration with an edge safety factor below two \cite{Strumberger2019}. As such, the stellarator was known to be unstable to violent kink instabilities. CASTOR3D provides a detailed analysis of these instabilities, characterising the mode structure, and the influence of several stabilising effects, such as resistive walls. While these studies give a detailed description of the linear dynamics, there are remaining questions regarding the nonlinear dynamics. In particular, it is unclear at what amplitude the observed kink instabilities will saturate, and whether they can trigger internal modes, leading to a disruption, as has been observed in kink driven disruptions in tokamaks \cite{buratti2012}.

Nonlinear, stellarator capable, MHD codes are necessary to answer these questions with a high fidelity. There are presently only a few nonlinear codes that can be applied to stellarator geometries, such as MEGA \cite{todo2017comprehensive} and FLUXO \cite{hindenlang2017parallel}. The nonlinear MHD code, JOREK \cite{hoelzl2020jorek}, is currently being extended to model stellarators. A hierarchy of stellarator capable models have been derived \cite{nikulsin2019three, nikulsin2021}, and are currently being validated. Assessments of the nonlinear dynamics of global current driven instabilities require an evolution of the magnetic field in time. For advanced stellarators, where the magnetic field description requires many toroidal harmonics, simulations are expected to be computationally demanding.

As such, it follows to consider supplementing the high fidelity models being developed in JOREK with simplified simulations of the nonlinear dynamics of stellarators. A common simplification to choose is to modify configurations such that they can be modeled axisymmetrically. Axisymmetric approximations of stellarators have a long history, beginning with the first stellarator equilibrium expansion \cite{greene1961determination}. The ordering used in this expansion is similar to the high beta tokamak expansion, with the inclusion of a large helical field to represent the external rotational transform, leading to a modified Grad Shafranov equation for the equilibrium \cite{freidberg1987ideal}. Soon after the derivation of this expansion, it was applied to linear MHD stability \cite{greene1962equilibrium}. In the 1970s, common tools for assessing tokamak stability, such as the tearing mode equation, and reduced MHD models were adapted to look at stellarator instabilities, both linearly and nonlinearly \cite{matsuoka1977, wakatani1978,  Wakatani1983}. Such models have provided insight into MHD phenomena experimentally observed on W7-A and W7-AS, as well as being applied to predict the stability properties for NCSX \cite{team1980, weller2001, fredrickson2007tearing}.

This paper outlines an implementation of the axisymmetric stellarator approximation in JOREK,  called the virtual current model. The approach makes use of current equilibrium codes to provide an appropriate description of the external rotational transform, by representing it as a \textit{virtual current} inside the plasma that is parallel to the total equilibrium magnetic field. \textcolor{black}{This virtual current is not carried by the plasma itself, and so does not contribute to the current drive of MHD instabilities.} Past studies using a similar approach have attributed this additional current to a lower hybrid current drive \cite{fredrickson2007tearing}. Herein, we prefer to use the term virtual current to emphasise that this current is not a true plasma current, but a model to incorporate the effect of 3D coils on the rotational transform axisymmetrically. 

The virtual current model in JOREK has already been validated for simple tearing mode, and vertical displacement event test cases, showing the expected stabilisation of both instabilities \cite{Ramasamy2018Stell}. Linear growth rates of double tearing modes and resistive kinks in Wendelstein 7-X (W7-X) have been compared against CASTOR3D, using different axisymmetric approximations. The results of this validation are shown in the Appendix. 


In this study, the virtual current model is applied to external kinks in a compact, high beta QA configuration. These instabilities prove to be more challenging for the model to capture, due to the global nature of the modes, and the influence of 3D shaping and the magnetic well, which cannot be captured axisymmetrically. It is shown that there is reasonable agreement between the linear growth rates and eigenfunctions observed in the virtual current model, and CASTOR3D, for the equilibria considered. \textcolor{black}{Simulations are then continued into the nonlinear phase, comparing the results of the stellarator with an \textit{equivalent tokamak}, in a similar way as carried out in \cite{Strumberger2019} for the linear phase. The equivalent tokamak has a similar pressure and safety factor profile as the original stellarator, but with a larger plasma current, $I_p=11.5\ MA$, leading to stronger MHD activity. It is shown that while the nonlinear dynamics of the stellarator are milder than the equivalent tokamak, they remain violent, with a large plasma current spike, a characteristic feature of disruptive MHD activity.}

\textcolor{black}{For the simulated QAS, $\iota_{ext}$ is $\sim0.15$. This is a relatively small fraction of the overall rotational transform, as the edge $\iota$ is just above 0.5. As such, the plasma current of the virtual current approximation of the QA configuration remains high at $8.25\ MA$. To gain a first order understanding of how the nonlinear dynamics are influenced by larger fractions of external rotational transform, this equilibrium parameter is artificially increased for the simulated equilibrium. This is achieved by increasing the virtual current, while maintaining the equilibrium profile of the total current within the torus. In such a way, the safety factor profile of the original QA configuration is maintained, while $I_p$ decreases, reducing the drive for current driven MHD instabilities, until the equilibrium becomes stable at $I_p\approx2.75\ MA$.}

\textcolor{black}{The additional cases produced by this method are conceptual, and do not have a direct correspondence to a real stellarator equilibrium. They are only used to gain an initial understanding of how the nonlinear dynamics might evolve for a stellarator with a larger external rotational transform than the stellarator that has been used as the basis for this study.} Although the external kink is stabilised by increasing $\iota_{ext}$, internal modes are triggered over much of the parameter space, leading to a loss of confinement across the middle and outer region of the plasma. \textcolor{black}{While it is acknowledged that additional non-axisymmetric effects such as the magnetic well and 3D shaping can modify the dynamics, the results give a first indication of the potential dynamics that might occur in QA configurations that can be used to inform more detailed, future studies.}



The rest of the paper is arranged as follows. In Section \ref{sec:virt_model} the implementation of the virtual current model in JOREK is described. The equilibria considered for this study are discussed in Section \ref{sec:equilibrium}. The virtual current model is validated in the linear regime for external kink cases in Section \ref{sec:linear_validation}. The nonlinear dynamics of these external kinks is then assessed in Section \ref{sec:nonlinear_study}, and the paper concluded in Section \ref{sec:conclusion}.

\section{The virtual current model in JOREK} \label{sec:virt_model}

JOREK is a fully implicit, 3D, nonlinear extended MHD code, capable of being run with reduced and full MHD models of varying complexity \cite{hoelzl2020jorek}. In cylindrical coordinates, the poloidal planes are discretised using bi-cubic $C^1$ Bezier finite elements, with a real Fourier series in the toroidal direction \cite{czarny2008bezier}. A Crank-Nicholson or Gears time stepping scheme is used to advance the physics variables, leading to a large sparse matrix that is iteratively solved using GMRES \cite{saad1986gmres}. The solution to the block diagonal matrices formed by individual toroidal harmonics is typically solved using PASTIX \cite{henon2002pastix}, or STRUMPACK \cite{ghysels2017}, and used as a pre-conditioner for the solution to the full system of equations.

\textcolor{black}{The application of the virtual current model in JOREK requires a method for constructing an axisymmetric equilibrium that approximates the original stellarator, and a modification of the time evolution equations to account for virtual currents. The approach used in this study is outlined in the following section.}

\subsection{Equilibrium construction} \label{sec:equil_construct}
\begin{figure*}
    \centering
    \includegraphics[width=\textwidth]{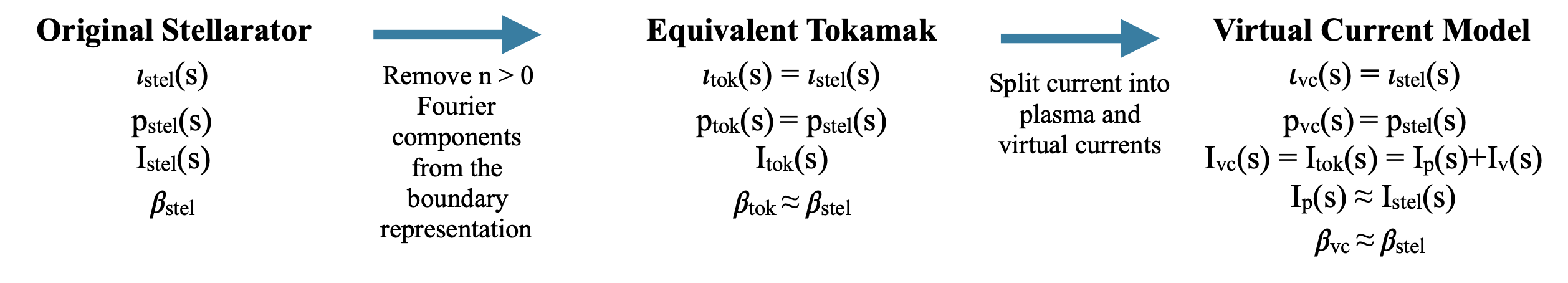}
    
    \scriptsize{(a) Equivalent tokamak and virtual current model construction for internal modes}
    \vspace{0.25cm}
    
    \includegraphics[width=\textwidth]{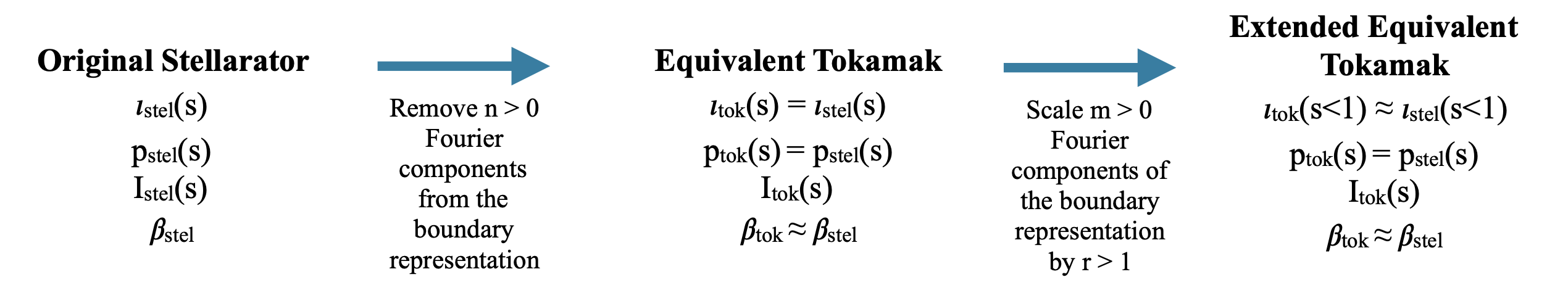}
    
    \scriptsize{(b) Equivalent tokamak construction for external modes in JOREK}
    \vspace{0.25cm}
    
    \includegraphics[width=\textwidth]{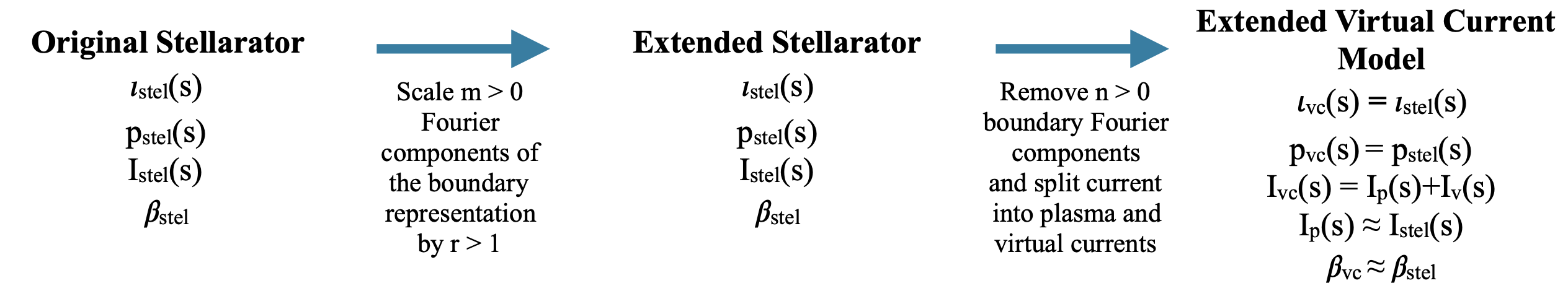}
    
    \scriptsize{(c) Virtual current model construction for external modes in JOREK}
    \caption{\textcolor{black}{Workflow for constructing the equivalent tokamak and virtual current model of a stellarator using VMEC for internal (a) and external (b-c) mode studies in JOREK. There are two main steps in the approach. The first is to modify the definition of the plasma boundary to be axisymmetric, preserving the pressure, and $\iota$ profiles of the original stellarator. This leads to an equivalent tokamak approximation. The virtual current model is constructed by splitting the toroidal current in the new axisymmetric equilibrium into plasma and virtual current components. For external modes, an additional step is necessary to extend the equilibrium into the vacuum region. This is necessary because of the JOREK boundary conditions, and to capture the correct rotational transform outside the plasma in the virtual current model.}}
    \label{fig:vmec_workflow}
\end{figure*}

Modern axisymmetric approximations of stellarators normally take a numerical approach, which represents the magnetic field corresponding to the external rotational transform as a toroidal current \cite{Strumberger2019, fredrickson2007tearing, yu2020numerical}. As described in detail in \cite{Strumberger2019}, this can be achieved by mapping the stellarator equilibrium onto an axisymmetric domain, using the Variational Moments Equilibrium Code (VMEC) \cite{hirshman1983steepest}. The main steps in the procedure are shown in Figure \ref{fig:vmec_workflow} (a) for simulating internal modes. The $n=0$ Fourier components of the stellarator boundary are used to generate an axisymmetric equilibrium with the same pressure and rotational transform profiles as the original stellaror, $p_{stel}$ and $\iota_{stel}$, respectively. The poloidal field generated externally is then replaced by an increase in the toroidal current in the new axisymmetric device, $I_{tok}$, when compared with the original stellarator, $I_{stel}$. 

In \cite{Strumberger2019}, the enhanced destabilising current density gradients of this approximation led to more unstable kink modes with similar structure. This type of axisymmetric approximation is refered to as the \textit{equivalent tokamak} approximation. In the following study, to replicate the physics of an external rotational transform, the additional current in the equivalent tokamak is removed from the plasma, as virtual currents, leading to a model refered to as the \textit{virtual current} approximation\textcolor{black}{, where}

\begin{equation} \label{eq:I_vc_split}
    I_{vc}(s) = I_p(s) + I_v(s),
\end{equation}

\textcolor{black}{defining $I_p$ and $I_v$ as the plasma and virtual current within a given toroidal configuration, respectively, and $s$ is used in equations as the normalised toroidal flux, $\hat \Phi$.}

In order to \textcolor{black}{split $I_{vc}$ between $I_p$ and $I_v$} in JOREK, the solution to the Grad Shafranov equilibrium equation is modified, such that

\begin{eqnarray}
    \eqalign{
    \Delta^* \psi_t  = j_{vc}(\psi_t,R) &= - \mu_0 R ^ 2 p'\left(\psi_t\right) - FF'\left(\psi_t\right) \\
                    &= - \mu_0 R ^ 2 p'\left(\psi_t\right) - \left[F_pF'_p\left(\psi_t\right) + F_vF'_v\left(\psi_t\right)\right],
    }
\end{eqnarray}

\textcolor{black}{where $j_{vc}(\psi_t,R)$ describes the total current density in the virtual current approximation}. The $\Delta^*$ operator is defined in cylindrical coordinates by

\begin{eqnarray}
    \eqalign{
    \Delta^* \psi_t &= R^2 \nabla \cdot \left( \frac{\nabla{\psi_t}}{R^2}  \right) = R\frac{\partial}{\partial R}\left(\frac{1}{R}\frac{\partial \psi_t}{\partial R} \right) + \frac{\partial^2 \psi_t}{\partial Z^2}.
    }
\end{eqnarray}

\textcolor{black}{and the plasma and virtual current densities are defined by}

\begin{equation}
    \mu_0 R \mathbf{J_p}\cdot \hat e_\phi = j_p = - \mu_0 R ^ 2 p'\left(\psi_t\right) - F_pF'_p\left(\psi_t\right)
\end{equation}

\begin{equation}
    \mu_0 R \mathbf{J_v}\cdot \hat e_\phi = j_v = - F_vF'_v\left(\psi_t\right).
\end{equation}

\begin{figure}
\centering
    \includegraphics[width=0.45\textwidth]{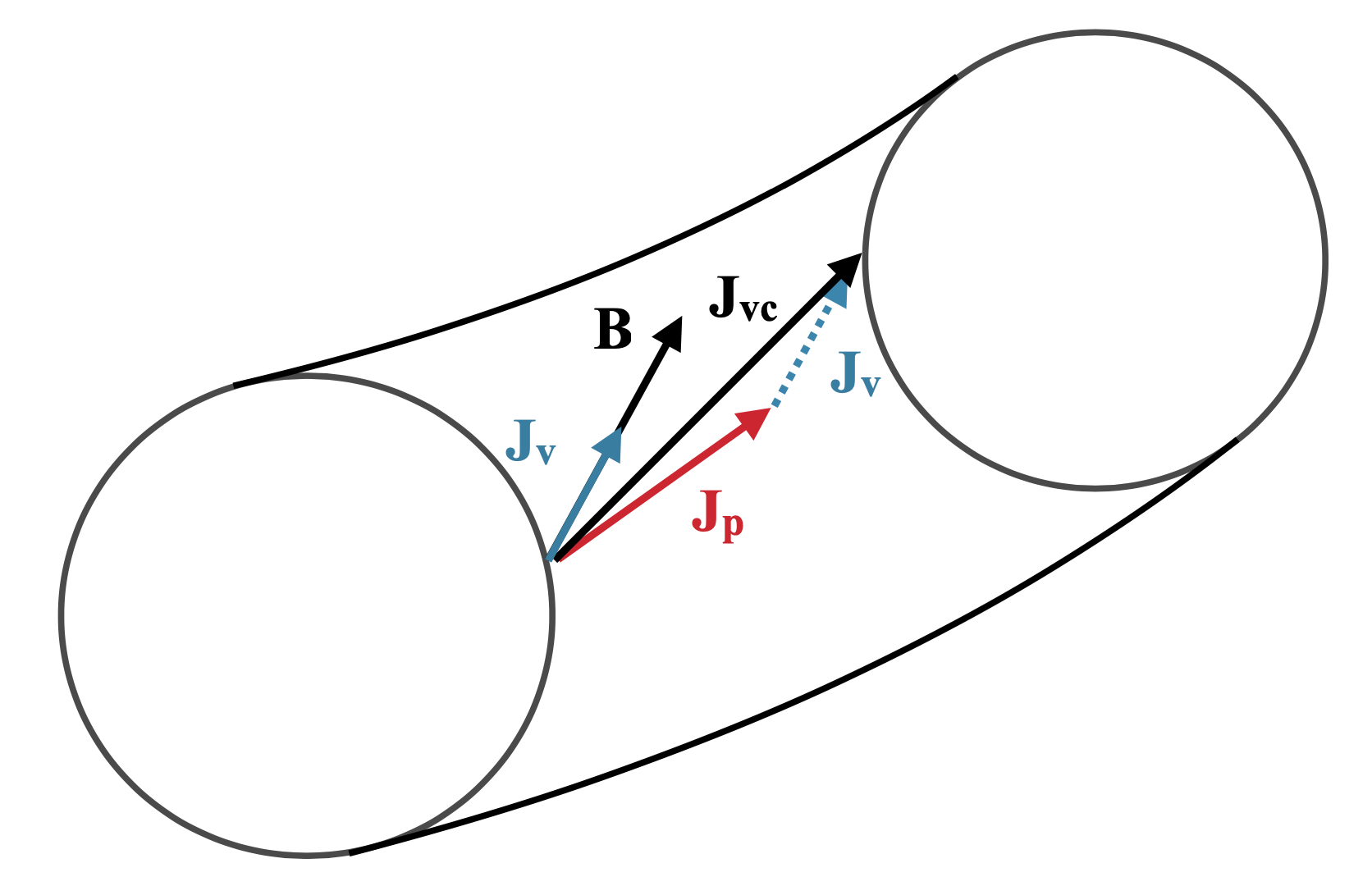}
\caption{\textcolor{black}{Sketch of virtual currents in a circular axisymmetric approximation of a stellarator. The total current density, $\mathbf{J_{vc}}$, has contributions from virtual, $\mathbf{J_v}$, and plasma, $\mathbf{J_p}$, currents. The virtual current is along the total equilibrium magnetic field, $\mathbf{B}$, and as such does not modify the solution to the Grad Shafranov equation.}}
\label{fig:virt_curr_model}
\end{figure}

These modifications to the Grad Shafranov equation do not violate the equilibrium force balance. Once the solution for $\psi_t$ is found, the current is split into plasma and virtual current components, both of which remain functions of $\psi_t$. As shown in Figure \ref{fig:virt_curr_model}, the diamagnetic current represented by $FF'$ is parallel to the total equilibrium magnetic field, and as such the virtual current removed from this term can be considered force free.

In order to be a meaningful approximation of the stellarator, \textcolor{black}{the plasma current density of the virtual current model should correspond to the plasma current density of the original stellarator. This can be achieved by calculating $j_v$ - a flux surface quantity - such that}

\begin{equation} \label{eq:virt_curr_def}
    j_v = \mu_0 \langle R\rangle \left[\langle \frac{j_{vc}(\psi_t,R)}{R}\rangle - \frac{dI_{stel}}{dA_{vc}}\left(\psi_t\right) \right],
\end{equation} 

\textcolor{black}{where $\langle\ \rangle$ refers to a flux surface averaged quantity, and $A$ is the toroidally averaged cross sectional area of a given flux surface. To first order in the aspect ratio, this ensures that the stellarator and the virtual current model correspond to the same plasma current distribution, $I_p(s)=I_{stel}(s)\left(1+O\left(\epsilon^2\right)\right)$.} The righthand side of the above equation can be computed numerically from VMEC and JOREK. \textcolor{black}{Equation \ref{eq:virt_curr_def} can be used to define $I_v(s)$ in equation \ref{eq:I_vc_split}, using}

\begin{equation}
    I_v(s) = \int_0^{A_{vc}(s)} \frac{j_v}{R} dA_{vc}.
\end{equation}

As shown in Section \ref{sec:equilibrium}, the toroidal current profile of the QAS considered in this study is approximately preserved with this method, deviating only by $2.5\ \%$.

\textcolor{black}{The workflow used to construct the axisymmetric approximations for the linear validation in Section \ref{sec:linear_validation} is outlined in Figure \ref{fig:vmec_workflow} (b) and (c). An additional step in the equilibrium construction is required for the study of external modes in JOREK as the vacuum region needs to be included in the simulation domain. There are two reasons for this. The first is that JOREK applies a Dirichlet boundary condition on the poloidal velocity at the boundary. This boundary condition needs to be sufficiently far from the plasma-vacuum interface to not interfere with the growth of external modes.}

\textcolor{black}{Secondly, the vacuum magnetic shear of a tokamak and stellarator are different, due to the contribution of the helical coils. To capture the rotational transform of the stellarator axisymmetrically, virtual currents need to flow in the vacuum region to approximate the effect of these helical coils. This is important as the magnetic shear of the \textcolor{black}{vacuum} has a stabilising effect on the external kink. }

For these preliminary studies, the original fixed boundary computations are extended smoothly into the vacuum region, linearly scaling the edge toroidal flux with the new area of the extended domain. This provides an approximate description of the stellarator equilibrium, which is then transformed to a new axisymmetric approximation in the same way as has been previously described. This extended domain is only used in JOREK. Numerical difficulties were encountered when solutions of extended equilibria into the vacuum region were used in CASTOR3D. For this reason, the 3D equilibria are only modeled in CASTOR3D up to the plasma boundary, coupling the vacuum and MHD equations at this interface.

\subsection{Time evolution}
During the time evolution, the virtual current is fixed in space and time. JOREK's single fluid, reduced MHD equations \cite{franck2015energy} are used, removing the parallel momentum equation to improve numerical stability. \textcolor{black}{The nonlinear dynamics of the simulated instabilities are dominated by particle transport due to perpendicular convection of the kink motion, and parallel heat transport along magnetic field lines. As the particle transport along magnetic field lines occurs on the time scale of the ion sound speed, it is much slower than the parallel heat transport. For this reason, only limited differences would be expected by including the parallel momentum equation. This is also supported by the fact that particle losses in somewhat comparable edge localized mode crashes are usually strongly dominated by perpendicular convection \cite{huijsmans2013non}}.

The total poloidal flux, electric potential, density and temperature, are evolved in time, subject to the governing equations

\begin{eqnarray} \label{eq:induction_strong}
    \frac{\partial \psi_t}{\partial t} = R[\psi_t, u] + \eta \left(\textcolor{blue}{j_{vc} - j_v}\right) - F_0 \frac{\partial u}{\partial \phi}
\end{eqnarray}

\begin{eqnarray} \label{eq:density_strong}
    \frac{\partial {\rho}}{\partial t} = -\nabla \cdot \left( \rho \mathbf{v} \right) + \nabla \cdot \left(D_{\bot}\nabla_{\bot}\rho + D_{\parallel}\nabla_{\parallel}\rho \right) + S_\rho
\end{eqnarray}

\begin{eqnarray} \label{eq:temperature_strong}
    \eqalign{
    \frac{\partial \left(\rho T\right)}{\partial t} &= -\mathbf{v}\cdot\nabla\left(\rho T\right) - \gamma \rho T \nabla \cdot \mathbf{v} \\
    &+ \nabla \cdot\left(\kappa_{\bot}\nabla_{\bot}T + \kappa_{\parallel}\nabla_{\parallel}T \right) \\
    &+ \frac{2}{3 R^2} \eta\left(T\right) \left(\textcolor{blue}{j_{vc}-j_v}\right)^2 + S_p 
    }
\end{eqnarray}

\begin{eqnarray} \label{eq:momentum_strong}
    \eqalign{
    R \nabla \cdot \left(R^2 \rho \nabla_{\bot}\frac{\partial u}{\partial t}\right) &= \frac{1}{2} \left[ R^2 \left|\nabla_\bot u\right|^2, R^2\rho \right] \\
    &+ \left[ R^4 \rho \omega, u \right] + \left[ \psi_t, \textcolor{blue}{j_{vc}-j_v}\right] \\
    &- \frac{F_0}{R}\frac{\partial j_{vc}}{\partial \phi} + \left[\rho T, R^2\right] \\
    &+ R \mu\left(T\right) \nabla^2\omega 
    }
\end{eqnarray}

\begin{eqnarray} \label{eq:current_strong}
    j_{vc} = \Delta^* \psi_t = \textcolor{blue}{j_p + j_v}
\end{eqnarray}

\begin{eqnarray}
    \omega = \frac{1}{R}\frac{\partial}{\partial R}\left( R \frac{\partial u}{\partial R} \right) + \frac{\partial^2 u}{\partial Z^2}
\end{eqnarray}


where the Poisson bracket, $\left[a,b\right]=\nabla \phi\cdot\left( \nabla a \times \nabla b \right)$, has been used, and the reduced MHD ansatz is assumed for the plasma velocity, $\mathbf{v}$, and magnetic field, $\mathbf{B}$

\begin{eqnarray} \label{eq:field_ansatz}
    \mathbf{B} = \nabla \psi_t \times \nabla \phi + F_0 \nabla \phi
\end{eqnarray}

\begin{eqnarray}
    \mathbf{v} = -\frac{R^2}{F_0}\nabla u \times \nabla \phi
\end{eqnarray}

where $\psi_t$ is the total poloidal flux. $F_0=R B_\phi$ represents the toroidal field, which is fixed in time. The necessary modifications for the virtual current model are marked in blue, and amount to the inclusion of a virtual current, which is subtracted from the terms responsible for releasing free magnetic energy. The change to equation \ref{eq:induction_strong} is equivalent to the addition of a current source, fixing the background virtual current. In equations \ref{eq:temperature_strong} and \ref{eq:momentum_strong}, the virtual current is removed from the Ohmic heating term and Lorentz force, as the plasma should only experience the influence of its own current. 


As external kinks are modeled in this study in the no wall limit, keeping the $n=0$ component fixed, the influence of virtual currents on the boundary conditions also needs to be considered. The free boundary condition for $n>0$ toroidal harmonics is included by coupling JOREK to the STARWALL code \cite{merkel2015linear, hoelzl2012coupling}. This coupling enforces that

\begin{eqnarray} \label{eq:starwall_coupling_B_norm}
    \mathbf{n}\cdot\mathbf{B}_\mathrm{JOREK}=\mathbf{n}\cdot\mathbf{B}_\mathrm{STARWALL}
\end{eqnarray}

\begin{eqnarray} \label{eq:starwall_coupling_B_tan}
    \mathbf{n} \times \mathbf{B}_{\mathrm{JOREK}}=\mathbf{n} \times \mathbf{B_{\mathrm{STARWALL}}}
\end{eqnarray}

\begin{eqnarray} \label{eq:starwall_coupling_J_norm}
    \mathbf{n}\cdot\mathbf{J}_{\mathrm{JOREK}}=\mathbf{n}\cdot\mathbf{J}_{\mathrm{STARWALL}}
\end{eqnarray}

on the JOREK simulation boundary. Because the virtual current is a function of the initial $\psi_t$, and the simulation boundary is initially a flux surface, equation (\ref{eq:starwall_coupling_J_norm}) is independent of the virtual currents in the domain.
In JOREK, the boundary conditions are implemented in terms of the total magnetic field \cite{artola2018free}, and so there is no need to change the coupling between the two codes. 

As shown in Section \ref{sec:equilibrium}, it is important to note that the correct rotational transform profile of a stellarator in the vacuum region can only be achieved axisymmetrically by having virtual currents in the vacuum region, to represent the external rotational transform. As the STARWALL coupling assumes that there is no current in the vacuum region outside the simulation domain, the rotational transform in this region will not match the modeled stellarator. This limitation is not considered to be significant, because the simulation domain is already extended beyond the plasma boundary, as discussed in Section \ref{sec:equil_construct}. Therefore, the correct external rotational transform is captured in the region of interest.

The changes outlined above are simple to implement, and can just as easily be applied to more advanced models, allowing for a fast, albeit qualitative assessment of stellarator nonlinear dynamics that are intended to be compared against more accurate simulations including the full geometry, once such models are available.

\textcolor{black}{Before moving forward with the reduced MHD model outlined above, its application to the external modes considered in this paper needs to be justified. In a recent study, the reduced MHD model in JOREK has been compared against full MHD models in JOREK and CASTOR3D for a variety of benchmarks \cite{pamela2020extended}. Therein, it is shown that the reduced MHD model is in reasonable agreement with full MHD models for most cases. Currently, the main limitation of the reduced MHD model that has been identified is finite beta internal kinks, which are not relevant for the current study. These results, and the linear validation carried out in Section \ref{sec:linear_validation}, justify the use of the reduced model.}

\section{Equilibria of interest} \label{sec:equilibrium}
\begin{figure*}
    \centering
    \begin{minipage}{0.325\textwidth}
      \includegraphics[width=0.98\textwidth]{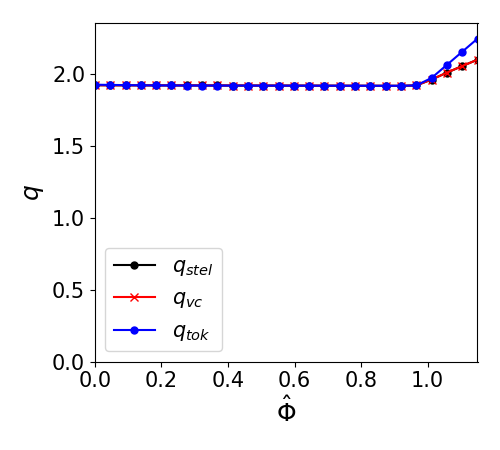}
      \centering
      \scriptsize(a)
    \end{minipage}
    \begin{minipage}{0.325\textwidth}
      \includegraphics[width=0.98\textwidth]{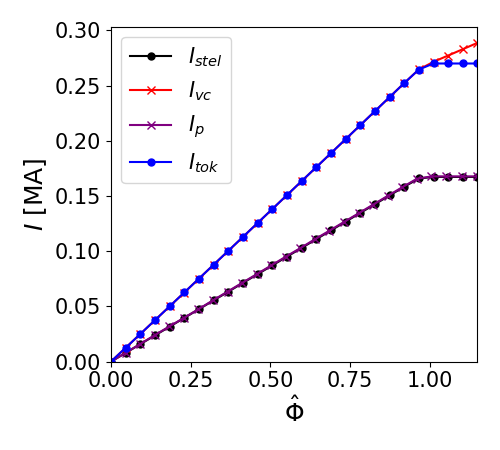}
      \centering
      \scriptsize{(b)}
    \end{minipage}
    \begin{minipage}{0.325\textwidth}
      \includegraphics[width=0.98\textwidth]{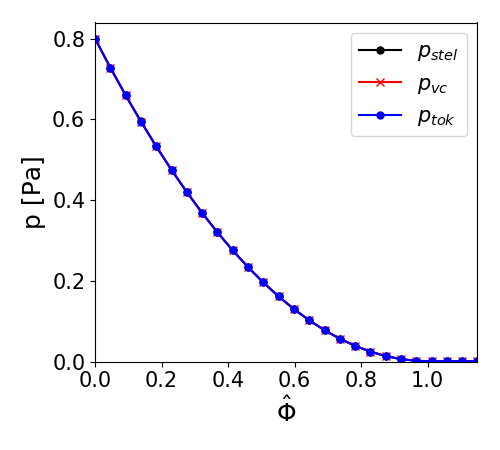}
      \centering
      \scriptsize{(c)}
    \end{minipage}

    \centering
    \begin{minipage}{0.325\textwidth}
      \includegraphics[width=0.98\textwidth]{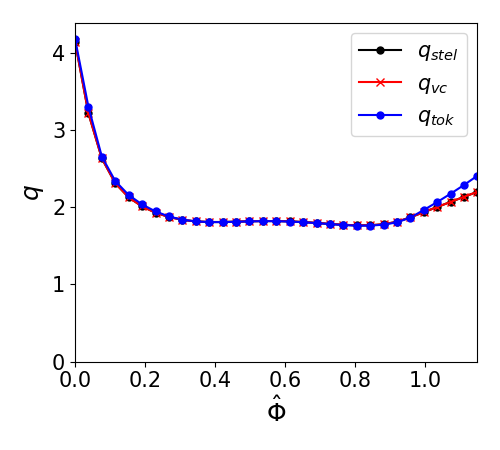}
      \centering
      \scriptsize(d)
    \end{minipage}
    \begin{minipage}{0.325\textwidth}
      \includegraphics[width=0.98\textwidth]{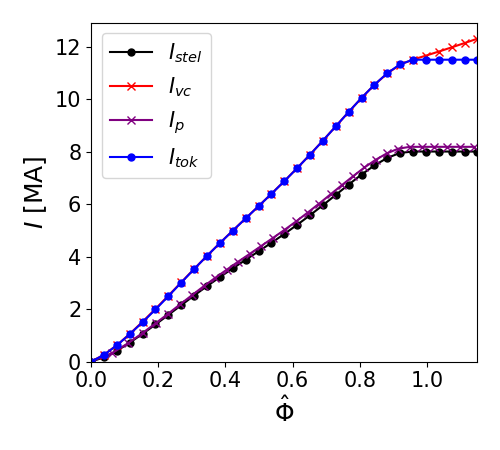}
      \centering
      \scriptsize{(e)}
    \end{minipage}
    \begin{minipage}{0.325\textwidth}
      \includegraphics[width=0.98\textwidth]{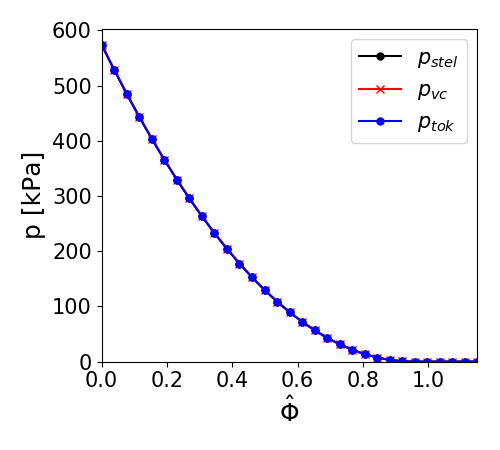}
      \centering
      \scriptsize{(f)}
    \end{minipage}
    \caption{Equilibrium profiles for the $l=2$ stellarator (a-c) and QA device (d-f) test cases, and their axisymmetric approximations. The normalised toroidal flux, $\hat \Phi$, is used as a radial coordinate. The equivalent tokamak (blue) has a larger total toroidal current and vacuum magnetic shear than the stellarator (black). The virtual current approximation (red) has the same pressure, and safety factor profile as the stellarator. \textcolor{black}{$I_{vc}$ continues to increase in the vacuum region, beyond $\hat \Phi=1$. \textcolor{black}{Due to the approximations made in Section \ref{sec:equil_construct},} $I_p$ is only approximately equal to $I_{stel}$. For the QA device, the maximum deviation is less than $2.5\ \%$.}}
    \label{fig:equilibrium_profiles}
\end{figure*}

Two stellarator equilibria are modeled in this study. A simple $l=2$ stellarator is considered in the linear validation, alongside the main QA stellarator that is continued into the nonlinear phase. Their profiles are shown in Figure \ref{fig:equilibrium_profiles}, alongside the profiles of their corresponding axisymmetric approximations.

The $l=2$ stellarator was designed as a simple linear test case for external kinks, to validate the virtual current model in a high aspect ratio, low beta stellarator. A flat current density profile is used, such that the case would be strongly unstable to external kinks. The QA configuration is a slightly modified version of the two field period equilibrium previously modeled in \cite{Strumberger2019}. \textcolor{black}{In particular, within the plasma region, the pressure profile was modified to be}

\begin{equation}
    p(s)=\cases{(p_0 - p_{vac}) (1 - 2s + s^2) + p_{vac} &for $s \le 1$\\
                p_{vac}                                  &for $s>1$\\}
\end{equation}

\textcolor{black}{where $p_0$ and $p_{vac}$ are the pressure on axis and in the vacuum region, respectively.} In such a way, the gradients of density and temperature at the plasma edge smoothly transition to the vacuum conditions. The final safety factor profile has also been modified such that the edge safety factor is closer to the $q=2$ rational surface. To obtain a reasonable core temperature of approximately $10\ keV$, the core number density is set to $3.295\times10^{20}\ m^{-3}$. The form of both the density and temperature profiles used are proportional to the square root of the pressure profile. 

Comparing the extended equilibrium stellarator profiles with their tokamak and virtual current approximations, as shown in Figure \ref{fig:equilibrium_profiles}, the pressure profile is consistent between the three cases. The safety factor profile is approximately the same for the stellarator and its equivalent tokamak approximation within the plasma. In the vacuum region, it can be seen that the magnetic shear in the equivalent tokamak is larger than the stellarator, for the reasons discussed in Section \ref{sec:virt_model}. The safety factor profile of the virtual current approximation remains identical to the stellarator across the entire simulated domain, as a result of the virtual current flowing in the vacuum region to capture the field generated from external coils.

\begin{figure*}
    \centering
    \begin{minipage}{0.2425\textwidth}
      \includegraphics[width=0.98\textwidth]{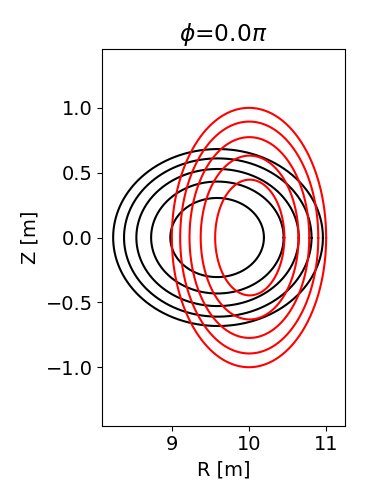}
      \centering
      \scriptsize(a)
    \end{minipage}
    \begin{minipage}{0.2425\textwidth}
      \includegraphics[width=0.98\textwidth]{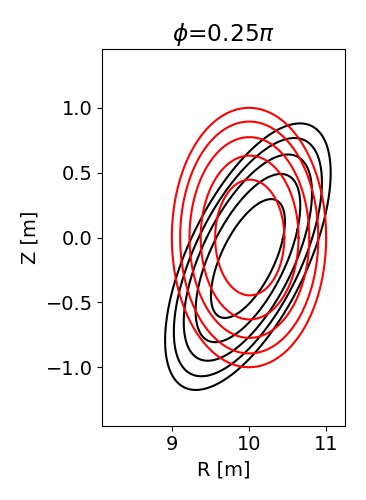}
      \centering
      \scriptsize{(b)}
    \end{minipage}
    \begin{minipage}{0.2425\textwidth}
      \includegraphics[width=0.98\textwidth]{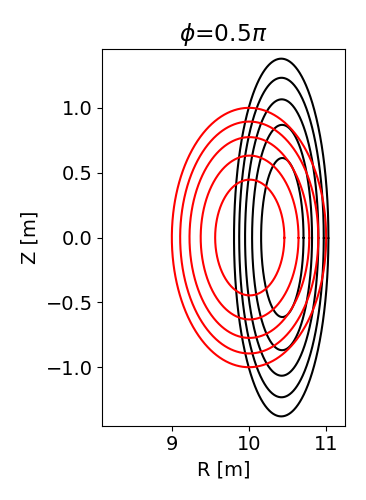}
      \centering
      \scriptsize{(c)}
    \end{minipage}
    \begin{minipage}{0.2425\textwidth}
      \includegraphics[width=0.98\textwidth]{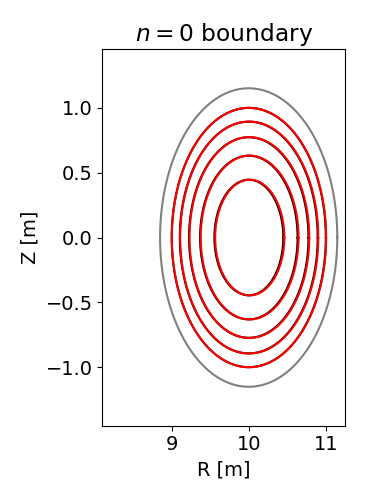}
      \centering
      \scriptsize{(d)}
    \end{minipage}
    
    \begin{minipage}{0.2425\textwidth}
      \includegraphics[width=0.98\textwidth]{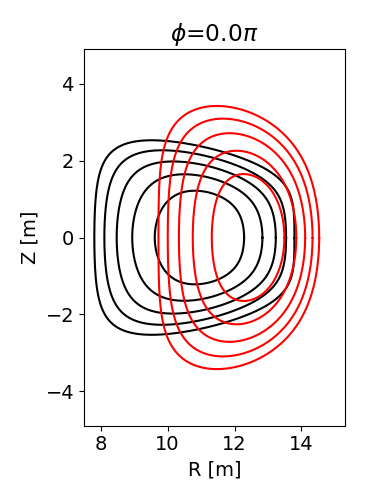}
      \centering
      \scriptsize(e)
    \end{minipage}
    \begin{minipage}{0.2425\textwidth}
      \includegraphics[width=0.98\textwidth]{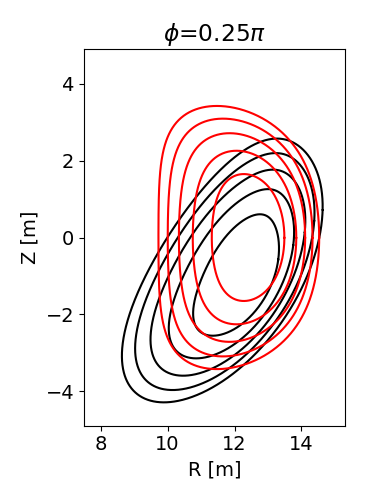}
      \centering
      \scriptsize{(f)}
    \end{minipage}
    \begin{minipage}{0.2425\textwidth}
      \includegraphics[width=0.98\textwidth]{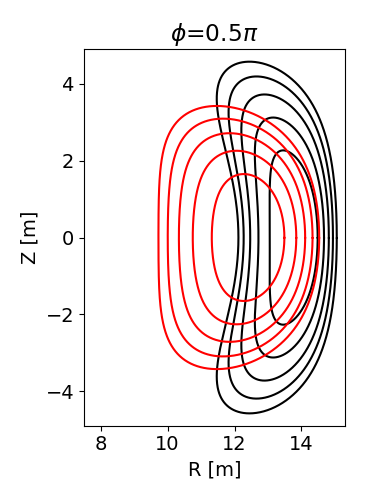}
      \centering
      \scriptsize{(g)}
    \end{minipage}
    \begin{minipage}{0.2425\textwidth}
      \includegraphics[width=0.98\textwidth]{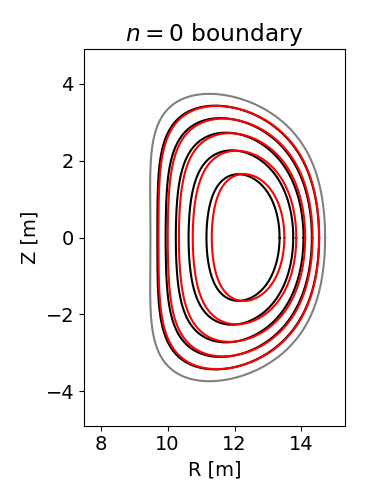}
      \centering
      \scriptsize{(h)}
    \end{minipage}

    \caption{Flux surface contours for the $l=2$ stellarator (a-c) and QA configuration (e-g) test cases up to the plasma boundary. Surfaces of the stellarator equilibria (black) and their virtual current approximations (red) are overlayed at different toroidal planes. The $n=0$ Fourier components of the $l=2$ stellarator (d) are aligned with its axisymmetric approximation. A similar comparison for the QA configuration (h) shows a deviation in the Shafranov shift. \textcolor{black}{In (d) and (h), the boundary of the JOREK simulation domain (grey) is also plotted.}}
    \label{fig:qas_flux_surfaces}
\end{figure*}

A comparison of the flux surfaces of the generated equilibria for poloidal cross sections at $\phi=0.0\pi,\ 0.25\pi$ and $0.5\pi$ are shown in Figure \ref{fig:qas_flux_surfaces}. This shows that after the extension of the plasma, the QA equilibrium has not changed considerably from \cite{Strumberger2019}. In addition the $n=0$ Fourier harmonics of the QA equilibrium are plotted against corresponding surfaces from the axisymmetric approximation. Here, it can be seen that the Shafranov shift of the axisymmetric approximation is larger than the 3D case. For the $l=2$ stellarator, the pressure has been set to an artificially low value, removing this discrepancy. This shows one of the limitations of the axisymmetric approximation, which is that the magnetic well of a stellarator cannot be captured. As a result, the plasma can shift further towards the low field side of the device, modifying the magnetic shear, and local current density gradient, which can have an effect on the MHD stability. \textcolor{black}{The influence of the error this effect introduces needs to be quantified on a case by case basis. As shown in the Section \ref{sec:linear_validation}, the shift of the equilibrium flux surfaces does not lead to a significant difference for the simulated case.}

\section{Linear validation} \label{sec:linear_validation}

\begin{figure}
    \centering
    \includegraphics[width=0.4\textwidth]{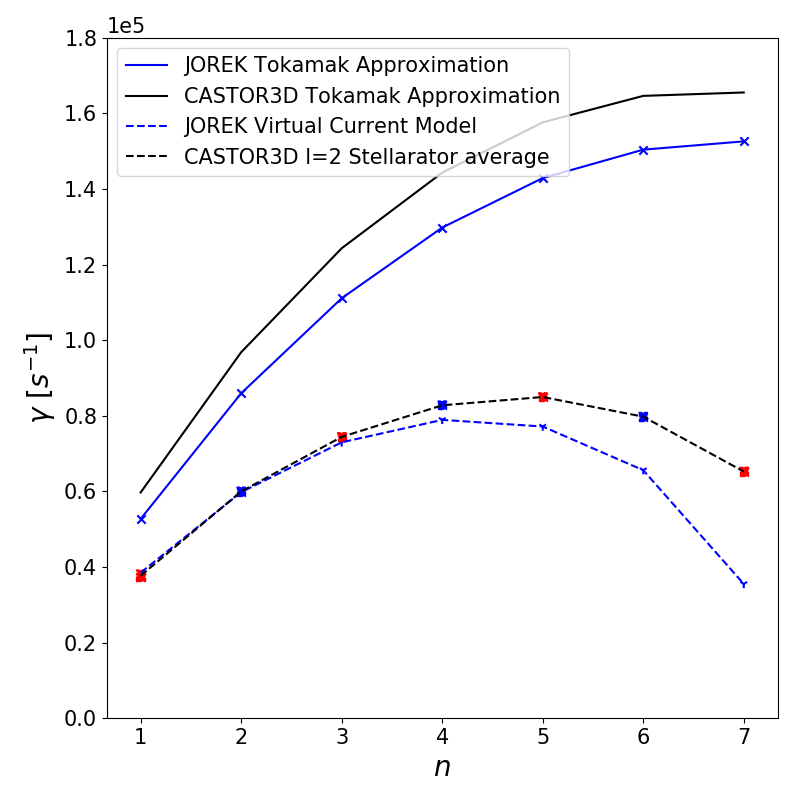}
    
    \scriptsize{(a)}
    
    \includegraphics[width=0.4\textwidth]{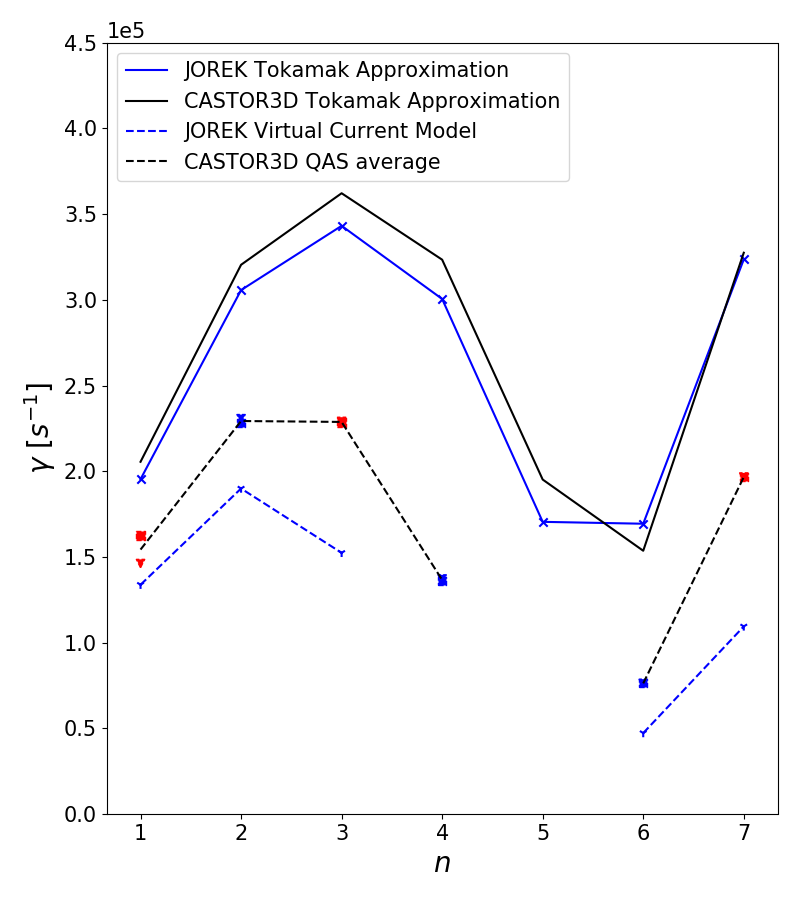}
    
    \scriptsize{(b)}
    
    \caption{Linear growth rates of external kinks observed in the $l=2$ (a) and QA (b) stellarators, corresponding to the equilibria in Figures \ref{fig:equilibrium_profiles} and \ref{fig:qas_flux_surfaces}. \textcolor{black}{The stellarator average and QAS average lines (black dashed) refer to the average of the sinusoidal (triangles) and cosinusoidal (squares) orthogonal solutions of the odd (red) and even (blue) mode families. This average of the stellarator modes is used to compare against the virtual current model (blue dashed).}}
    \label{fig:linear_growth_rates}
\end{figure}

The results of the linear growth rate comparison are shown in Figure \ref{fig:linear_growth_rates}. \textcolor{black}{In this comparison, the same full MHD version of CASTOR3D is used as in \cite{Strumberger2019}, without any flows such that the conditions are comparable to the reduced MHD model in JOREK. As such, CASTOR3D can be considered a fully tested and reliable basis for validating the use of the reduced model in JOREK for the simulated instabilities.} As in this previous study, the mode number of the instabilities from CASTOR3D corresponds to the dominant toroidal mode number of the respective eigenfunction, and the results are split into odd and even mode families. The sinusoidal and cosinusoidal pairs for each $n$ of the stellarator instabilities are also plotted. These orthogonal solutions are only non-degenerate at low toroidal mode numbers in the QA case. For the $l=2$ stellarator, with a simpler 3D geometry, the two mode families do not diverge significantly. 

External kinks are challenging to validate because of the sensitivity of these modes to the edge safety factor, and the differing boundary conditions between JOREK and CASTOR3D. JOREK applies a Dirichlet boundary condition on the normal plasma velocity component, such that the vacuum region needs to be modeled to observe external kinks correctly, without an artificial influence from the boundary conditions. CASTOR3D assumes an infinitely resistive, perfect vacuum outside the simulation boundary. To get a good match between the codes, this demanding vacuum condition, which is harder to match than realistic configurations, needs to be captured in JOREK.

\begin{figure}
    \centering
    \includegraphics[width=0.4\textwidth]{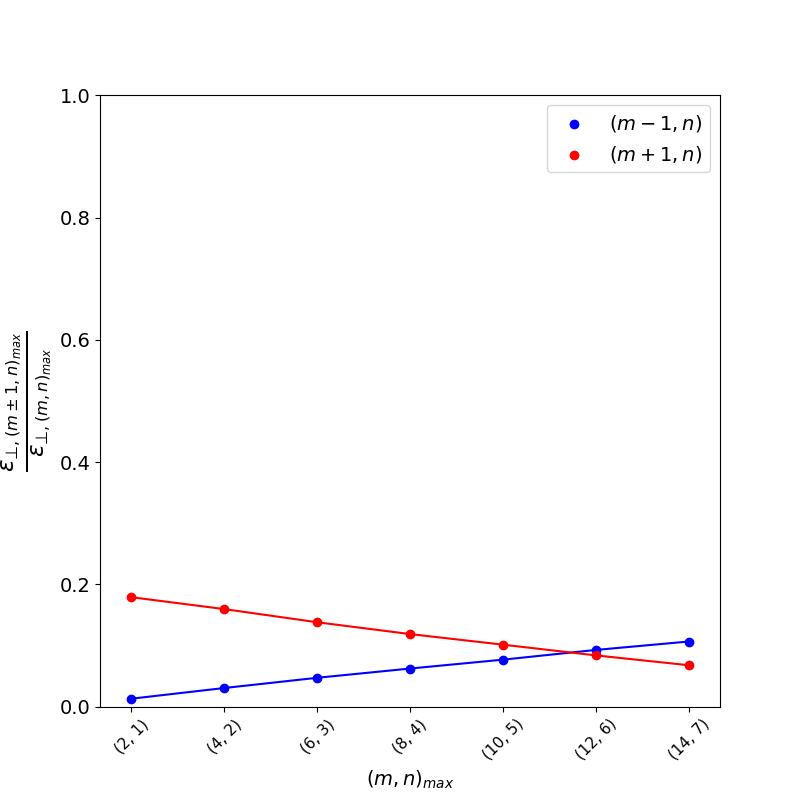}
    
    \scriptsize{(a) Equivalent l=2 tokamak}
    
    \includegraphics[width=0.4\textwidth]{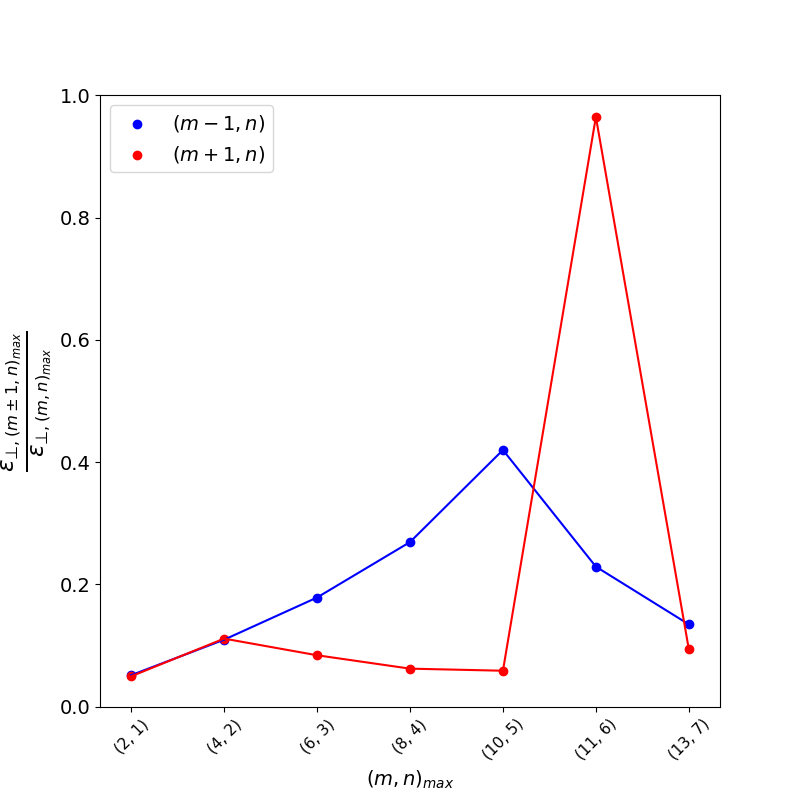}
    
    \scriptsize{(b) Equivalent QAS tokamak}
    
    \caption{\textcolor{black}{Relative magnitude of the maximum radial displacement, $\epsilon_{\bot}$, contributions of poloidal sidebands $(m\pm1,n)_{max}$ compared to the dominant pair, $(m, n)_{max}$. The large relative amplitude of the $n=5$ and 6 modes for the equivalent tokamak of the QAS case indicate more poloidal mode coupling, because the edge safety factor is closer to the $q=9/5$ and $12/6$ rational surfaces, inside and outside of the plasma respectively.}}
    \label{fig:pol_coupling}
\end{figure}

Comparing the equivalent tokamak approximation in JOREK and CASTOR3D, the external kink growth rates show very good agreement for both the $l=2$ and QA configuration. In the QA case, the dip in the growth rates close to $n=5$ and $n=6$ is attributed to poloidal mode coupling between ($m\pm1$, n) pairs inside and outside of the plasma near $\hat\Phi=1$ that provide a stabilising effect. \textcolor{black}{This effect is shown more clearly by plotting the relative contributions to the radial displacement eigenfunction, $\epsilon_\bot$, of the sideband poloidal modes, compared with the dominant (m, n) pair, as shown in Figure $\ref{fig:pol_coupling}$. The results in Figure $\ref{fig:pol_coupling}$ (b) show that there is stronger poloidal mode coupling for the $n=5$ and $n=6$ cases.} 

The (9, 5) and (11, 6) modes are very close to the plasma edge. Small deviations in the $q$ profile, and resistivity profile, which defines the plasma boundary, can strongly influence the contribution of the sideband modes close to the plasma edge, and lead to larger deviations between the two codes. In the $l=2$ stellarator case, the flat $q$ profile was intentionally chosen to be above the closest lower sidebands of the $q=2$ surface for toroidal modes up to $n=7$. \textcolor{black}{Figure \ref{fig:pol_coupling} (a) shows that the poloidal coupling is reduced.} In general, the comparison of the equivalent tokamak approximations show that the reduced MHD model used in this study is appropriate for modeling these external kinks.

Comparing the virtual current model with the stellarator results from CASTOR3D, we see that the general trend of the instabilities is captured in both cases. Particularly in the QA configuration, the deviations are largest where the dominant (m, n) mode is again expected to be stabilised by its ($m\pm1$, n) poloidal sidebands. In particular, the $n=4$ mode found in CASTOR3D for the QA configuration was not observed in JOREK. The reason for the modification of the mode structure cannot be identified from the growth rates alone, but is expected to be due to the modification of the toroidally averaged flux surfaces between the two cases, as shown in Figure \ref{fig:qas_flux_surfaces}.

\begin{figure*}
    \centering
    \begin{minipage}{0.32\textwidth}
      \includegraphics[width=\textwidth]{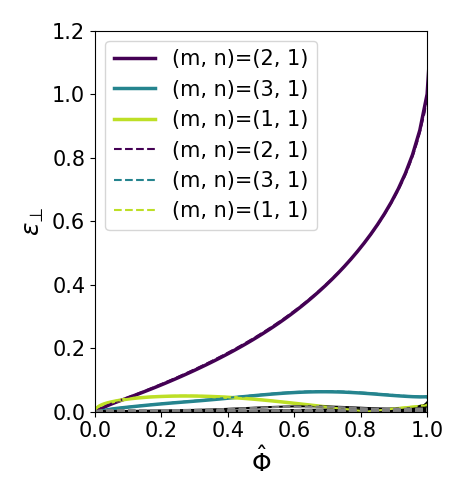}
      \centering
      \scriptsize{(a)}
    \end{minipage}
    \begin{minipage}{0.32\textwidth}
      \includegraphics[width=\textwidth]{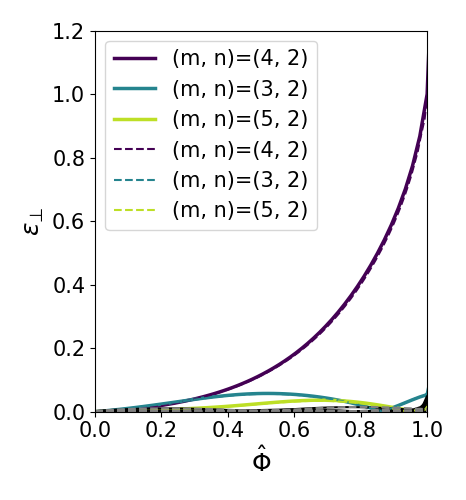}
      \centering
      \scriptsize{(b)}
    \end{minipage}
    \begin{minipage}{0.32\textwidth}
      \includegraphics[width=\textwidth]{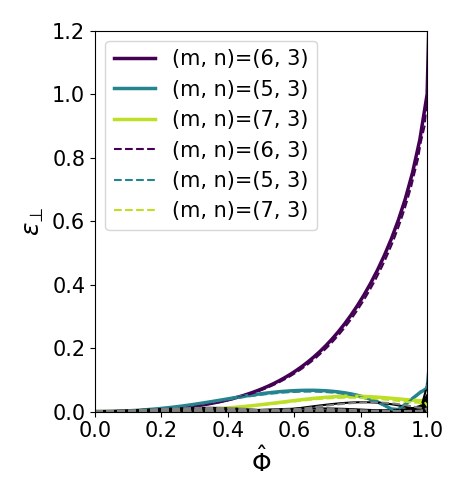}
      \centering
      \scriptsize{(c)}
    \end{minipage}
    
    \begin{minipage}{0.32\textwidth}
      \includegraphics[width=\textwidth]{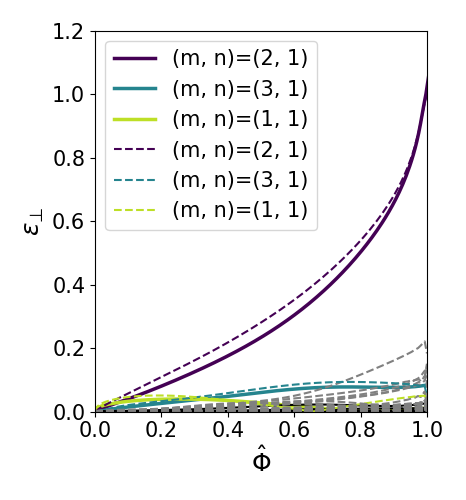}
      \centering
      \scriptsize{(d)}
    \end{minipage}
    \begin{minipage}{0.32\textwidth}
      \includegraphics[width=\textwidth]{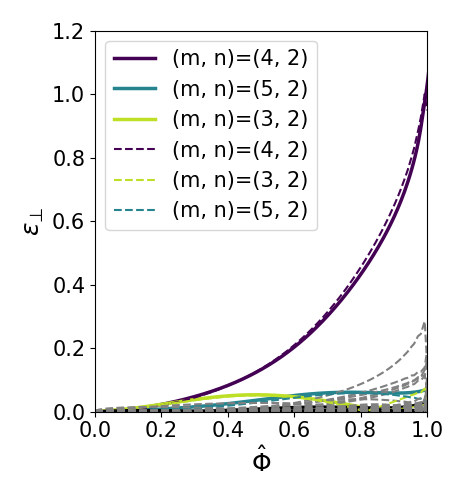}
      \centering
      \scriptsize{(e)}
    \end{minipage}
    \begin{minipage}{0.32\textwidth}
      \includegraphics[width=\textwidth]{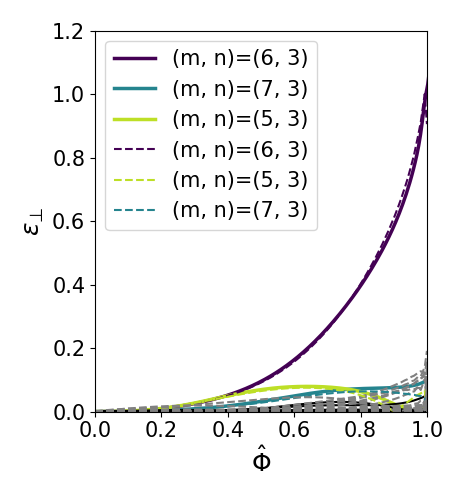}
      \centering
      \scriptsize{(f)}
    \end{minipage}
    \caption{Radial displacement eigenfunction comparisons of JOREK (solid) with CASTOR3D (dashed) in Boozer coordinates. The equivalent tokamak approximations (a-c) show very good agreement. The virtual current model (d-f) instabilities have similar contributions from the leading toroidal harmonic of the equivalent QA instability in CASTOR3D.}
    \label{fig:eigenfunction_comparison}
\end{figure*}

The eigenfunctions for the radial displacement in the QA case are shown in Figure \ref{fig:eigenfunction_comparison}. The results in JOREK and CASTOR3D are generated using Boozer coordinates \cite{boozer1981plasma}. The radial structure of the instabilities is remarkably similar for the equivalent tokamak approximation. Comparing the virtual current model with 3D results, the structure of the captured contributions from the leading toroidal mode number of the instability compare very well. The grey dashed lines that grow in amplitude towards the plasma edge are from toroidal sidebands of the stellarator instability observed in CASTOR3D. An axisymmetric model does not capture the toroidal mode coupling of a stellarator, because the equilibrium magnetic field is not a function of the toroidal coordinate \cite{schwab1993ideal}.

Given the complexity of the case being modeled, the linear mode structure of the virtual current approximation, and the true stellarator agree surprisingly well. In particular, the growth rate of the $n=1$ instability is captured. As it will be shown in the next section, the $n=1$ mode is dominant in the nonlinear dynamics. The virtual current model is therefore considered a reasonable tool for a first study of these modes to provide an understanding of the nonlinear dynamics.


\section{Nonlinear Dynamics of External Kinks} \label{sec:nonlinear_study}

\subsection{Simulation set up and numerical parameters}

In this Section, the equivalent tokamak and virtual current approximations are continued into the nonlinear phase. The initial stellarator considered for this study, has a relatively small external rotational transform, such that the plasma current, $I_p$, of its virtual current approximation remains high at $8.25\ MA$. \textcolor{black}{It has already been shown in \cite{Strumberger2019} that this stabilises the $n=0$ mode, which leads to qualitatively different nonlinear dynamics. The equivalent tokamak needs to be actively controlled in order to avoid a vertical displacement event (VDE). Vertical stability is not considered in this article, as the focus is on the suppression of the external kink that can be achieved as a result of the external rotational transform.}

\begin{figure}
    \centering
    \includegraphics[width=0.45\textwidth]{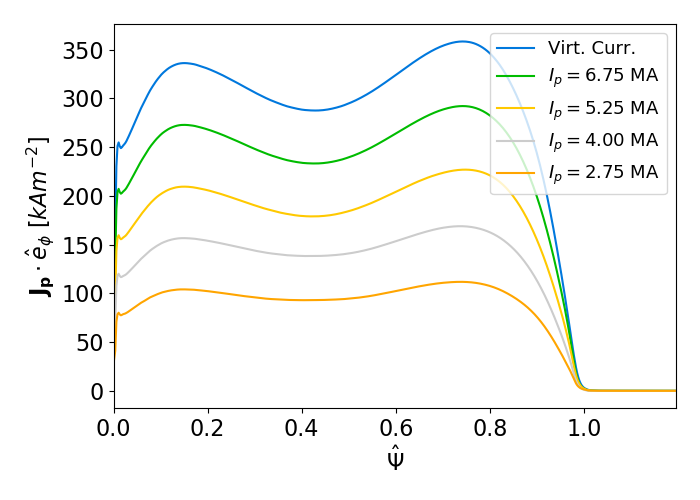}
    \caption{\textcolor{black}{Plasma current density of the simulated cases. The $I_p=8.25\ MA$ (Virt. Curr.) case corresponds to the approximation of the original stellarator equilibrium from Section \ref{sec:equilibrium}. The others are conceptual equilibria, that are used to vary the external rotational transform, while fixing all other equilibrium parameters. They maintain the same normalised plasma current density profile as in the original approximation.}}
    \label{fig:jp_equilibria}
\end{figure}

\textcolor{black}{It will be shown that, with respect to the external kink, the virtual current approximation of the stellarator is more stable than the equivalent tokamak, but the dynamics remain violent for the small external rotational transform, $\iota_{ext}\approx0.15$, that is applied. For this reason, several additional simulations were run at reduced plasma currents corresponding to an increased external rotational transform. The same equilibrium is used, preserving the safety factor, pressure and total current profiles from Figure \ref{fig:equilibrium_profiles} (d-f), while reducing $I_p$. Approximately the same normalised plasma current density profile is maintained as in the original stellarator. The plasma current density of the modified equilibria are shown in Figure \ref{fig:jp_equilibria}, comparing with the original virtual current approximation from Section \ref{sec:equilibrium} in blue.} Six cases have been simulated at plasma currents ranging from $11.5\ MA$, in the equivalent tokamak approximation, to $8.25\ MA$, representing the original QA configuration, and four conceptual cases with lower current at $6.75$, $5.25$, $4.00$ and $2.75\ MA$ respectively. 

In such a way, the full parameter space can be explored to give an intuition of when the external kink is nonlinearly stabilised by the external rotational transform. \textcolor{black}{It should be noted that the modified equilibria at lower plasma currents do not have a known corresponding 3D equilibrium. They are conceptual cases generated for the purpose of this study, to isolate the effect of the external rotational transform, keeping all other variables in the equilibrium the same.}

Table \ref{tab:diffusive_parameters} shows the numerical and diffusive parameters used in the following nonlinear simulations. The simulations carried out in the previous section were set up to match the boundary conditions of CASTOR3D, modeling a highly resistive vacuum region outside the plasma. For nonlinear simulations, these conditions are not realistic and numerically demanding, and so some numerical parameters were modified. \textcolor{black}{For this initial study of the nonlinear dynamics where the objective is to gain a general understanding of the MHD activity, the numerical parameters are considered reasonable.}

A Spitzer-like profile is used for resistivity, viscosity and their equivalent hyper-diffusive terms, in order to dissipate sub-grid resolution current sheets that form during the penetration of the kink into the hot core. \textcolor{black}{The parameters that have been chosen are as physically reasonable as was possible within the limits of numerical resolution. Using a core resistivity around $10^{-7}\ \Omega m$ is common in nonlinear MHD studies, because the physical values around $10^{-9}$ to $10^{-8}\ \Omega m$ are often too numerically demanding.}

A realistic, temperature dependent, parallel heat conductivity is used. The parallel conductivity is set based on the Spitzer-Haerm conductivity, such that $\chi_{\parallel}=2.18\times10^9\ m^2s^{-1}$ at the magnetic axis. Because the simulation domain does not include a limiter, or X-point, there are closed flux surfaces outside the plasma, which do not replicate the transport physics of the vacuum region expected in a real experiment. For this reason, a relatively large perpendicular diffusion is used in the vacuum region to approximate the physics of a limiter around the plasma where transport out of the simulation domain should be high. In the plasma region, the perpendicular particle and thermal diffusion coefficients are spatially constant. Because the parallel momentum equation is neglected in this study for simplicity and to improve numerical stability, a parallel particle diffusivity is used as a proxy for the parallel mass transport that would otherwise be neglected. Heat and particle sources are used to maintain the density and temperature profiles for the timescale of the simulations.

\begin{table} 
\caption{\label{tab:diffusive_parameters} Diffusive parameters used in nonlinear external kink simulations. The $\psi$ dependent profiles sharply transition from the lower to upper limit of their ranges at the plasma boundary.} \begin{indented} 
\item[]\begin{tabular}{@{}lll}
\br
Parameter & Range & Dependence \\
\mr
        $\chi_{\parallel}\ [m^2s^{-1}]$ & $68.9 - 2.18 \times 10^9$                        & Spitzer-Haerm \\
        $\chi_{\perp}\ [m^2s^{-1}]$     & $1.45 - 2.179 \times 10^3$                       & Step function \\
        $D{\parallel}\ [m^2s^{-1}]$     & $847.0$                                          & Constant \\
        $D_{\perp}\ [m^2s^{-1}]$        & $0.847 - 847.0$                                  & Step function \\
        $\eta\ [\Omega m]$              & $1.06\times 10^{-7} - 3.38 \times 10^{-4}$       & Spitzer-like \\
        $\eta_{num}$                    & $1.06\times 10^{-12} - 3.38 \times 10^{-9}$      & Spitzer-like \\
        $\mu\ [kg\ m^{-1}s^{-1}]$       & $9.37\times 10^{-8} - 2.96 \times 10^{-4}$       & Spitzer-like \\
        $\mu_{num}$                     & $9.37 \times 10^{-13} - 2.96 \times 10^{-9}$     & Spitzer-like \\
\br
\end{tabular}
\end{indented}
\end{table}

\begin{figure}
    \centering
    \includegraphics[width=0.4\textwidth]{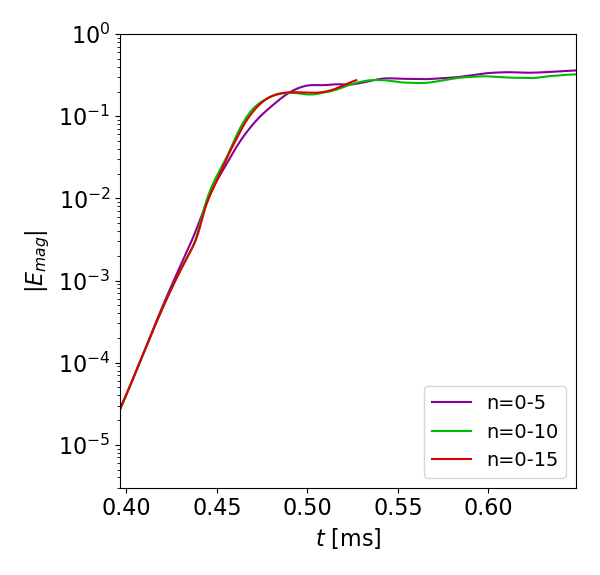}
    \caption{Toroidal resolution scan for the equivalent tokamak approximation, since it has the most violent dynamics. The $n=1$ mode, which dominates the instability, is sufficiently converged using 10 toroidal harmonics.}
    \label{fig:tokamak_resolution_scan}
\end{figure}

The instabilities are modeled with a polar grid, discretised using 150 poloidal and 120 radial elements. Free boundary conditions are applied to all toroidal harmonics, except for the $n=0$ component. A toroidal resolution scan considering the evolution of the magnetic energies in the equivalent tokamak approximation is shown in Figure \ref{fig:tokamak_resolution_scan}. For the reduced MHD simulations carried out in this study, only the poloidal magnetic energy needs to be considered in the evolution of MHD instabilities, as the toroidal magnetic field is fixed in time. \textcolor{black}{The poloidal magnetic energy of a given toroidal mode number, $n$, is defined in normalised units as }

\begin{equation}
    E_{mag} = \frac{\mu_0}{2} \oint_V \frac{1}{R^2} \left( \frac{\partial \psi_n}{\partial R}^2 + \frac{\partial \psi_n}{\partial Z}^2 \right) dV,
\end{equation}

\textcolor{black}{where $\psi_n$ is the poloidal flux of the $n^{th}$ toroidal harmonic.} Figure \ref{fig:tokamak_resolution_scan} shows that the magnetic energy of the $n=1$ toroidal harmonic, which leads the instability, is sufficiently resolved using 10 toroidal harmonics. 

\begin{table} 
\caption{\label{tab:time_normalisation} Growth rates and saturation times for the different simulated cases. \textcolor{black}{The equivalent tokamak and virtual current simulations are the same equilibria as used in Section \ref{sec:linear_validation}. The growth rates have been modified by changes in the resistivity and viscosity profiles for nonlinear simulations.}} \begin{indented} 
\item[]\begin{tabular}{@{}llll}
\br
Case &$\gamma_{n=1}$ \textcolor{black}{[$s^{-1}$]} &$t_{sat}$ [$ms$] &$t_{offset}$ [ms] \\
\mr
        Equiv. Tok.     & $4.367 \times 10^4$              &   0.445    &  0.075 \\
        Virt. Curr.     & $1.840 \times 10^4$              &   1.134     &  0.192  \\ 
        $I_p=6.75\ MA$  & $8.760 \times 10^3$              &   2.492     &  0.483  \\
        $I_p=5.25\ MA$  & $3.427 \times 10^3$              &   6.709     &  1.403  \\
        $I_p=4.00\ MA$  & $3.413 \times 10^2$              &   67.083    &  13.849 \\
        $I_p=2.75\ MA$  &  -                               &  -         &  -  \\
\br
\end{tabular}
\end{indented}
\end{table}


The observed linear growth rates, and times to saturation are shown in Table \ref{tab:time_normalisation}. With the modifications described above, the structure of the growth rates in the linear validation is damped, but the external kink modes remain violently unstable. The growth rates decrease significantly \textcolor{black}{as $I_p$ decreases}, until for the $I_p=2.75\ MA$ case, the virtual current approximation becomes linearly stable to external kinks. \textcolor{black}{This is shown in Table 2, using the growth rate of the $n=1$ kink instability, $\gamma_{n=1}$}. Some of the plots in this section use a normalised time, $\hat t$, allowing all cases to be compared, despite their varying time scales. $\hat t$ is normalised by $\gamma_{n=1}$, such that

\begin{eqnarray}
    \hat t = 1.1769\times10^{-6}\ \gamma_{n=1} \left(t - t_{offset}\right)
\end{eqnarray}

where the prefactor is the time normalisation to JOREK units, and $t_{offset}$ is used so that at $\hat t=0$, all simulated cases have the same magnetic energy in the $n=1$ mode, which has become linearly unstable.

The Ohmic decay of the plasma current, $\eta (j_{vc} - j_v)$ in equation \ref{eq:induction_strong}, is applied only after the initial kink instability has begun to saturate, in order to preserve the current profile during the linear phase. A more realistic bootstrap current profile cannot be evolved from the initial conditions because the initial pressure and bootstrap current density profiles have not been prescribed self-consistently. Allowing the current density profile to decay exacerbates resistive instabilities, and as a result can be considered a worst case scenario for the dynamics, as enhanced resistivity is likely to lower the stability threshold for nonlinearly triggered modes \cite{holties1996stability}. 

\subsection{Evolution overview}
\begin{figure}
    \centering
    \includegraphics[width=0.4\textwidth]{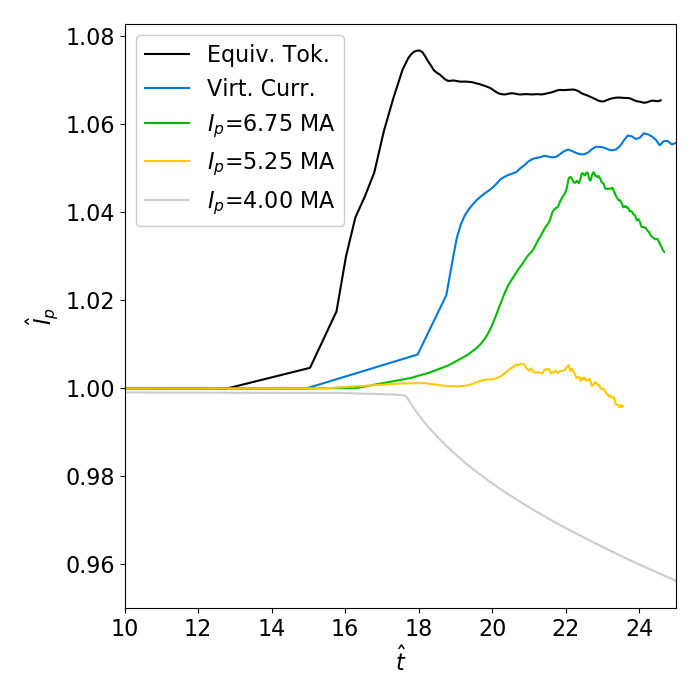}
    
    
    
    
    
    \caption{\textcolor{black}{Normalised plasma current, $\hat I_p$, with respect to the initial value, $I_p$, during the evolution of the external kink.}} 
    \label{fig:current_evolution}
\end{figure}

In this section, the overall dynamics of the simulated cases are summarised using the averaged toroidal magnetic energy spectrum during the nonlinear phase, and the evolution of the total plasma current and thermal energy in time. In the early nonlinear phase, the external kink should lead to similar current dynamics as in a constrained relaxation process \cite{taylor1986relaxation, bhattacharjee1982energy}. The current profile flattens, redistributing into the vacuum region. As a result, the inductance of the plasma decreases, leading to a current spike. The magnitude of the spike is a reasonable metric for the violence of the initial instability. Figure \ref{fig:current_evolution} shows the evolution of the normalised plasma current, $\hat I_p$, as a function of time. As expected, it can be seen that the relative spike amplitude increases with the total plasma current. Eventually for the $I_p=4.00\ MA$ case, the initial kink instability becomes so mild that a current spike is not observed, and the current resistively decays. 




\begin{figure}
    \centering
    \includegraphics[width=0.4\textwidth]{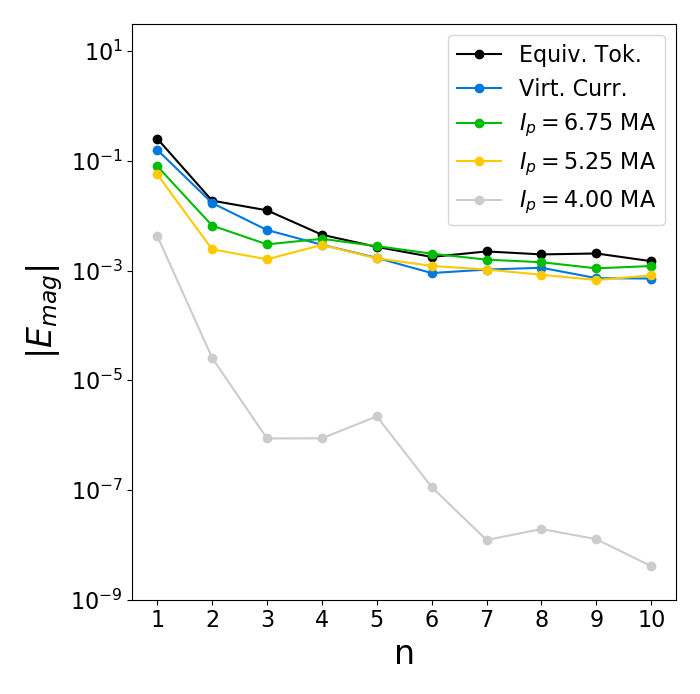}
    \caption{Magnetic energy spectrum during the nonlinear phase in which the external kink has saturated.}
    \label{fig:energy_spectrum}
\end{figure}

Figure \ref{fig:energy_spectrum} shows that most of the simulated cases have a very broad magnetic energy spectrum during the nonlinear phase. This is to be expected given that the edge $q$ being modeled is below 2, and is therefore highly unstable. In the equivalent tokamak case especially, all toroidal harmonics are linearly unstable. As $I_p$ decreases, more modes are stabilised, such that the $n=1$ component becomes the driving mode in the nonlinear saturation. While the energies of the $n=1$, $2$ and $3$ modes reduce with the plasma current, the higher toroidal harmonics remain relatively constant during the nonlinear phase, indicating that these modes are not just driven by the lower toroidal harmonics, and the stability threshold of internal modes has been crossed. This is in particular true for the $n=4$ and $5$ modes, as discussed in Section \ref{sec:internal_modes}. The $I_p=4.00\ MA$ case shows a significantly lower saturated energy than the higher current cases. This seems to be because the initial kink instability is not violent enough to trigger significant MHD activity at internal surfaces.

\textcolor{black}{The thermal energy in SI units is defined as }

\begin{equation}
    E_{therm} = \frac{1}{\gamma - 1} \oint_V  p  dV,
\end{equation}

\begin{figure}
    \centering
    \includegraphics[width=0.4\textwidth]{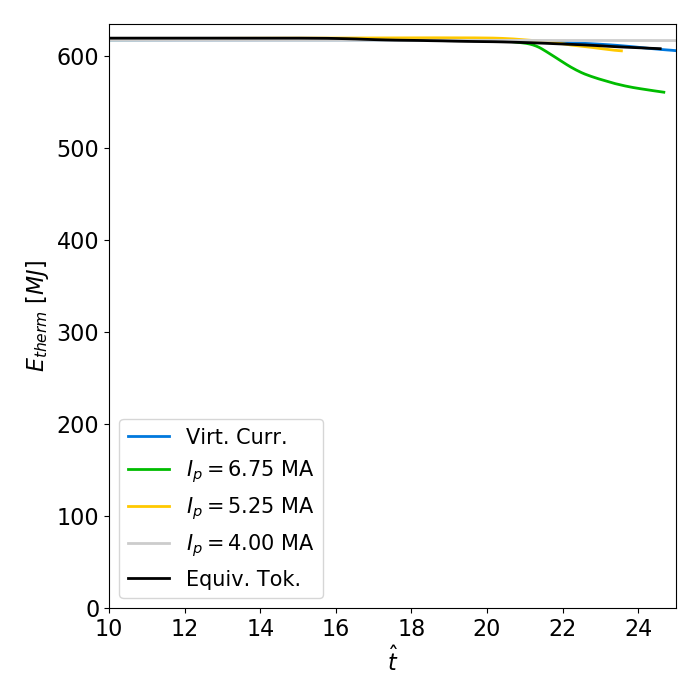}
    \caption{Thermal energy during the evolution of the external kink. The $I_p=6.75\ MA$ case has a significantly increased transport of thermal energy out of the plasma.}
    \label{fig:w_therm}
\end{figure}

The change in thermal energy stored in the plasma volume over time is plotted for all cases in Figure \ref{fig:w_therm}. In most cases the thermal energy does not change significantly in the simulated time frame. This is because the dynamics are strongest in the outer region of the plasma, while most of the thermal energy is stored in the core, because of the assumed parabolic pressure profile. Somewhat surprisingly, the $I_p=6.75\ MA$ case loses the most thermal energy. The increase in thermal losses are too large to be due to the longer timescale of the simulated instability alone. The behaviour at $I_p=6.75\ MA$ is expected to be because of enhanced transport inside the plasma due to overlapping internal modes as discussed in Section \ref{sec:internal_modes}. At $I_p=4.00\ MA$, the thermal energy does not decay significantly. It is shown in Section \ref{sec:ergodisation} that confinement is maintained over most of the plasma region in this case.

\subsection{MHD instability dynamics} \label{sec:kink_dynamics}
\begin{figure*}
    \centering
    \begin{minipage}{0.32\textwidth}
      \includegraphics[width=\textwidth]{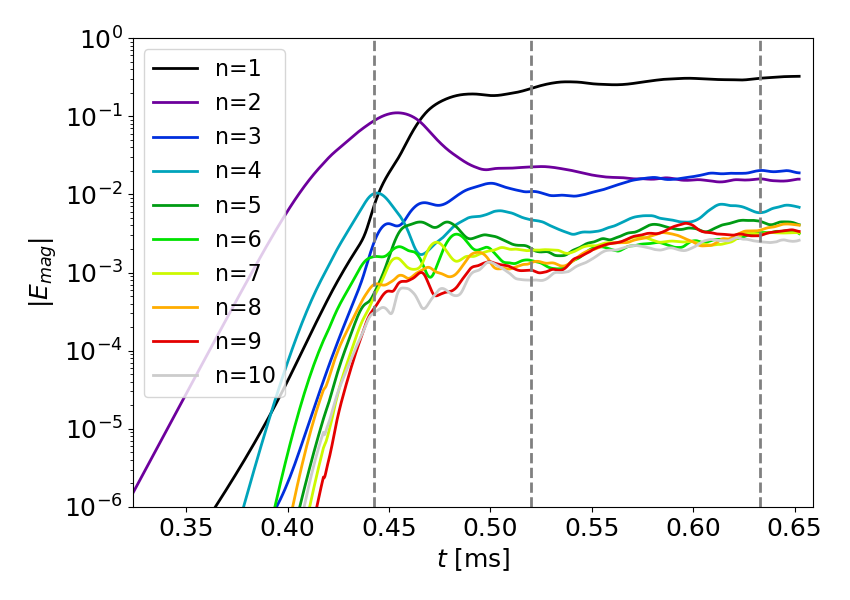}
      \centering
      \scriptsize{(a) \textcolor{black}{Equiv. Tok.}}
    \end{minipage}
    \begin{minipage}{0.19\textwidth}
      \includegraphics[width=\textwidth]{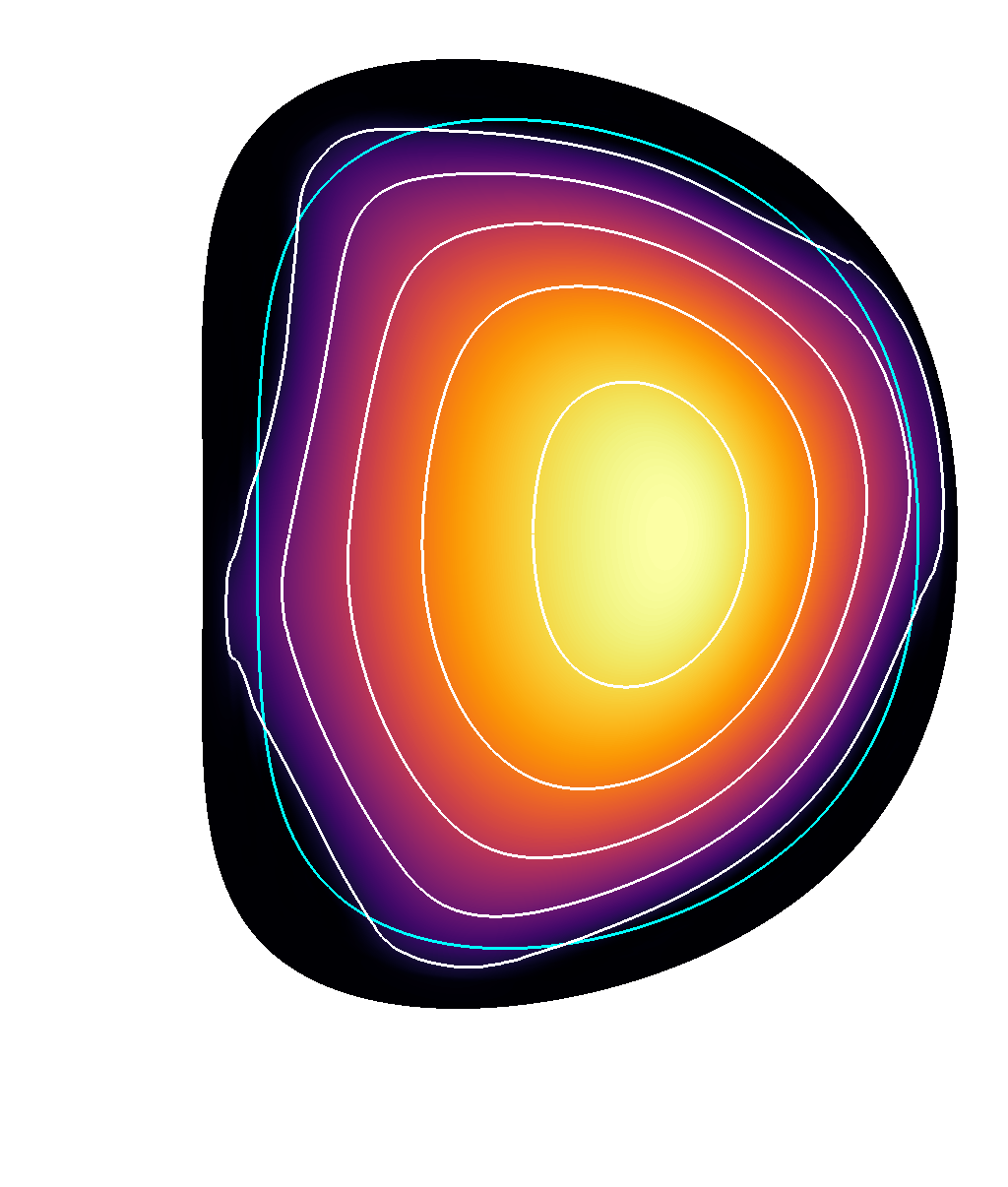}
      \centering
      \scriptsize{(b) \textcolor{black}{$t=0.442$ ms}}
    \end{minipage}
    \begin{minipage}{0.19\textwidth}
      \includegraphics[width=\textwidth]{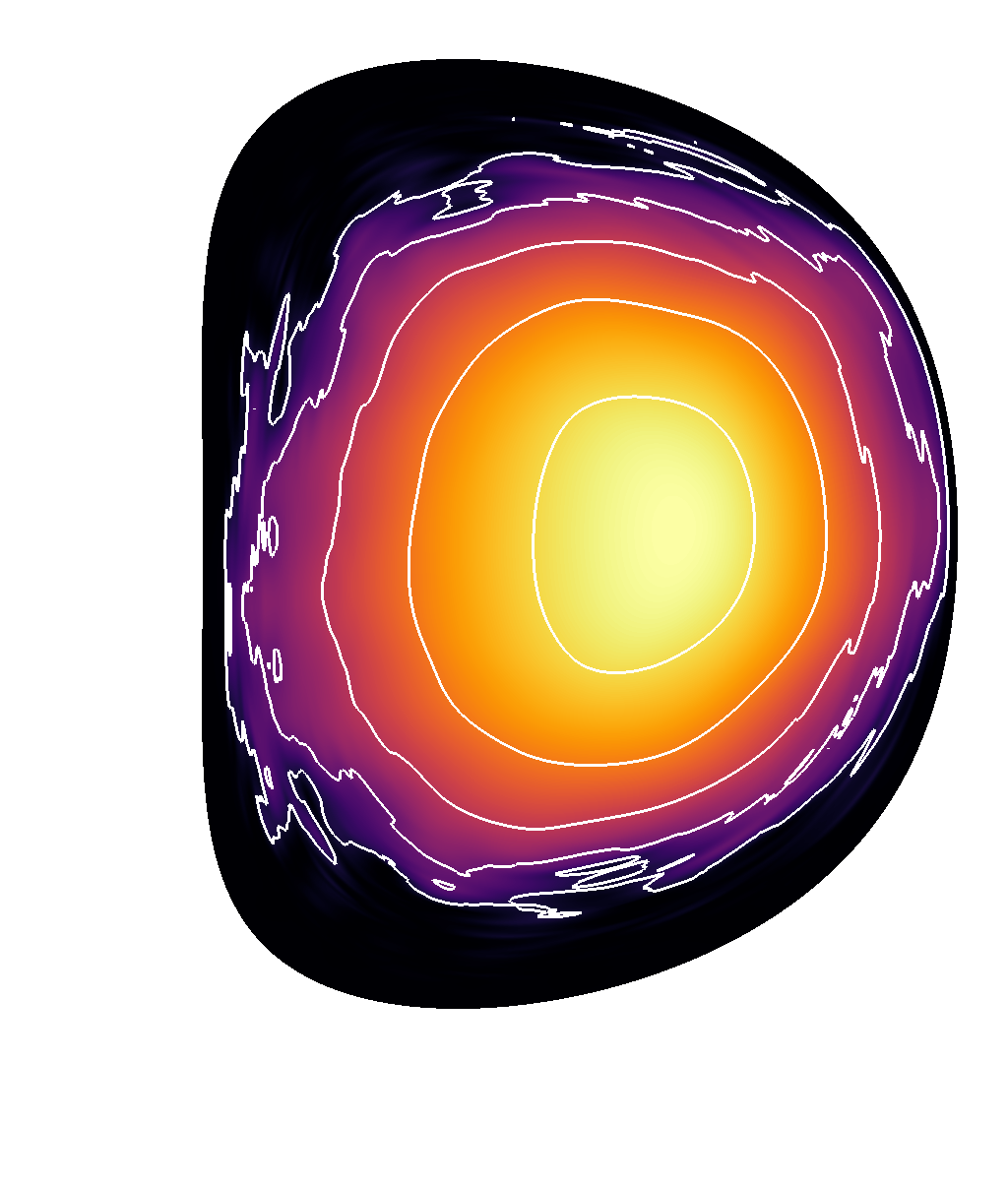}
      \centering
      \scriptsize{(c) $t=0.520$ ms}
    \end{minipage}
    \begin{minipage}{0.19\textwidth}
      \includegraphics[width=\textwidth]{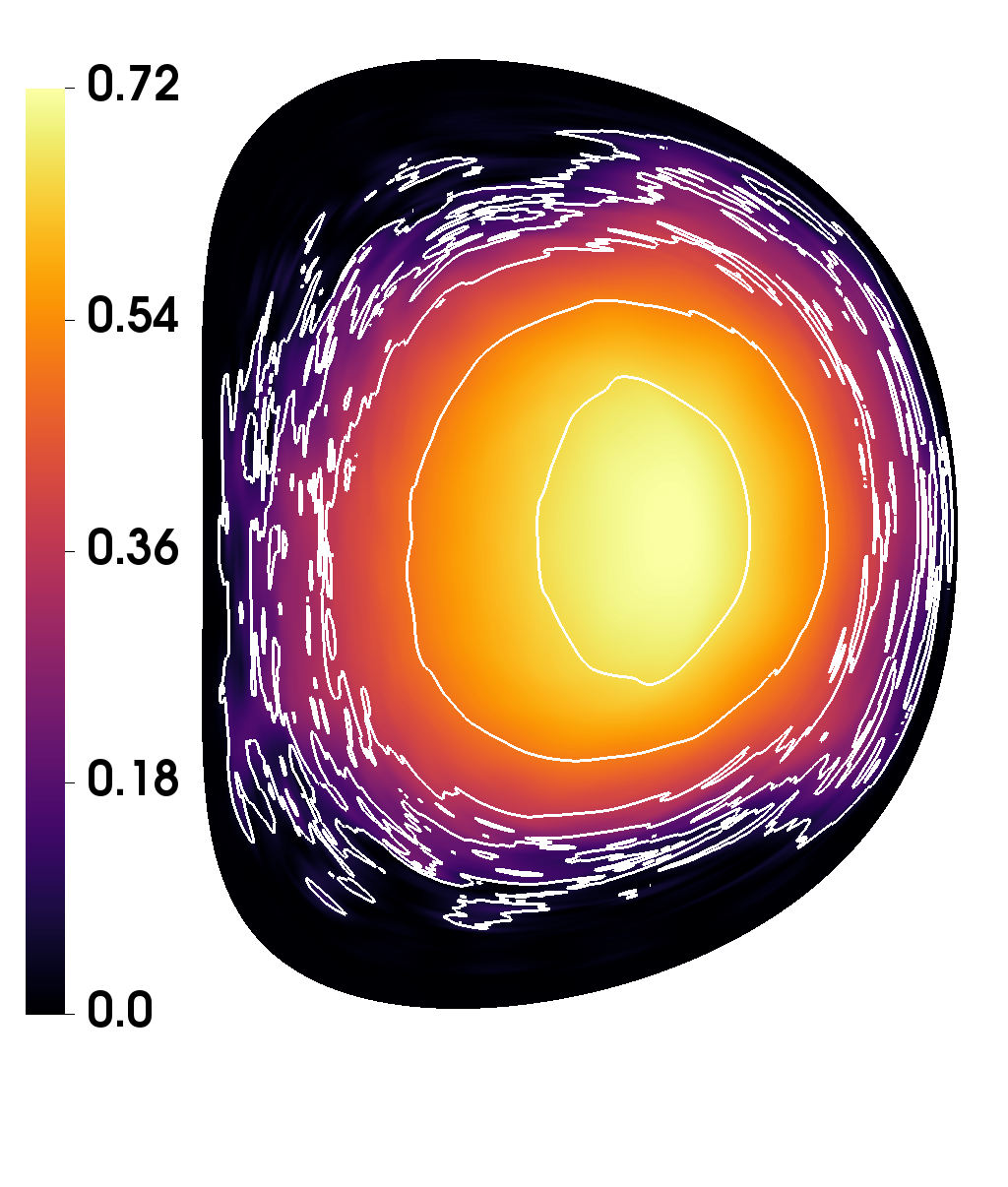}
      \centering
      \scriptsize{(d) $t=0.633$ ms}
    \end{minipage}
    
    \centering
    \begin{minipage}{0.32\textwidth}
      \includegraphics[width=\textwidth]{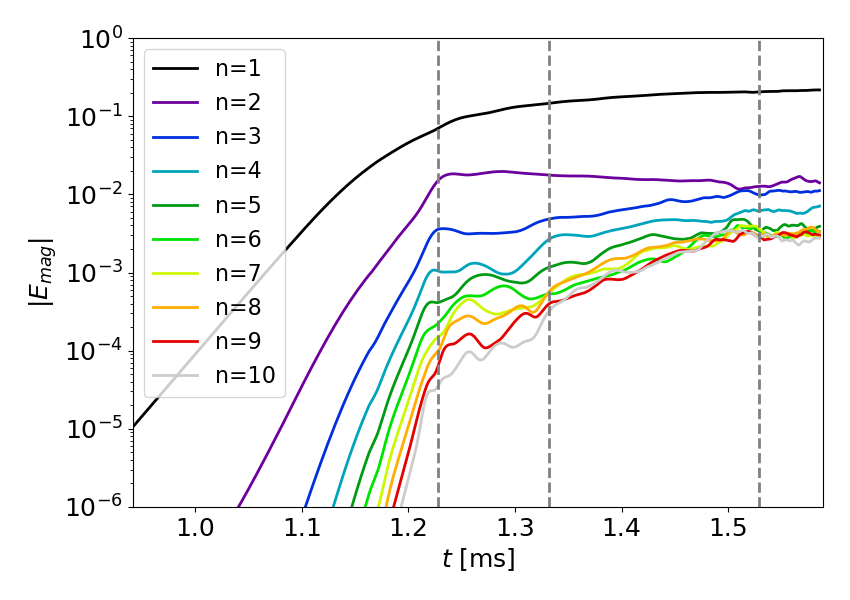}
      \centering
      \scriptsize{(e) \textcolor{black}{Virt. Curr.}}
    \end{minipage}
    \begin{minipage}{0.19\textwidth}
      \includegraphics[width=\textwidth]{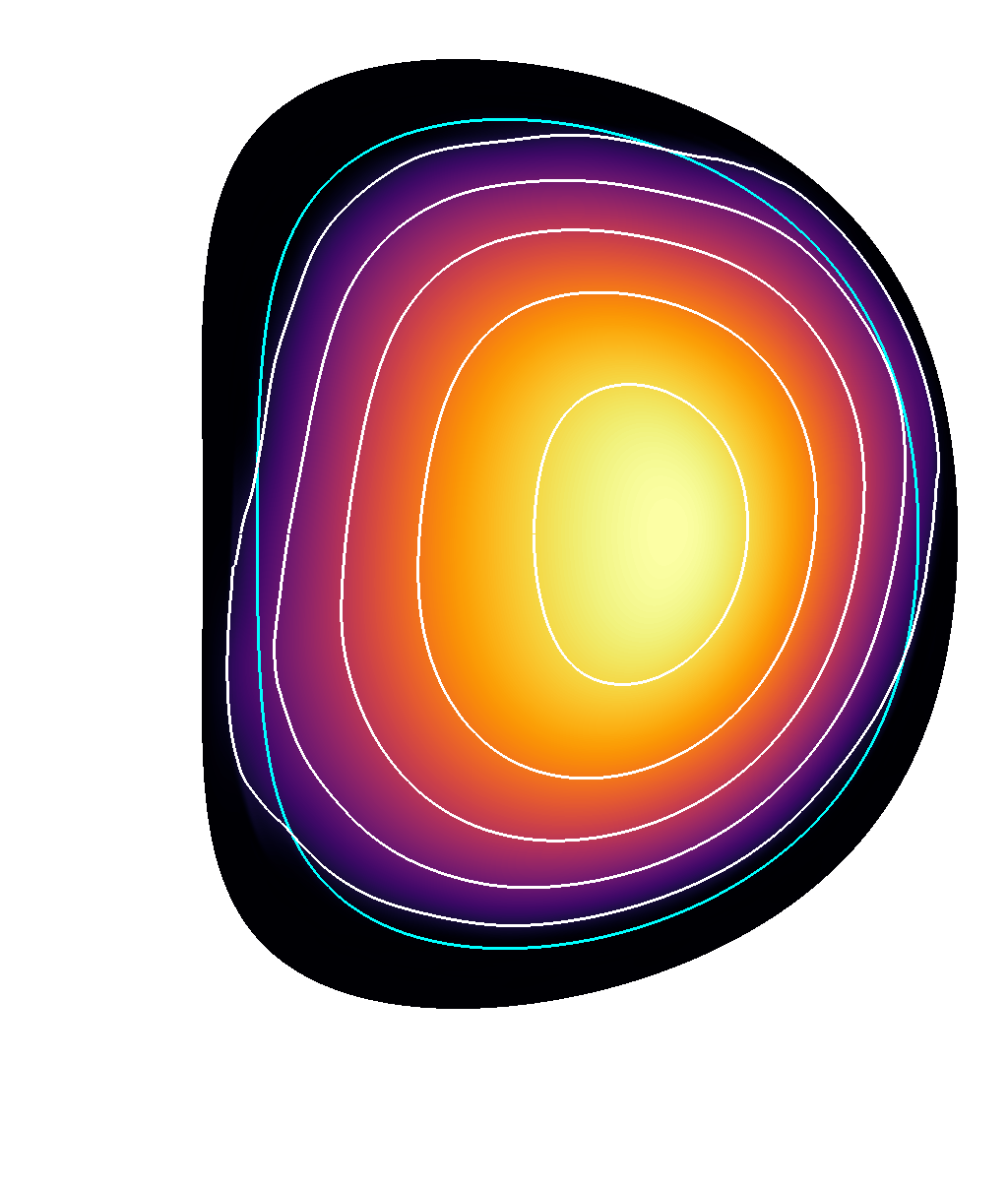}
      \centering
      \scriptsize{(f) $t=1.236$ ms}
    \end{minipage}
    \begin{minipage}{0.19\textwidth}
      \includegraphics[width=\textwidth]{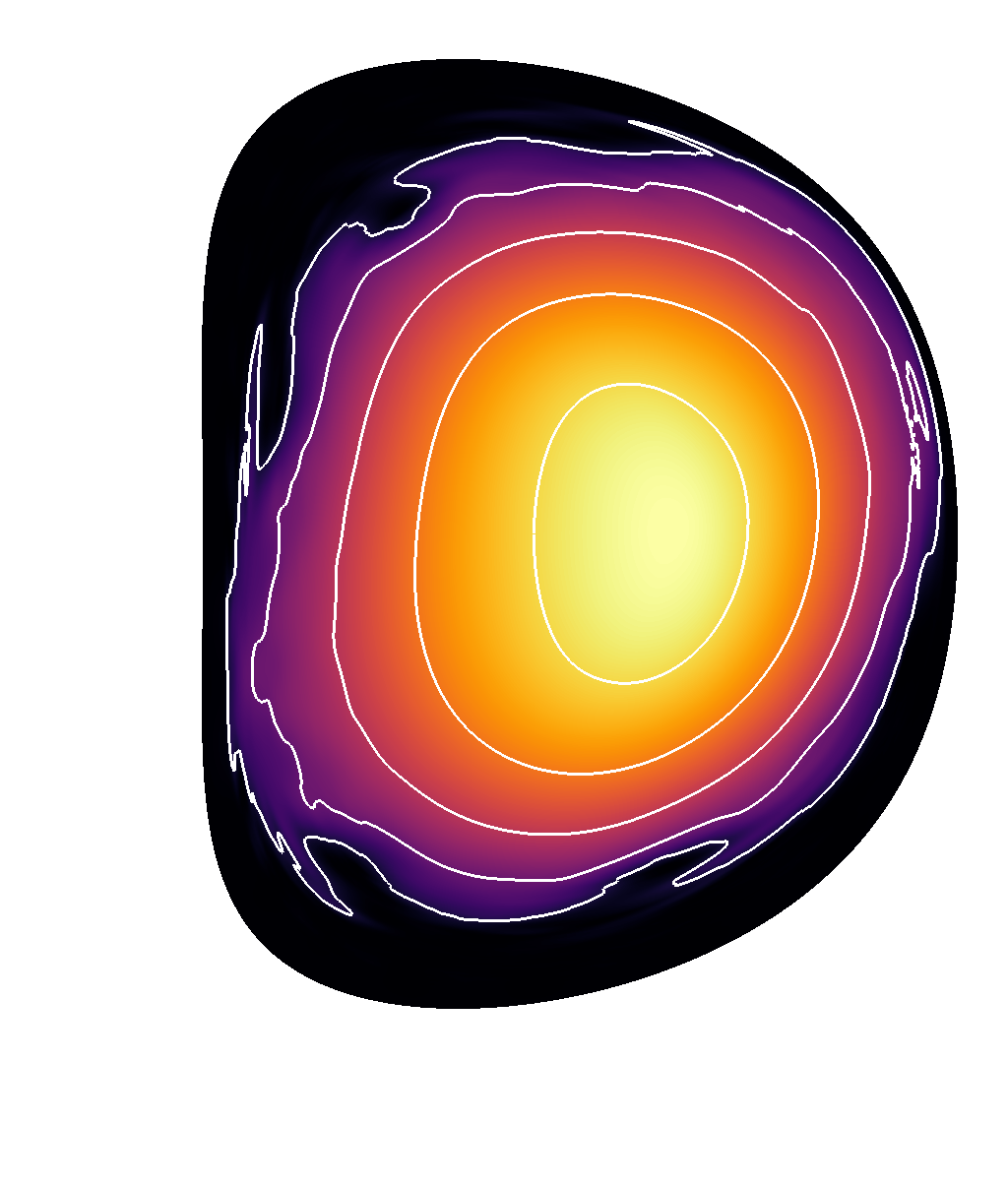}
      \centering
      \scriptsize{(g) $t=1.295$ ms}
    \end{minipage}
    \begin{minipage}{0.19\textwidth}
      \includegraphics[width=\textwidth]{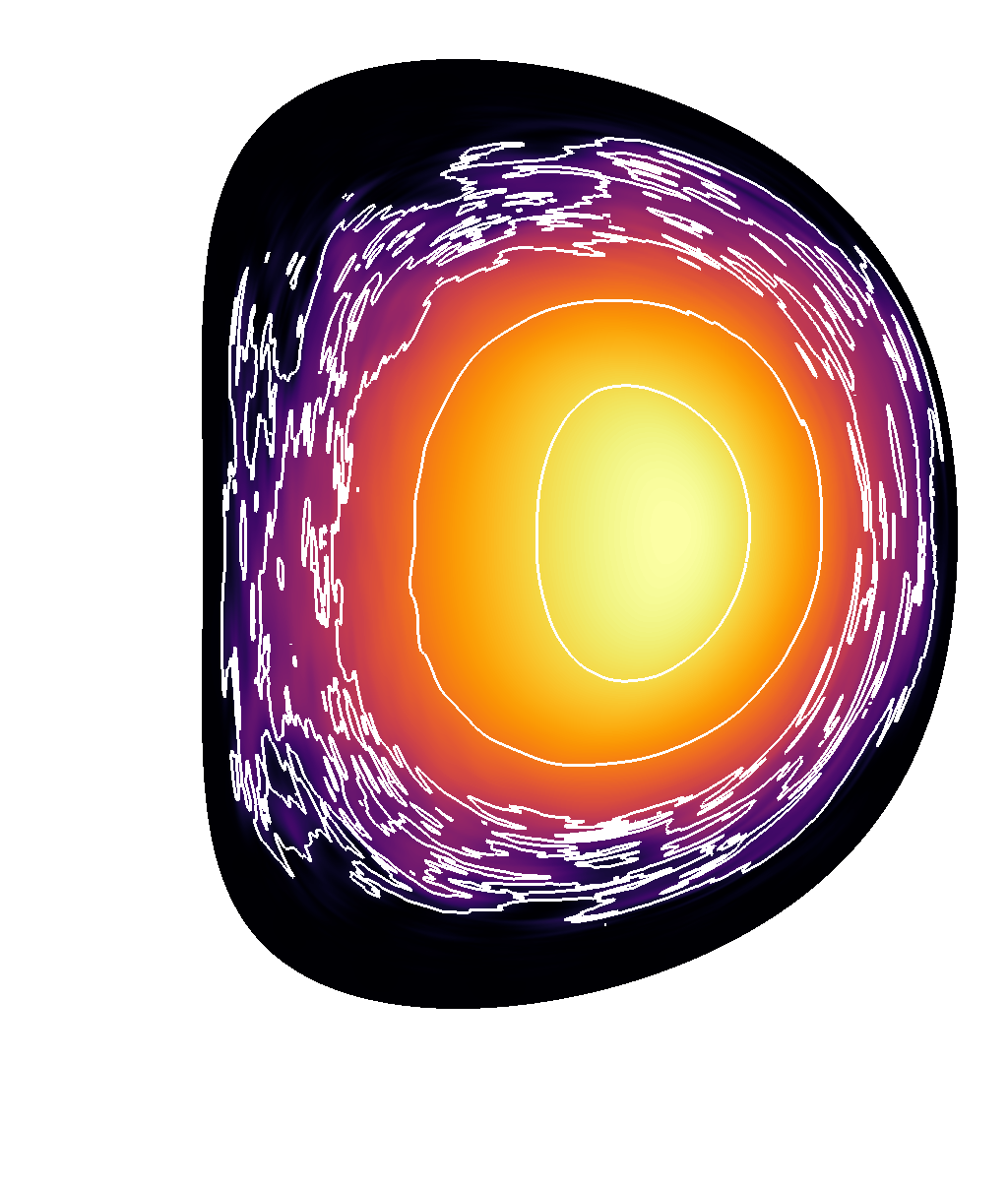}
      \centering
      \scriptsize{(h) $t=1.530$ ms}
    \end{minipage}
    
    \centering
    \begin{minipage}{0.32\textwidth}
      \includegraphics[width=\textwidth]{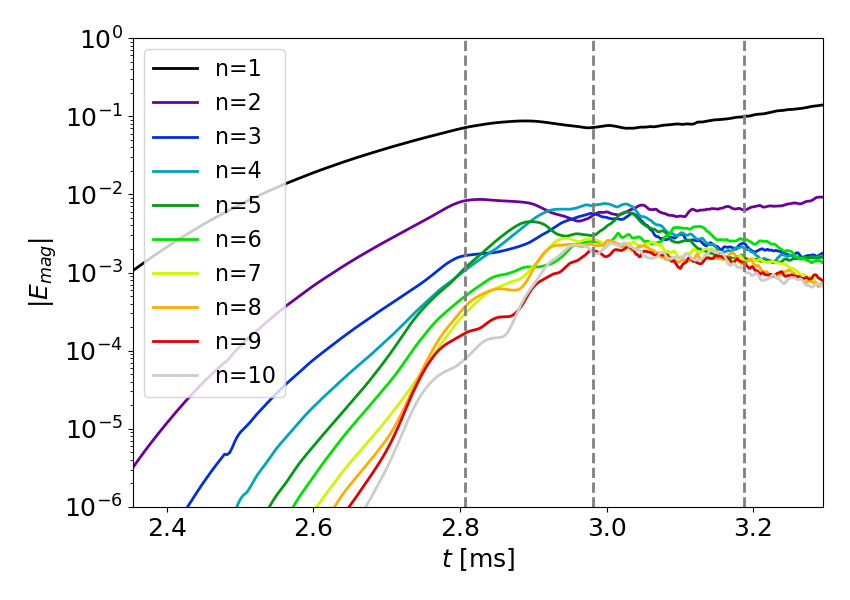}
      \centering
      \scriptsize{(i) $I_p=6.75\ MA$}
    \end{minipage}
    \begin{minipage}{0.19\textwidth}
      \includegraphics[width=\textwidth]{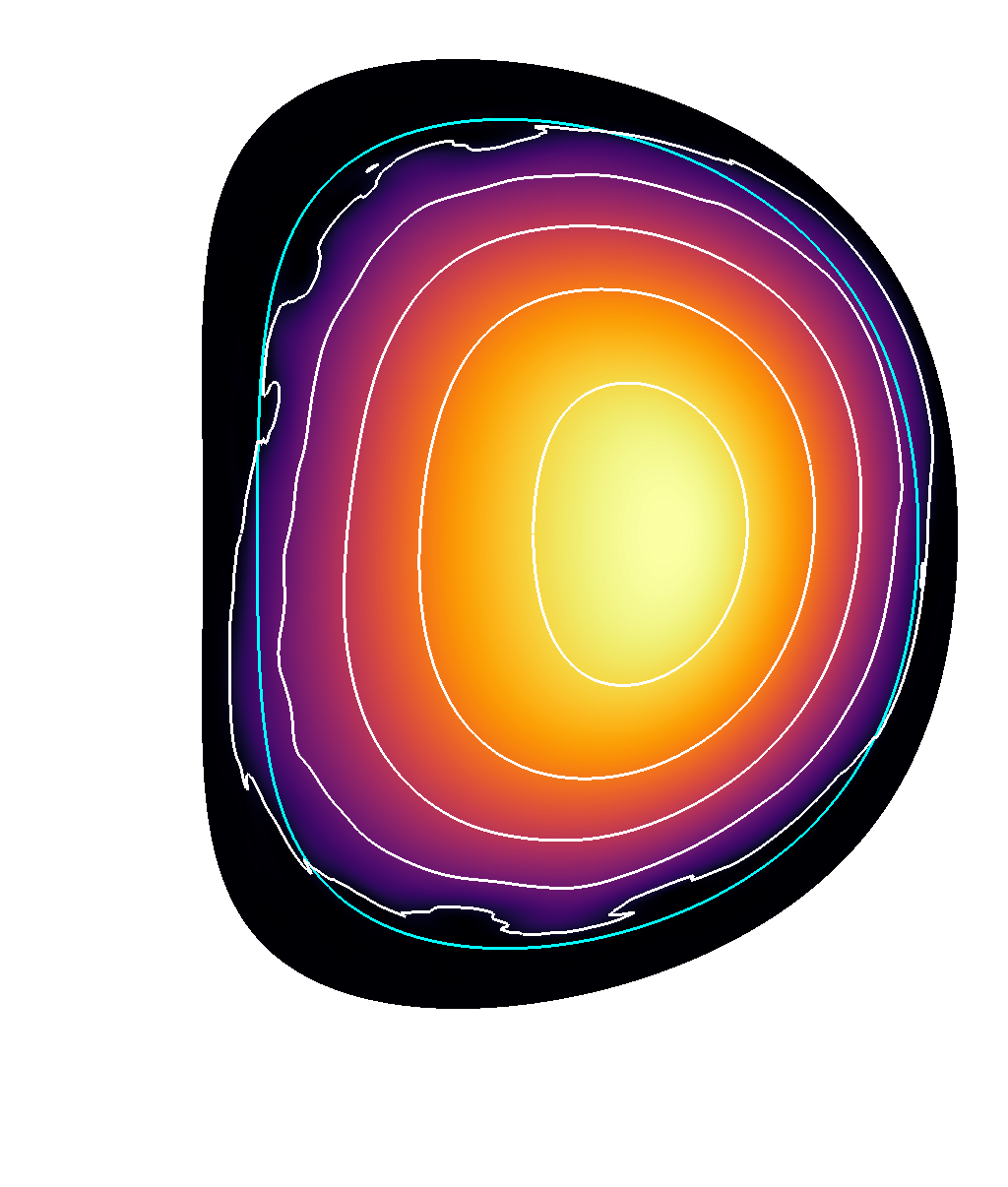}
      \centering
      \scriptsize{(j) $t=2.801$ ms}
    \end{minipage}
    \begin{minipage}{0.19\textwidth}
      \includegraphics[width=\textwidth]{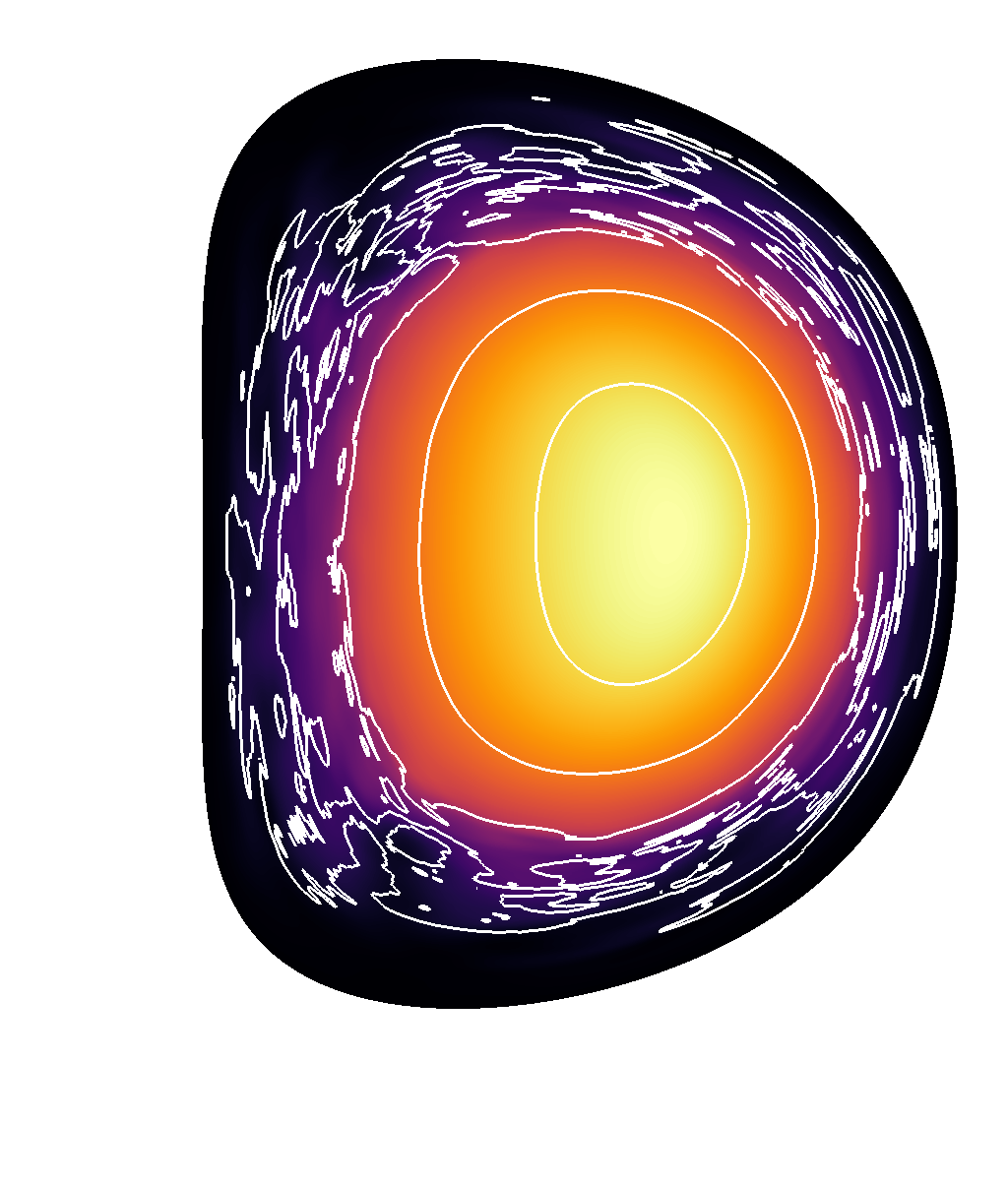}
      \centering
      \scriptsize{(k) $t=3.001$ ms}
    \end{minipage}
    \begin{minipage}{0.19\textwidth}
      \includegraphics[width=\textwidth]{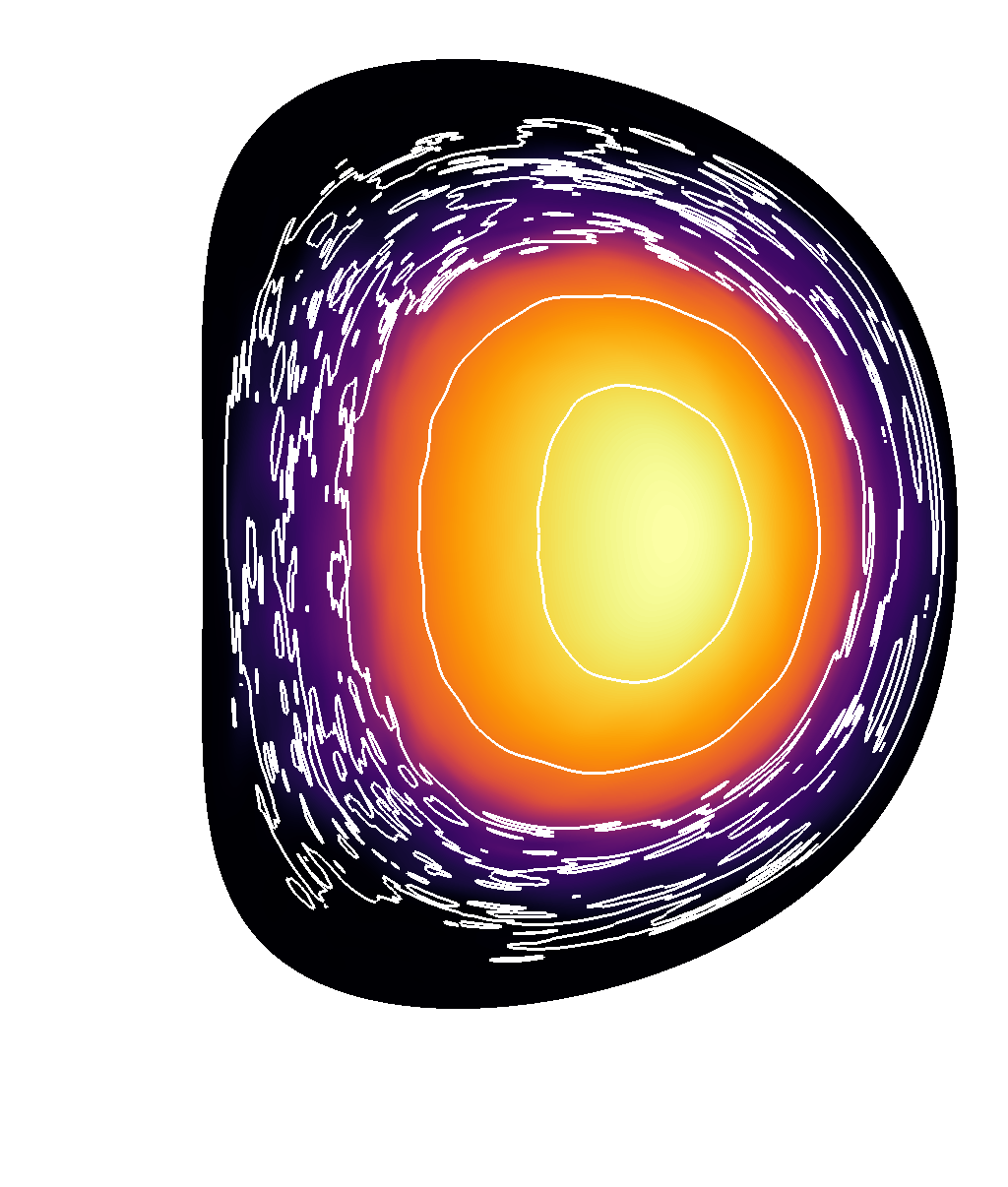}
      \centering
      \scriptsize{(l) $t=3.178$ ms}
    \end{minipage}
    
    \centering
    \begin{minipage}{0.32\textwidth}
      \includegraphics[width=\textwidth]{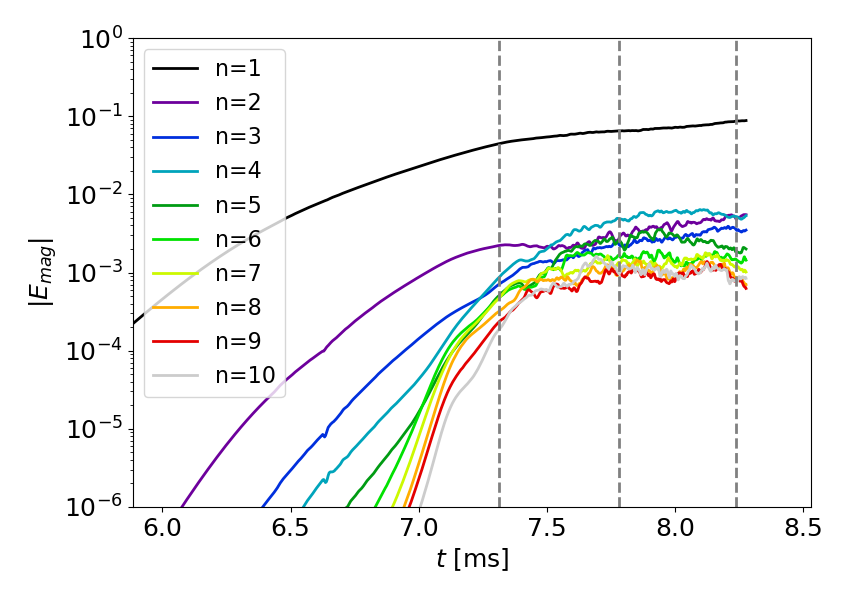}
      \centering
      \scriptsize{(m) $I_p=5.25\ MA$}
    \end{minipage}
    \begin{minipage}{0.19\textwidth}
      \includegraphics[width=\textwidth]{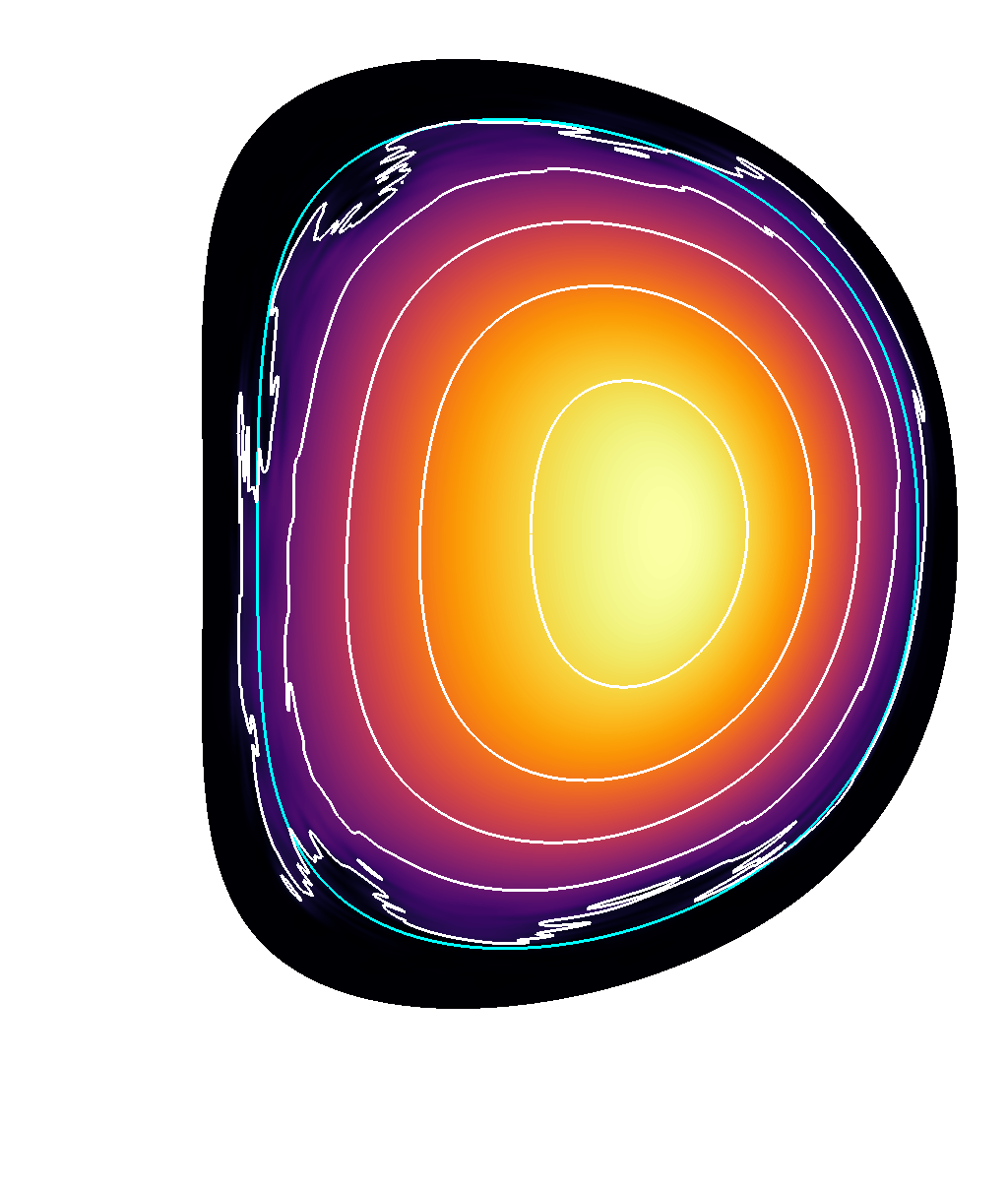}
      \centering
      \scriptsize{(n) $t=7.309$ ms}
    \end{minipage}
    \begin{minipage}{0.19\textwidth}
      \includegraphics[width=\textwidth]{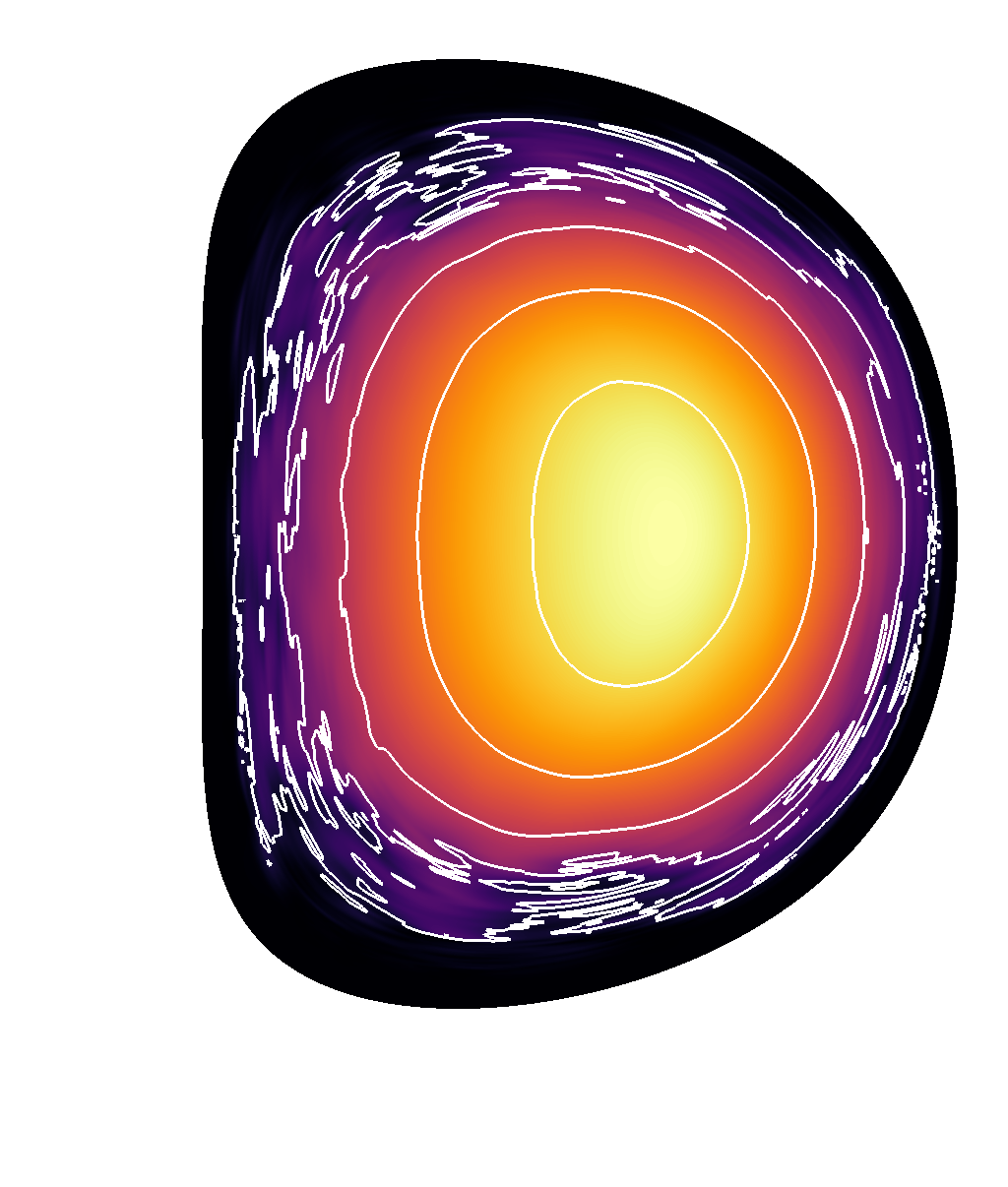}
      \centering
      \scriptsize{(o) $t=7.899$ ms}
    \end{minipage}
    \begin{minipage}{0.19\textwidth}
      \includegraphics[width=\textwidth]{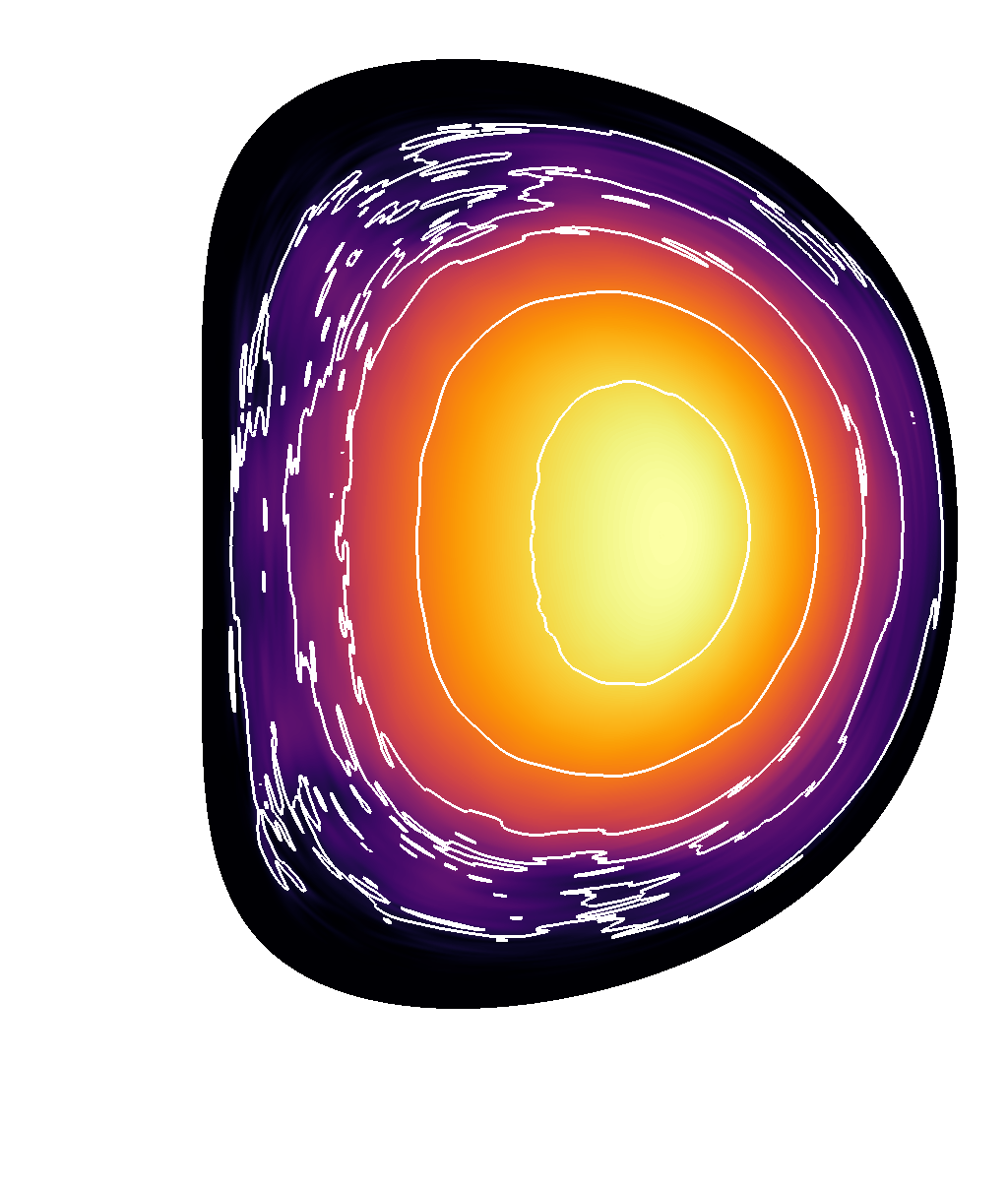}
      \centering
      \scriptsize{(p) $t=8.238$ ms}
    \end{minipage}
    
    \centering
    \begin{minipage}{0.32\textwidth}
      \includegraphics[width=\textwidth]{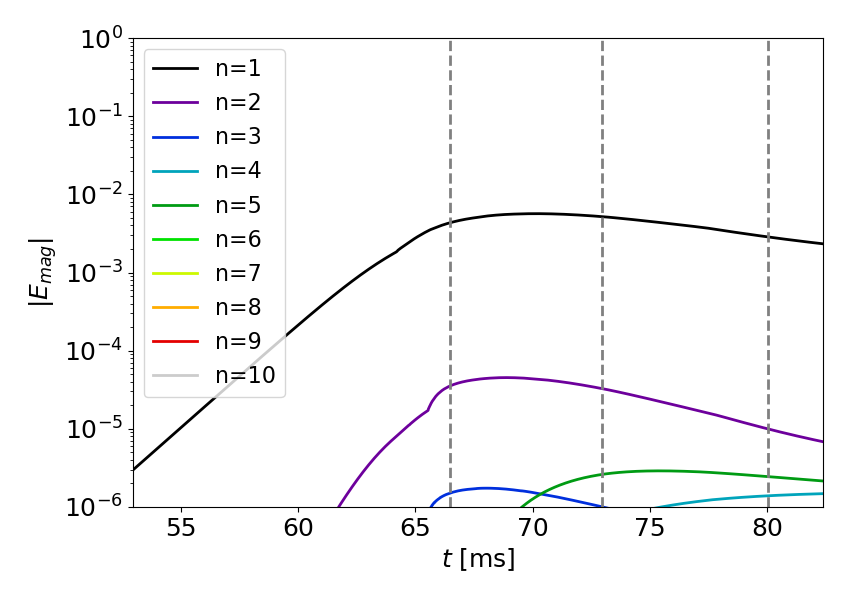}
      \centering
      \scriptsize{(q) $I_p=4.00\ MA$}
    \end{minipage}
    \begin{minipage}{0.19\textwidth}
      \includegraphics[width=\textwidth]{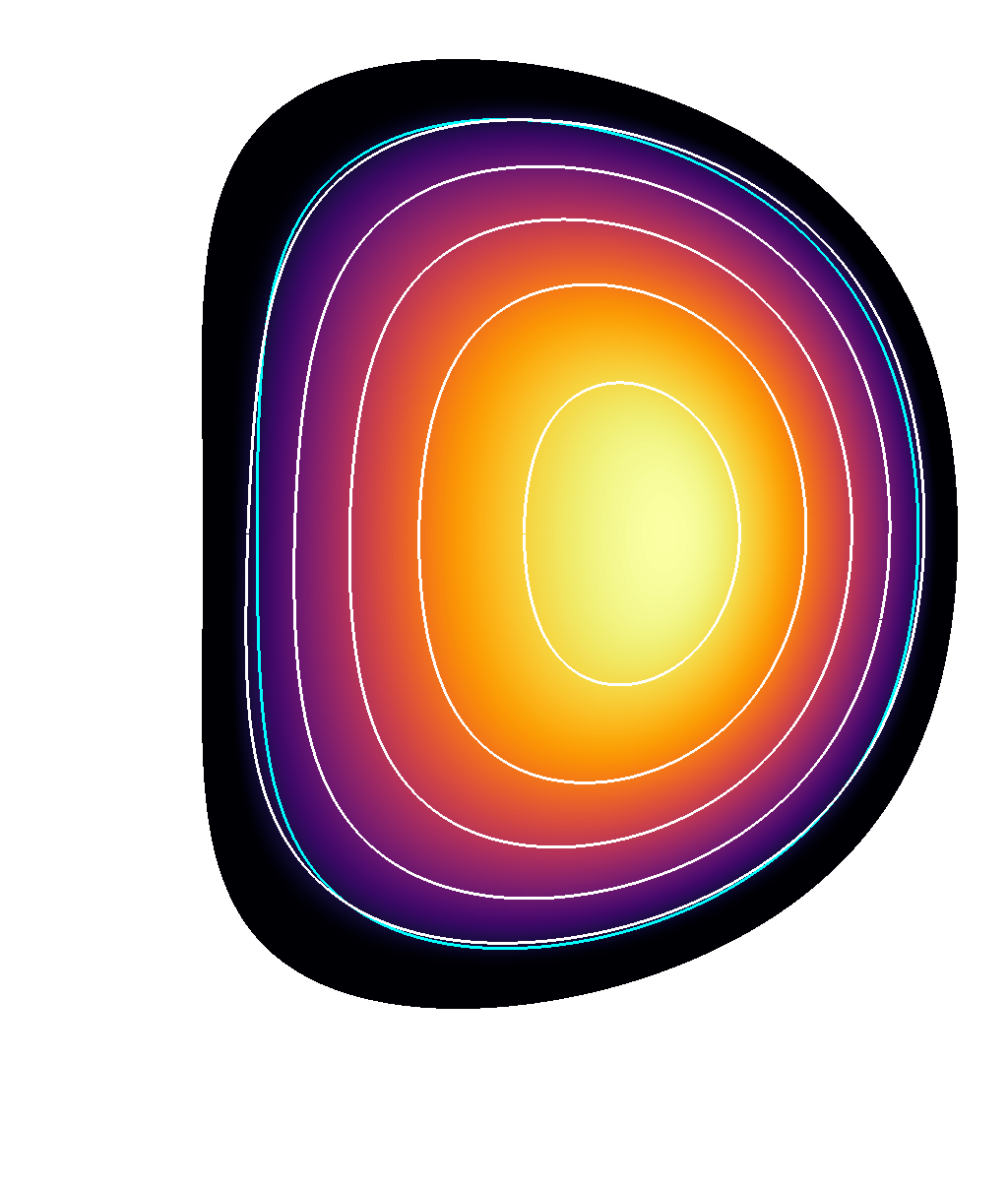}
      \centering
      \scriptsize{(r) $t=66.495$ ms}
    \end{minipage}
    \begin{minipage}{0.19\textwidth}
      \includegraphics[width=\textwidth]{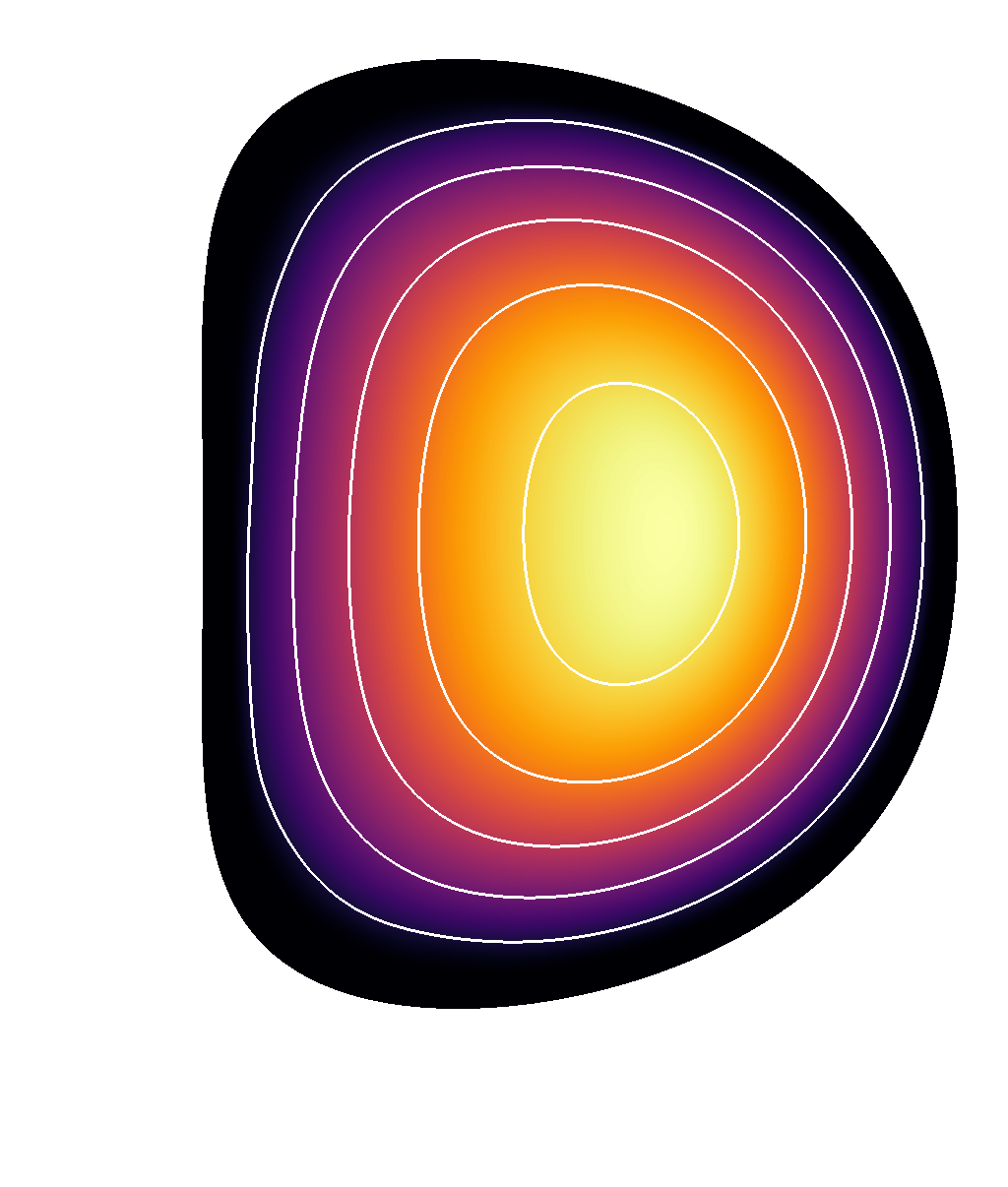}
      \centering
      \scriptsize{(s) $t=72.980$ ms}
    \end{minipage}
    \begin{minipage}{0.19\textwidth} 
        \hspace{\textwidth}
        \includegraphics[width=\textwidth]{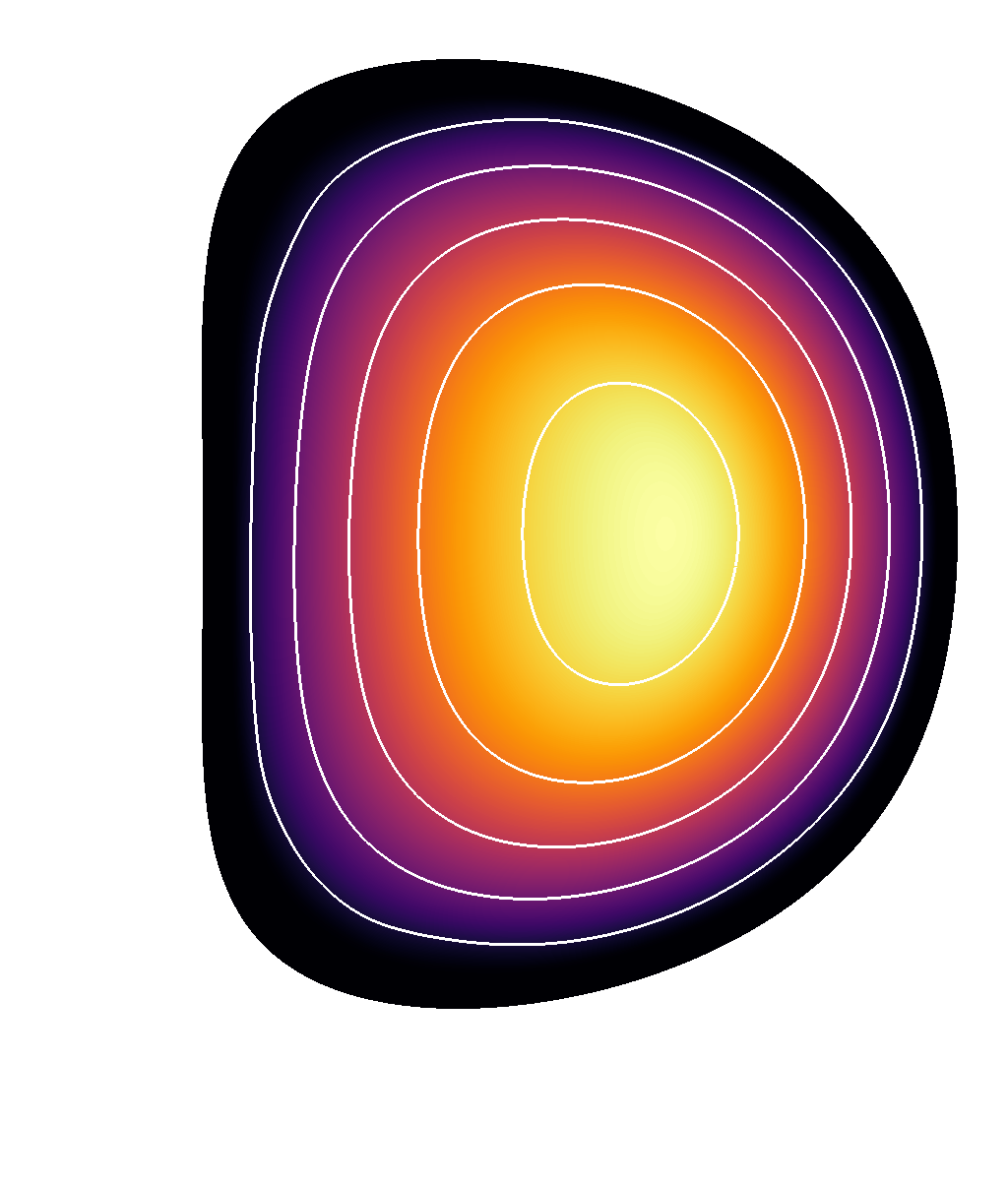}
        \centering
        \scriptsize{(s) $t=80.029$ ms}
    \end{minipage}
    
    \caption{Energy evolution of the toroidal harmonics, alongside pseudocolour plots of normalised temperature, during the dynamics. To show the approximate deformation of the plasma, contour lines of the particle number density at $n_{p}=0.1$, $0.3$, $0.5$, $0.7$, and $0.9$ in normalised units are also plotted. The initial $n_{p}=0.1$ contour is given in light blue in (b), (f), (j), (n) and (r) to show the initial deformation of the equilibrium.}
    \label{fig:kink_dynamics}
\end{figure*}

The general dynamics of the external kink are shown in Figure \ref{fig:kink_dynamics} using time traces of the energy, and pseudocolour plots of the temperature, in normalised units. Contours of density are also overlayed on the temperature plots. The low toroidal harmonics lead the instability, resulting in an initial saturation of the external kink, while higher toroidal harmonics continue to rise after the saturation of the dominant modes. As these higher harmonics saturate, the outer region of the plasma becomes increasingly chaotic, leading to a loss of thermal energy.

In the equivalent tokamak approximation, the $n=2$ mode leads the instability, before being suppressed by the $n=1$ mode, \textcolor{black}{as can be seen from the change in the density contours in Figure \textcolor{black}{\ref{fig:kink_dynamics}} (b) and (c), which trace the (4, 2) and (2, 1) mode structures that form.} The saturation of the $n=1$ mode is followed by a relatively stationary phase where the higher toroidal harmonics continue to grow, and saturate. The combination of multiple toroidal harmonics that are all linearly unstable leads to significant loss of confinement, and heat transport through perpendicular convection, and parallel conduction in the outer region of the plasma.

As $I_p$ decreases, the drive for the linear instabilities is reduced, and higher toroidal harmonics become increasingly stable, such that the $n=1$ mode initially drives the other harmonics. This can be seen to lead to similar, but milder dynamics in the virtual current case with $I_p=8.25\ MA$\textcolor{black}{, when compared to the equivalent tokamak. The $I_p=6.75\ MA$ and $5.25\ MA$ cases follow the same initial trend, such that the $n=1$ kink is again milder than the higher $I_p$ cases.} However the $n=4$ and $n=5$ modes play a larger role after the initial saturation, especially in the $I_p=6.75\ MA$ case. Here, it can be seen that there is a transient peaking of the $n=4$ and $n=5$ modes around $t=2.9\ ms$. This indicates that the kink triggers further internal instabilities led by these modes. A combination of this effect and the longer timescale of the instability lead to larger thermal losses by the end of the simulation, as can be seen from the constricted temperature plot in the $I_p=6.75\ MA$ case at $t=3.178\ ms$. Despite the longer time scale, the thermal losses in the $I_p=5.25\ MA$ case are smaller. This indicates that the combined internal and external instabilities are stronger at $I_p=6.75\ MA$.

The dynamics at $I_p=4.00\ MA$ are much milder. \textcolor{black}{Comparing the density contours in Figure \ref{fig:kink_dynamics} (r) with the initial contour, shown in light blue, it can be seen that the initial (2, 1) kink saturates with a relatively small deformation of the plasma.} The internal perturbations led by the $n=5$ mode close to the plasma edge do not penetrate significantly into the plasma region, or lead to secondary deformations of the plasma edge. The time scale of the $I_p=4.00\ MA$ case is orders of magnitude longer than at $I_p=6.75\ MA$, with much milder thermal losses that are unlikely to lead to a disruption without additional contributions to the destabilisation of the plasma that have not been modeled. 

\subsection{Internal Modes} \label{sec:internal_modes}

\begin{figure*}
    \centering
    \begin{minipage}{0.4\textwidth}
      \includegraphics[width=\textwidth]{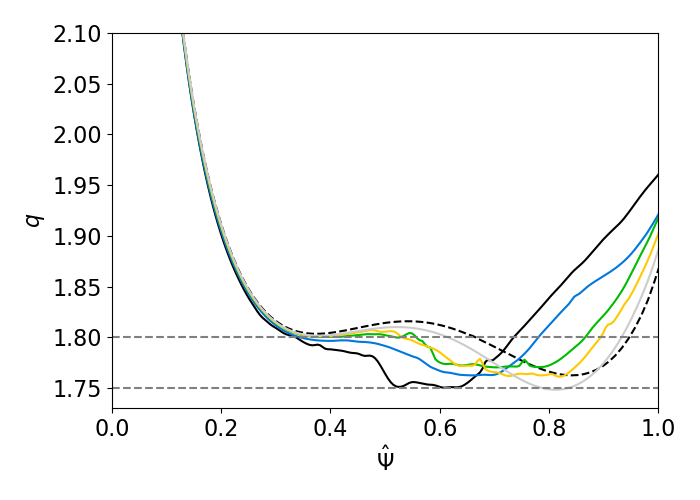}
      \centering
      \scriptsize{(a) $\hat t=21.5$}
    \end{minipage}
    \begin{minipage}{0.4\textwidth}
      \includegraphics[width=\textwidth]{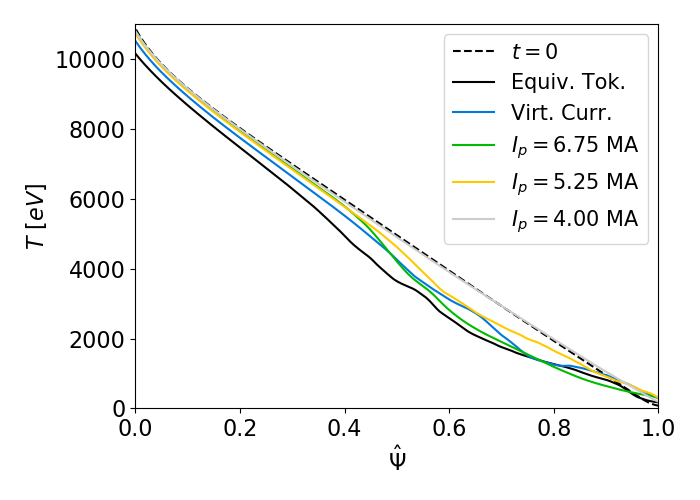}
      \centering
      \scriptsize{(b) $\hat t=21.5$}
    \end{minipage}
      
    \begin{minipage}{0.4\textwidth}
      \includegraphics[width=\textwidth]{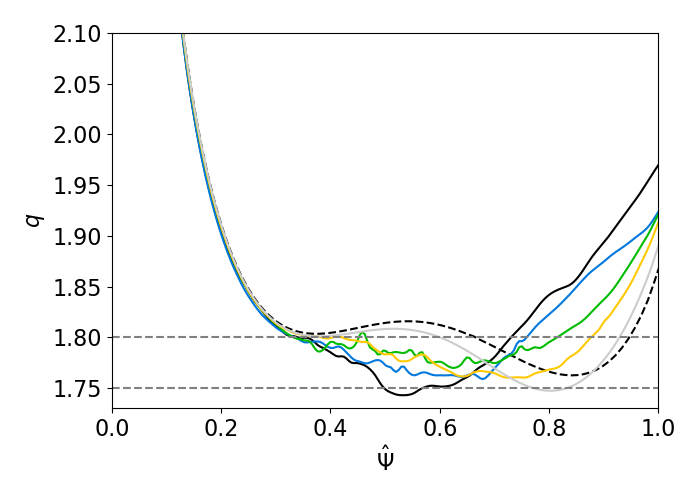}
      \centering
      \scriptsize{(c) $\hat t=23.0$}
    \end{minipage}
    \begin{minipage}{0.4\textwidth}
      \includegraphics[width=\textwidth]{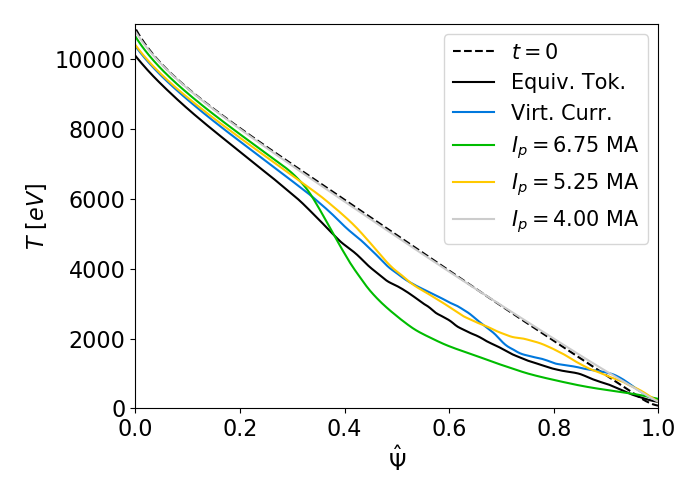}
      \centering
      \scriptsize{(d) $\hat t=23.0$}
    \end{minipage}
    
    \caption{Safety factor profile (a) and (c) and temperature profiles (b) and (d) averaged on the n=0 normalised poloidal flux, $\hat \Psi$, surfaces at $\hat t=21.5$ (a-b) and $\hat t=23.0$ (c-d). Island structures can be observed in the profiles, indicating internal modes have been triggered by the external kink. Flat regions are observed in q profiles, around the $q=1.8$ and $q=1.75$ surfaces. \textcolor{black}{Note that the $t=0$ line refers to $q_{vc}(t=0)$, which is the correct starting profile for all plotted cases, except for the equivalent tokamak, where it is approximately correct.}}
    \label{fig:temperature_radial}
\end{figure*}

The internal modes that are triggered can be observed in Figure \ref{fig:temperature_radial} as island structures in the surface averaged radial plots of temperature. Magnetic islands lead to large parallel heat transport, which leads to a local flattening of the temperature profile around rational surfaces \cite{meskat2001analysis}. Two time points are shown during and after the strong MHD activity observed for the $I_p=6.75\ MA$ case. The surfaces that are averaged over are defined by the $n=0$ flux surfaces. The q profile in particular is therefore only approximate, but gives an indication of where the rational surfaces are located. At the time slices shown, island structures can be inferred from the temperature plots around points where the profiles are partially flattened. The q profile suggests that the (9, 5) and (7, 4) rational surfaces are important in the dynamics. 

To explain why the transport of thermal energy out of the plasma is largest at $I_p=6.75\ MA$, it is observed that the q profile is flattest in this case both during and after the dynamics around the $q=1.8$ and $1.75$ rational surfaces. The external kink redistributes the plasma current, moving the peak in local current density at the plasma edge both into the plasma and out into the vacuum. Overall, this leads to a flattening of the q profile \textcolor{black}{across the outer plasma and vacuum region}. In the equivalent tokamak, current redistribution occurs up to the region of strong negative shear in the plasma core. \textcolor{black}{Because the q profile is approximately the inverse of the current density profile, this can be shown using Figure \ref{fig:temperature_radial} (a) and (c) where the q profile in the region of negative shear remains unchanged.} The radial extent of this process reduces with $I_p$, as the kink dynamics become milder. 

For the $I_p=6.75\ MA$ case, the redistributed current acts to mostly flatten the outer region of negative shear ($0.6<\hat \Psi<0.8$) in the q profile. This allows the $n=4$ and $5$ modes to dominate in the mid-region of the plasma, such that their island structures overlap. As shown in the temperature plots from a later point in time, the temperature profile relaxes significantly after this point, while in the other cases, the islands are too far apart, and so less thermal energy is transported out of the plasma. 

\begin{figure*}
    \centering
    \begin{minipage}{0.275\textwidth}
      \includegraphics[width=\textwidth]{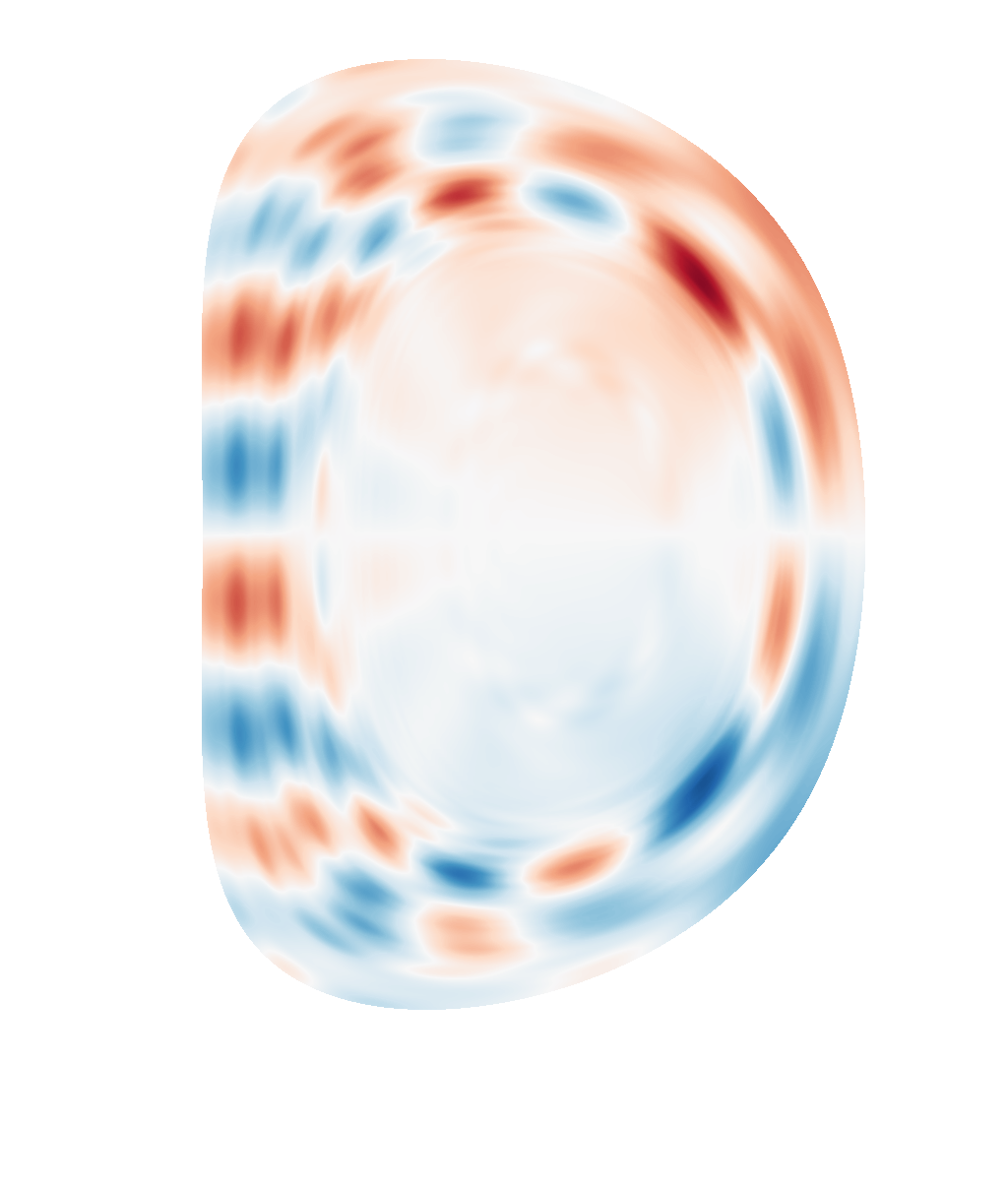}
      \centering
      \scriptsize{(a)}
    \end{minipage}
    \begin{minipage}{0.275\textwidth}
      \includegraphics[width=\textwidth]{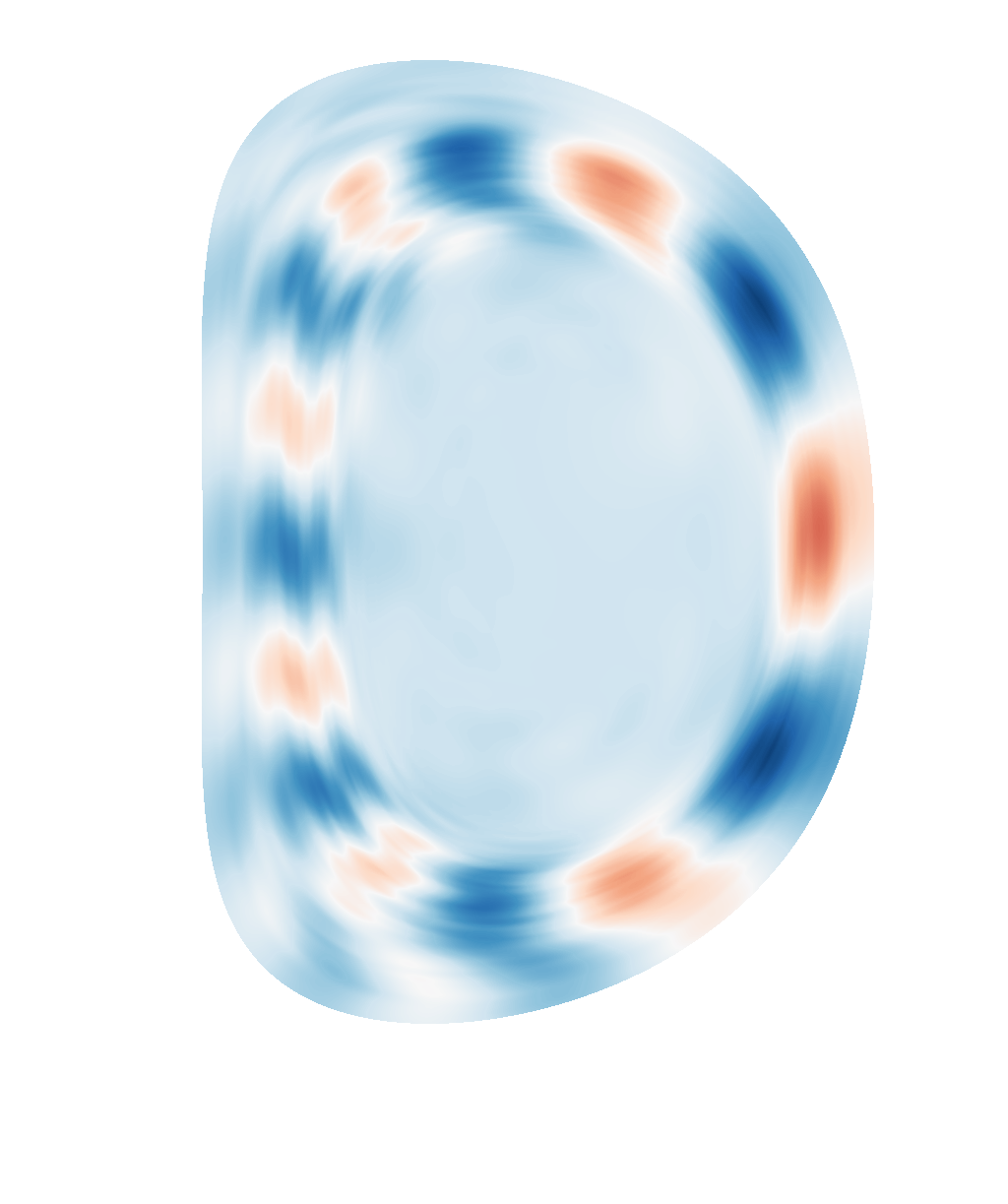}
      \centering
      \scriptsize{(b)}
    \end{minipage}
    \begin{minipage}{0.275\textwidth}
      \includegraphics[width=\textwidth]{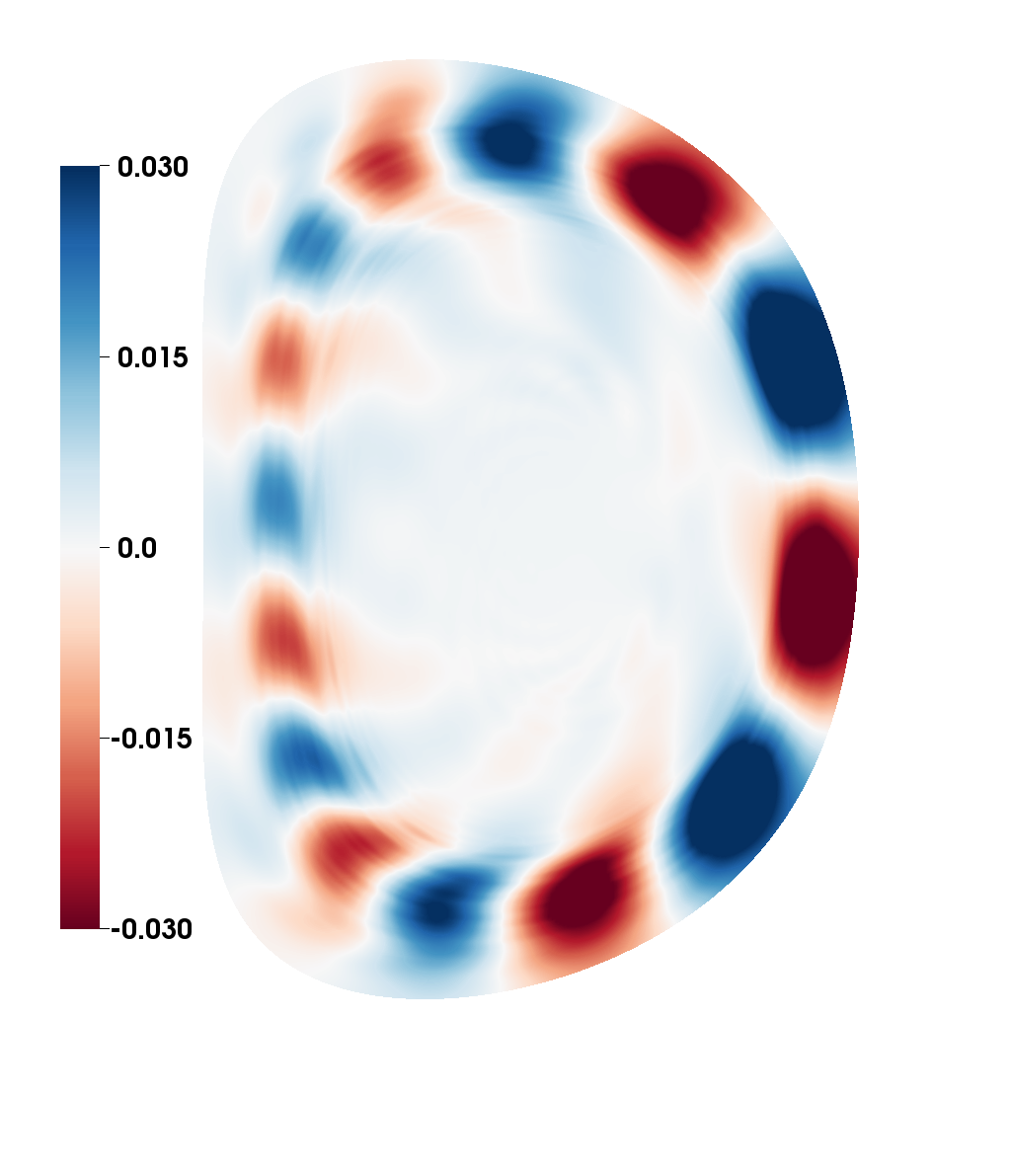}
      \centering
      \scriptsize{(c)}
    \end{minipage}
    
    \begin{minipage}{0.275\textwidth}
      \includegraphics[width=\textwidth]{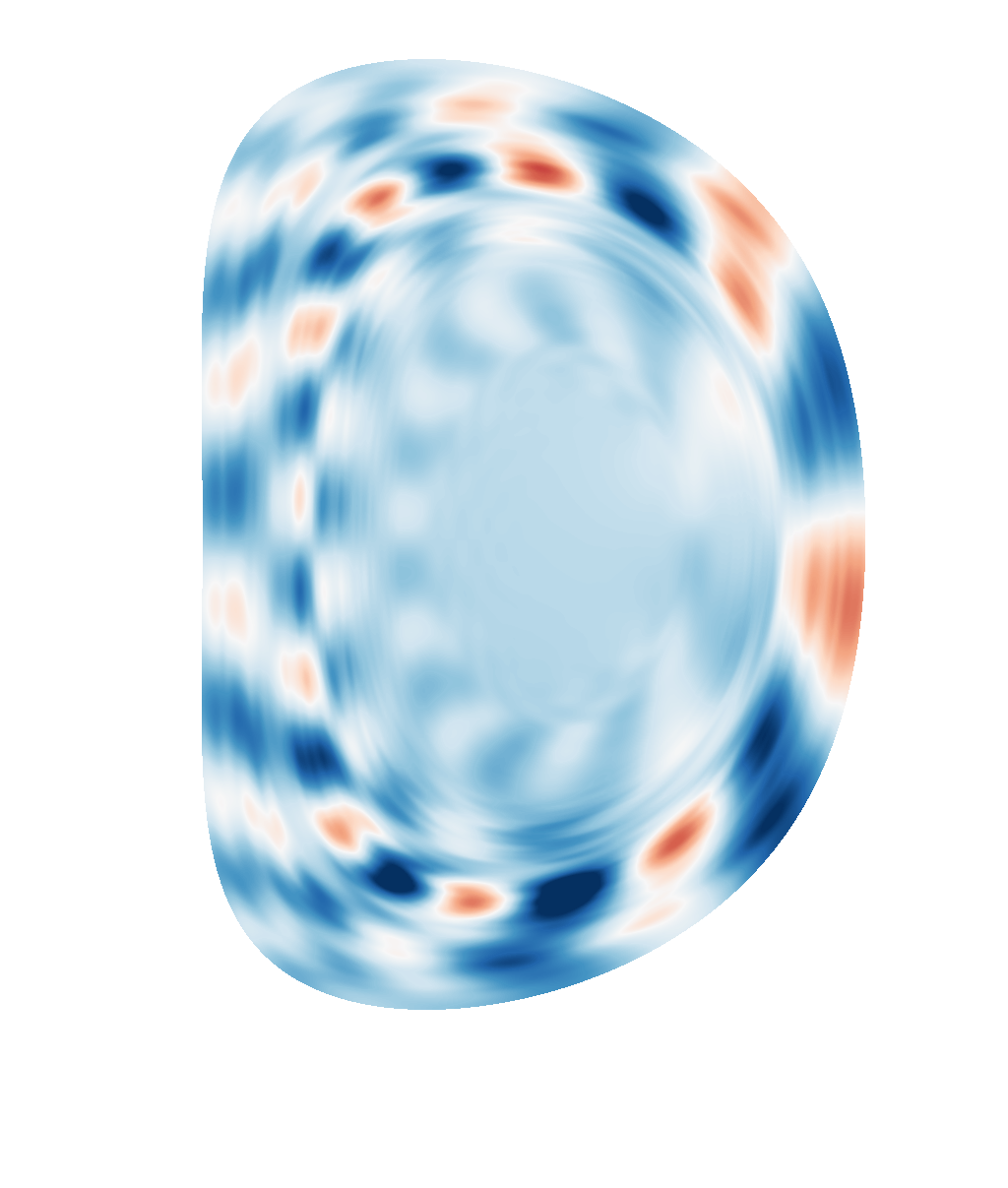}
      \centering
      \scriptsize{(d)}
    \end{minipage}
    \begin{minipage}{0.275\textwidth}
      \includegraphics[width=\textwidth]{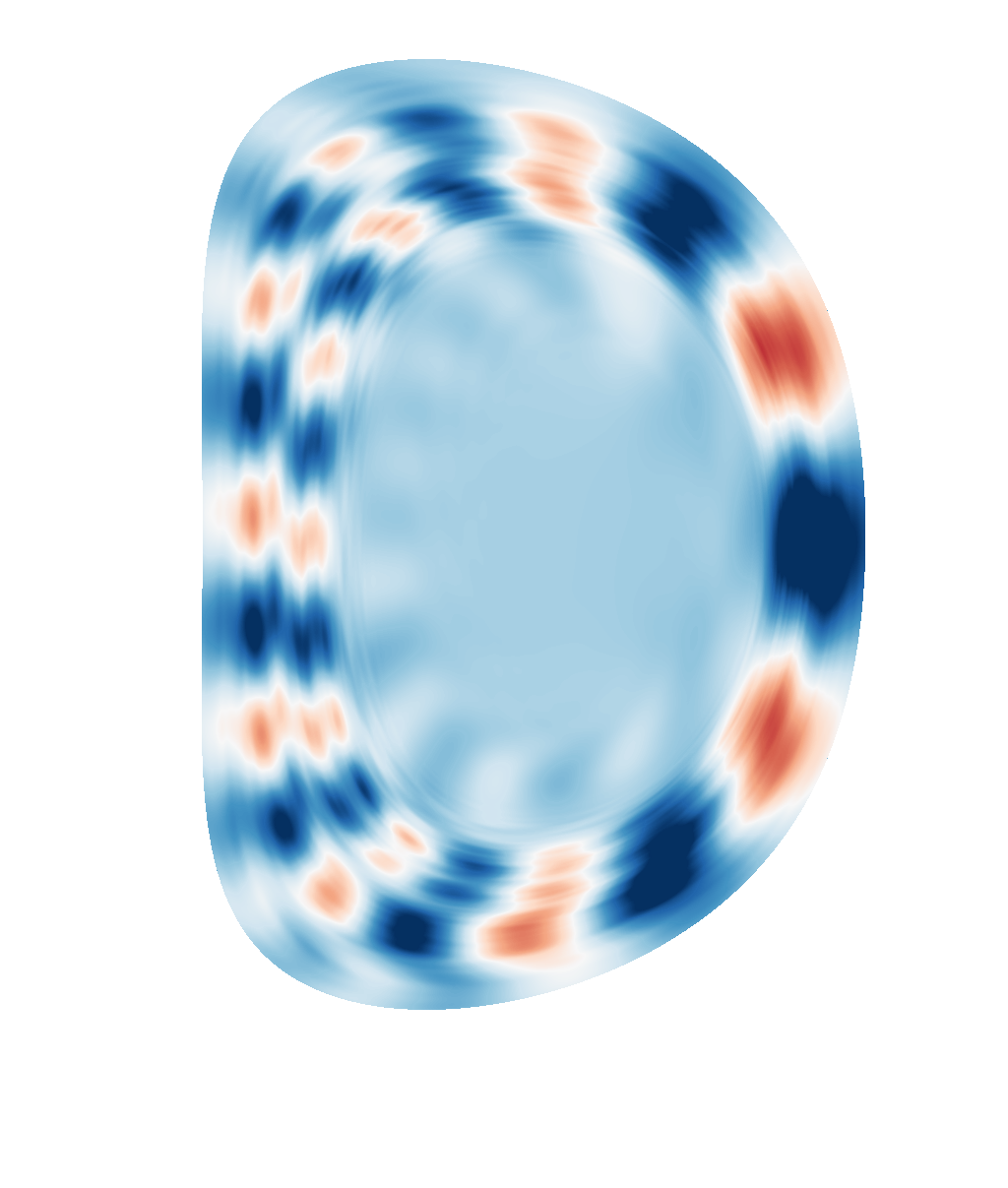}
      \centering
      \scriptsize{(e)}
    \end{minipage}
    \begin{minipage}{0.275\textwidth}
      \includegraphics[width=\textwidth]{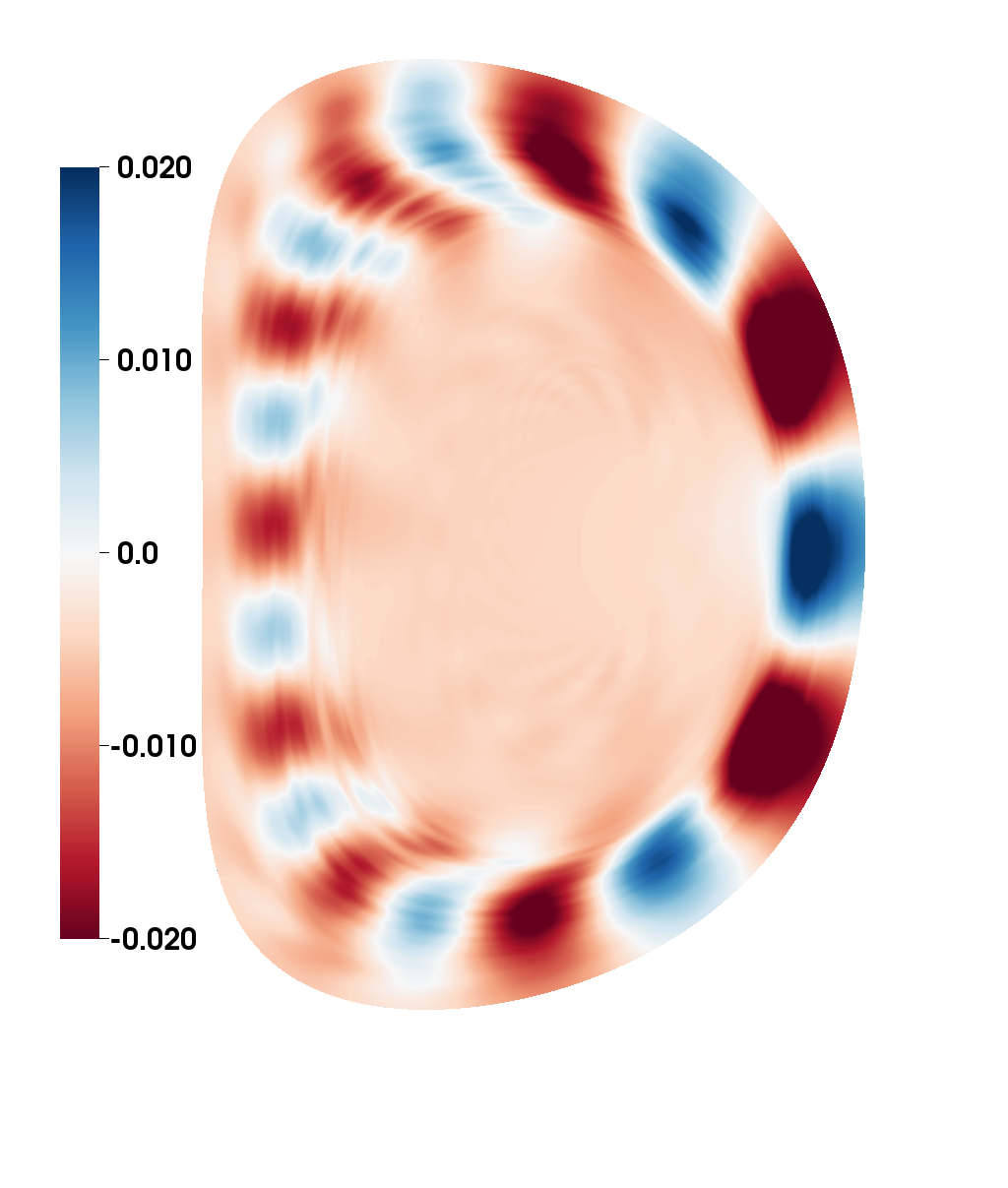}
      \centering
      \scriptsize{(f)}
    \end{minipage}
    
    
    \caption{Pseudocolour plots of the $n=4$ (a-c) and $n=5$ (d-f) components of the $m\neq0$ poloidal flux perturbation for the equivalent tokamak (a and d), $I_p=6.75\ MA$ (b and e) and $I_p=5.25\ MA$ (c and f) cases, at $\hat t=22.48$. (7, 4) and (9, 5) internal modes have been triggered in all cases.}
    \label{fig:flux_contours}
\end{figure*}

The significance of the (9, 5) and (7, 4) modes can also be shown using pseudocolour plots of the poloidal flux, as in Figure \ref{fig:flux_contours}. The contours show only the $n=4$ and $n=5$ components of the flux for the equivalent tokamak, $I_p=6.75\ MA$, and $I_p=5.25\ MA$ cases. All three cases show (7, 4) and (9, 5) structures, indicating that internal modes have been triggered. In the tokamak case, (8, 4) and (10, 5) structures can also be observed suggesting that the (2, 1) external kink still interferes with the internal modes. 

The modes that are triggered are likely to be either resistive double tearing modes, or infernal modes. There is typically a transition between the two modes depending on how close the plasma is to the ideal stability threshold \cite{charlton1989resistive}. The modes are normally distinguished by differing dependences of the linear growth rate on resistivity, which cannot be analysed for the nonlinearly triggered modes in this paper. It can be seen in Figure \ref{fig:kink_dynamics} that at $I_p=6.75\ MA$, the $n=4$ and $n=5$ energies fall during the period where significant thermal energy is lost, indicating that the modes might be pressure driven. This observation and the apparent occurrence of the (7, 4) mode in plasmas that do not seem to have a $q=1.75$ rational surface give some indication that the nature of these modes might be infernal.

\subsection{Plasma Ergodisation} \label{sec:ergodisation}
\begin{figure*}
    \centering
    \begin{minipage}{0.31\textwidth}
      \centering
      \includegraphics[width=\textwidth]{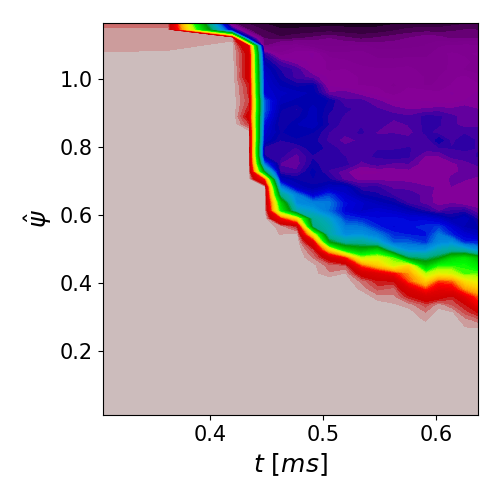}
      \scriptsize{(a) Equiv. Tok.}
    \end{minipage}
    \begin{minipage}{0.31\textwidth}
      \centering
      \includegraphics[width=\textwidth]{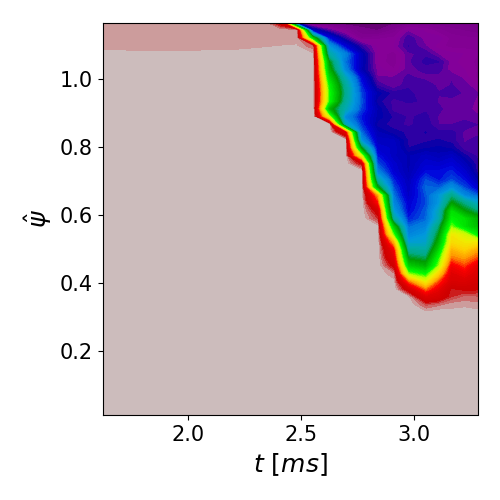}
      \scriptsize{(b) $I_p=6.75\ MA$}
    \end{minipage}
    \begin{minipage}{0.31\textwidth}
      \centering
      \includegraphics[width=\textwidth]{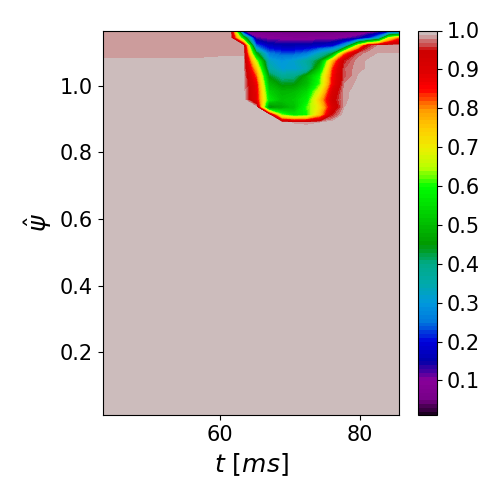}
      \scriptsize{(c) $I_p=4.00\ MA$}
    \end{minipage}

    \caption{Plots of normalised connection length to the simulation domain boundary at the end of the simulated timeframe.}
    \label{fig:connection_length}
\end{figure*}

The connection length of magnetic field lines to the simulation boundary can be used as a metric for the loss of confinement. Field lines are traced for a maximum of 100 toroidal turns around the torus, or until they connect with the simulation boundary. The connection length is determined for 1000 poloidal points on each of 51 radial surfaces, defined by the $n=0$ component of the poloidal flux. The normalised harmonic mean is calculated for each surface, and plotted over time in Figure \ref{fig:connection_length}. \textcolor{black}{The same normalisation is used in each panel, so that the quantitative value of the connection length is comparable.}

The dynamics is split into two phases. The initial ergodisation of the kink is characterised by a sudden loss in confinement near the plasma edge. As $I_p$ increases, this initial stochastisation driven by the external kink penetrates further into the core region. Later in time, there is a gradual loss of confinement further inside the plasma as MHD activity grows on the internal rational surfaces. After the strong internal modes in the $I_p=6.75\ MA$ case, it can be seen that there is a partial recovery of the internal surfaces, \textcolor{black}{from $t=3.0$ to $3.175\ ms$, where the contours of the normalised connection length between 0.4 and 0.8, shift outwards. It is expected that if the other simulations were continued further, a similar partial recovery may be observed once a comparable amount of thermal energy is lost, and the drive for the internal modes reduces. With respect to the overall dynamics, such a partial recovery is considered relatively minor, but gives some indication of how the internal modes have evolved.}

At $I_p=4.00\ MA$ the initial kink only penetrates up to $\hat\psi\approx 0.9$, and does not lead to a complete loss of confinement in this region. The subsequent internal modes break up the flux surfaces close to the plasma edge, but confinement is maintained over most of the plasma. Later in time, the edge current density gradient and total plasma current reduce due to resistive effects. This causes the edge safety factor to rise, and the drive for the instability to relax. As such, the energy in the toroidal perturbations decays, and the associated magnetic islands shrink, there is a partial of recovery of confinement in the simulation domain, such that losses are mostly in the vacuum region.

\begin{figure*}
    \centering
    \begin{minipage}{0.29\textwidth}
      \centering
      \includegraphics[width=\textwidth]{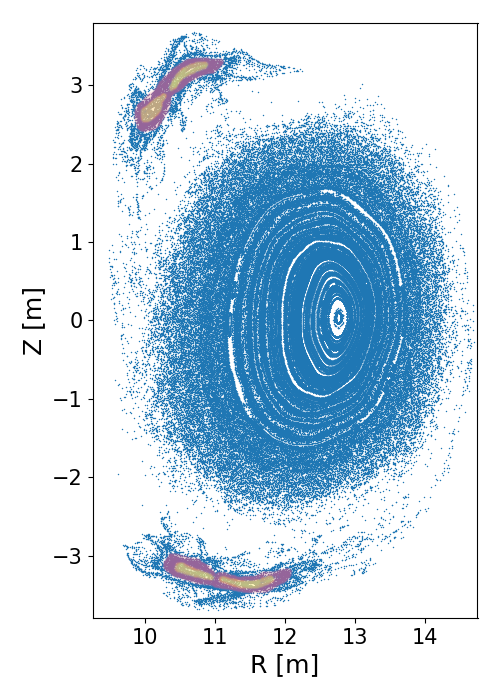}
      \scriptsize{(a) Equiv. tok.}
    \end{minipage}
    \begin{minipage}{0.29\textwidth}
      \centering
      \includegraphics[width=\textwidth]{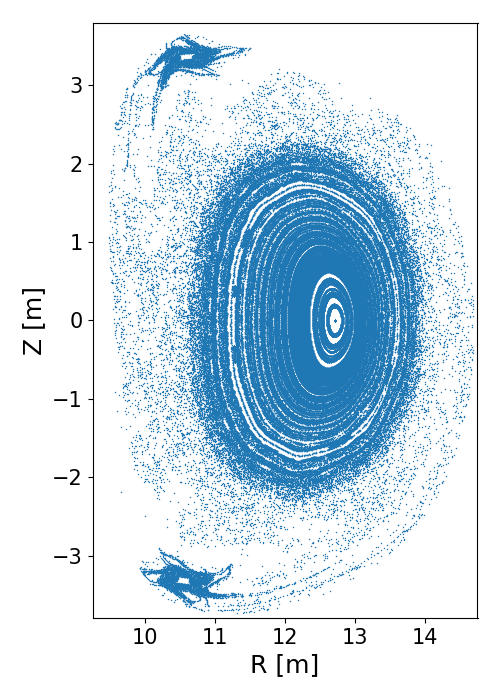}
      \scriptsize{(b) $I_p=6.75\ MA$}
    \end{minipage}
    \begin{minipage}{0.29\textwidth}
      \centering
      \includegraphics[width=\textwidth]{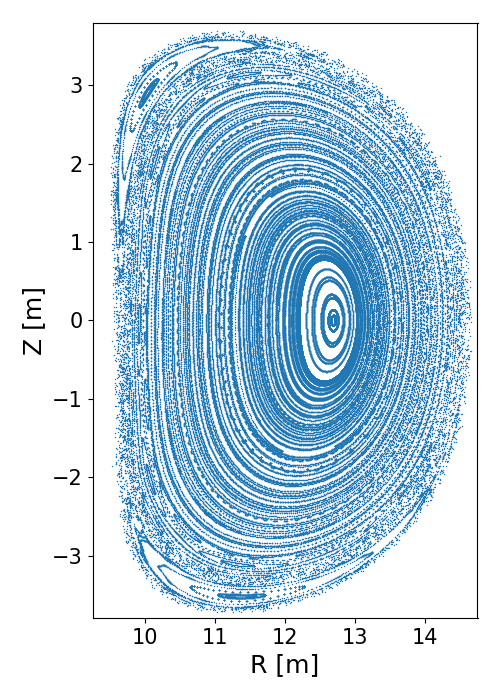}
      \scriptsize{(c) $I_p=4.00\ MA$}
    \end{minipage}
    
    \caption{Poincar\'e plots taken at $\hat t=23.0$ for the equivalent tokamak (a), $I_p=6.75\ MA$ (b) and $I_p=4.00\ MA$ case. \textcolor{black}{Each of the plots in this Figure use the same starting locations for field lines within the plasma. (a) and (b) use more field lines in the vacuum region to capture the island structures. The interaction between the dominant (2, 1) (red) and subdominant (4, 2) (yellow) islands is highlighted in (a).}}
    \label{fig:poincare_comparison}
\end{figure*}

Poincar\'e plots are shown in Figure \ref{fig:poincare_comparison} at the end of the simulated time. A dominant (2, 1) island structure is observed in all cases. In the equivalent tokamak approximation, the island structure is deformed by the remnants of the sub-dominant (4, 2) mode\textcolor{black}{, marked in yellow in Figure \ref{fig:poincare_comparison} (a)}. Relatively little structure can be observed in the region where confinement is lost. This is because field lines very quickly connect to the simulation boundary, via the large (2, 1) magnetic islands at the edge. \textcolor{black}{This fast connection has also been observed in other studies involving external kinks \cite{artola20203d}}. 


In the equivalent tokamak, the radial extent of the (2, 1) instability is larger than in cases with lower plasma current. This can also be seen from the visible kink in the core flux surfaces compared to the virtual current cases. The (2, 1) deformation of the plasma reduces with $I_p$. \textcolor{black}{In the $I_p=6.75\ MA$ case, it can be seen that there is more significant stochastisation in the mid-region of the plasma than in the equivalent tokamak, as a result of the internal modes discussed in Section \ref{sec:internal_modes}}.

\begin{figure}
    \centering
    \includegraphics[width=0.4\textwidth]{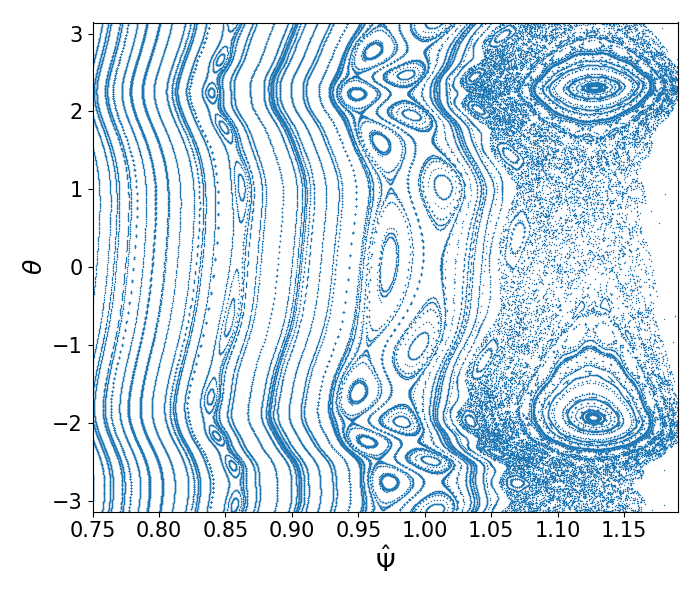}
    \caption{\textcolor{black}{Poincar\'e plots of the plasma edge taken at $\hat t=23.0$ for the $I_p=4.00\ MA$ case. Multiple (9, 5) and (7, 4) magnetic islands near the plasma edge can be observed.}}
    \label{fig:Ip_65_poinc}
\end{figure}

Similar to the connection length plots, the Poincar\'e plot of the $I_p=4.00\ MA$ case show relatively little deformation of the plasma. Large islands are still apparent, but they are much less ergodic, which leads to better confinement of the field lines within the external (2, 1) island structure. Smaller (7, 4) and (9, 5) island chains have been identified just inside the plasma region, as shown in Figure \ref{fig:Ip_65_poinc}, but they remain at low enough energies to only mildly degrade the confinement in the outer region of the plasma.

\section{Conclusion} \label{sec:conclusion}
External kinks have been modeled with JOREK nonlinearly in axisymmetric approximations of a QA configuration. This approach has been validated in the linear phase against CASTOR3D, showing reasonable agreement in the growth rates and eigenfunctions \textcolor{black}{of simple and advanced stellarators}, when compared with a 3D code. These results support the use of this model as a preliminary tool in understanding the dynamics of such configurations. 

\textcolor{black}{To explore the influence of larger fractions of external rotational transform on the nonlinear dynamics, this parameter was artificially increased.} In the nonlinear phase, the violence of the initial instability is reduced with increased contribution of virtual currents to the rotational transform, as measured by the current spike observed in the plasma\textcolor{black}{, and the initial loss in confinement due to the kink instability}. Nonlinearly triggered modes, in particular for this case the (7, 4) and (9, 5), mean that the loss in confinement can remain large, even after an appreciable external rotational transform has been introduced. This means that the variation of thermal losses with increasing external rotational transform can be non-monotonic. There are cases at large external rotational transform where internal modes are mild, indicating that the initial external modes could lead to a nonlinearly stable state.

Further extensions to this work should consider the dependence of the nonlinear dynamics on the edge safety factor and plasma pressure, as well as further destabilising effects such as vertical modes and impurities introduced from the plasma touching wall structures. Eventually a 3D code will be necessary to look at the detailed dynamics, especially near the stability threshold for the modeled external kinks. The results of this paper have been carried out to inform the direction of future studies with a stellarator capable 3D nonlinear code. When such a tool becomes available, the limitations of this work can be overcome.

\appendix

\section{W7-X Validation} \label{app:A}
\setcounter{section}{1}

As part of the validation of the models used in this paper, instabilities observed experimentally in W7-X have been modeled \textcolor{black}{resistively} and compared using CASTOR3D, JOREK and TM1 \cite{yu2020numerical}. These instabilities are induced by Electron Cyclotron Current Drive (ECCD) near the core of the device, which leads to the formation of two $q=1$ rational surfaces. Resistive kink and double tearing modes can be linearly unstable at these surfaces \cite{strumberger2020linear}.

\begin{figure}
    \centering
    \includegraphics[width=0.4\textwidth]{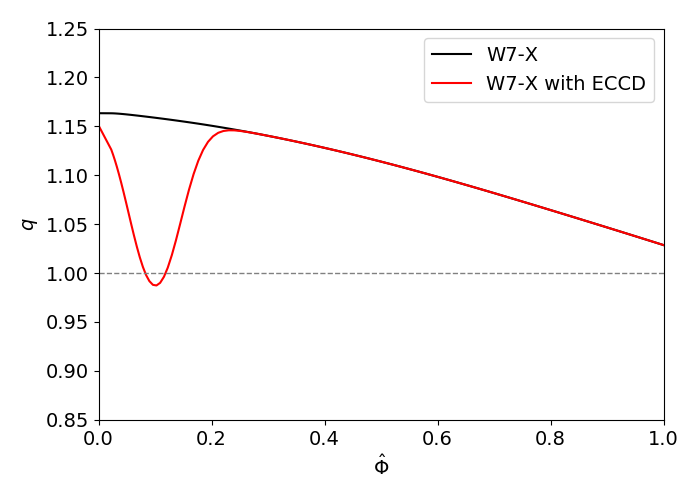}
    
    \scriptsize{(a)} 
    
    \includegraphics[width=0.4\textwidth]{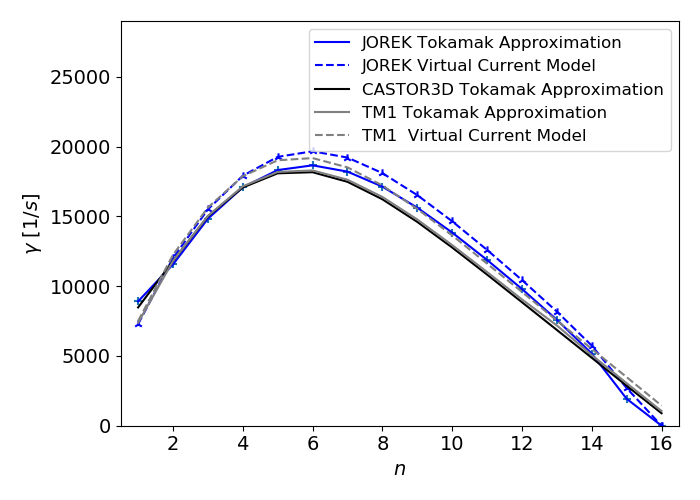}
    
    \scriptsize{(b) Circular}
    
    \includegraphics[width=0.4\textwidth]{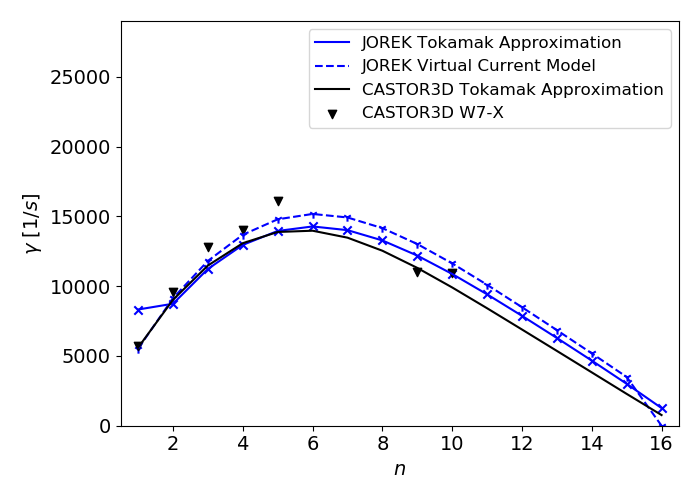}
    
    \scriptsize{(c) Axisymmetric and full 3D}
    \caption{Equilibrium q profiles (a) of W7-X equilibria with and without ECCD, and linear growth rates of W7-X resistive modes calculated with JOREK, CASTOR3D, and TM1, using a circular approximation (b) and the $n=0$ Fourier coefficients of the W7-X boundary (c). The full 3D growth rates are compared with the $n=0$ approximations of W7-X.}
    \label{fig:resistive_validation}
\end{figure}

Axisymmetric approximations of a $\beta=0$ W7-X equilibrium have been constructed, using both a circular boundary, and one defined by the $n=0$ Fourier harmonics of the last closed flux surface of the equilibrium. \textcolor{black}{The virtual current model has been constructed in such a way as to approximately preserve the plasma current of the original non-axisymmetric equilibrium.} The linear growth rates of resistive modes are compared against full 3D results from CASTOR3D, as shown in Figure \ref{fig:resistive_validation}. In this instability, the equivalent tokamak is more stable than the virtual current model, despite having a larger plasma current. The reason for this is because the current that represents the external rotational transform of a net zero current stellarator is monotonically increasing towards the plasma edge. When these currents are taken up by the plasma, as in the equivalent tokamak, they reduce the current density gradients around the rational surfaces, thereby reducing the drive for the instabilities. The sharp change in the $n=1$ growth rate of the equivalent tokamak in Figure \ref{fig:resistive_validation} (c) is because the (1, 1) mode can be either a resistive kink or double tearing mode. The results shown for CASTOR3D are for double tearing modes. In the equivalent tokamak, a resistive kink is observed in JOREK with a higher growth rate. 

The comparison shows good agreement between the three codes. \textcolor{black}{It should be noted that in CASTOR3D, the full non-axisymmetric instabilities were only modeled up to n=10. For higher n, computations become very numerically demanding as many toroidal and poloidal harmonics need to be taken into account. Due to the complicated Fourier spectra of the eigenmodes caused by the coupling of toroidal harmonics, and due to the numerical challenges posed by the considered case, the n=6, 7, and 8 modes in CASTOR3D could not be identified in this study.} 

The axisymmetric approximations and the full 3D calculations match well even at relatively high toroidal mode numbers. While geometric effects play only a minor role in the linear dynamics, using the $n=0$ Fourier components leads to a better match with the full 3D calculations than cylindrical approximations. Further studies were carried out for these resistive modes, exploring the full 3D linear stability in more detail, and the nonlinear dynamics in cylindrical geometry \cite{yu2020numerical, strumberger2020linear}. Results have shown that axisymmetric approximations work well, even for advanced stellarators, in the low beta, high aspect ratio limit.

\section*{Acknowledgements}
The authors would like to thank Qingquan Yu, for providing linear growth rates used in the validation of the models in this paper, and Florian Hindenlang, Nikita Nikulsin, Sophia Henneberg, and Carolin Nührenberg for their contributions to this work through valuable insights and fruitful discussions. This work was supported by the Max-Planck/Princeton Center for Plasma Physics. The authors acknowledge access to the EUROfusion High Performance Computer (Marconi-Fusion) and the JFRS-1 supercomputer in Japan to perform the presented simulations.

\section*{References}
\bibliographystyle{iopart-num}
\bibliography{references}

\end{document}